\documentclass[useAMS, usenatbib]{mn2e}
\usepackage{aas_macros}
\usepackage{amsmath}
\usepackage{amssymb} 
\usepackage{graphicx} 
\usepackage{float} 
\usepackage{amsfonts}
\usepackage{lmodern}
\usepackage{bm}
\pdfoutput=1

\newcommand{\bc}{\begin{center}}
\newcommand{\ec}{\end{center}}
\newcommand{\Msun}{M_{\rm \odot}}

\let\oldhat\hat
\renewcommand{\hat}[1]{\oldhat{\mathbf{#1}}}
\let\oldbullet\bullet \renewcommand{\bullet}[1][0pt]{%
\mathrel{\raisebox{#1}{$\oldbullet$}}%
}
\title[SN feedback in galaxy formation simulations]{Supernova feedback in numerical simulations of galaxy formation:
  separating physics from numerics} \author[Smith, Sijacki \& Shen]{Matthew C. Smith$^1$\thanks{E-mail: m.c.smith@ast.cam.ac.uk}, Debora
  Sijacki$^1$, Sijing Shen$^{1,2}$ \\
  $^1$ Institute of Astronomy and Kavli Institute for Cosmology,
  University of Cambridge, Madingley Road, Cambridge CB3 0HA, UK \\
  $^2$ Institute of Theoretical Astrophysics, University of Oslo, P.O. Box 1029, Blindern, N-0315, Oslo, Norway}

\begin{document}

\maketitle

\begin{abstract}
While feedback from massive stars exploding as supernovae (SNe) is thought to
be one of the key ingredients regulating galaxy formation, theoretically it is
still unclear how the available energy couples to the interstellar medium and
how galactic scale outflows are launched. We present a novel
implementation of six sub-grid SN feedback schemes in the moving-mesh code 
\textsc{Arepo}, including injections of thermal and/or kinetic energy, two
parametrizations of delayed cooling feedback and a `mechanical' feedback
scheme that injects the correct amount of momentum depending on the relevant
scale of the SN remnant resolved. All schemes make use of individually
time-resolved SN events. Adopting isolated disk galaxy setups at different
resolutions, with the highest resolution runs reasonably resolving the
Sedov-Taylor phase of the SN, we aim to find a physically motivated scheme
with as few tunable parameters as possible. As expected, simple injections of
energy overcool at all but the highest resolution. Our delayed cooling schemes
result in overstrong feedback, destroying the disk. The mechanical feedback scheme is
efficient at suppressing star formation, agrees well with the
Kennicutt-Schmidt relation and
leads to converged star formation rates and galaxy morphologies with increasing
resolution without fine tuning any parameters. However, we find it difficult
to produce outflows with high enough mass loading factors at all but the
highest resolution, indicating either that we have oversimplified the evolution of unresolved SN remnants, require other stellar feedback processes to be included, require a better star formation prescription or most likely some combination of these issues.  
\end{abstract}


\section{Introduction} \label{intro}
In the $\Lambda$CDM model of cosmology, dark matter dominates large scale
structure formation. Gas gathers in the potential wells of dark matter halos,
where it may radiatively cool and hence form stars. This baryonic matter makes
up the visible component of galaxies. This picture alone is not sufficient to
reproduce observations. A naive determination of the expected star formation
rate (SFR) based on a typical dynamical time yields excessive values. In fact,
star formation occurs on much longer timescales of the order of 20 - 100
dynamical times and has an efficiency of only a few percent (see for example
\citealt{Zuckerman1974, Williams1997, Kennicutt1998, Evans1999, Krumholz2007,
  Evans2009}). Thus, some form of feedback process or processes need to be
invoked to explain this discrepancy. At high halo masses, this may be provided
by an active galactic nucleus (AGN), but at lower masses stellar feedback
dominates, mainly from high mass stars in the form of stellar winds,
supernovae (SNe), photoionisation and radiation pressure. 

It is worth emphasising that it is not enough merely to halt the conversion of
gas to stars as some fraction of the accreted mass must be ejected out of the
galaxy. Without strong feedback, the baryon fraction of galaxy models are far
in excess of observations \citep[e.g.][]{White1991, Keres2009}. In addition,
the observed circumgalactic medium (CGM) is enriched with metals, requiring
baryons to have made it out from sites of star formation embedded within the
galaxies themselves \citep[e.g.][]{Aguirre2001, Pettini2003, Songaila2005a,
  Songaila2006, Martin2010}. Such outflows are observed, moving at hundreds of
$\mathrm{km\ s^{-1}}$ (see for example the review by
\citealt{Veilleux2005}). Observations suggest that the ratio of mass outflow
rate to SFR (i.e. the mass loading factor) must be at least unity or above
\citep[see e.g.][]{Bland-Hawthorn2007,Schroetter2015}. This is borne out by
theoretical models \citep{Oppenheimer2006, Sales2010, Genel2012, Shen2012, Dave2013, Puchwein2013,
  Vogelsberger2013, Hopkins2014a,Mitra2015,Christensen2016}.

While the observational evidence for SFR regularisation and outflow driving is
manifest, precisely how these mechanisms operate is as yet unclear. Here
numerical hydrodynamic simulations of galaxy formation are useful
tools. Unfortunately, the scales on which stellar feedback operates (parsecs
and below) are many orders of magnitude below the characteristic scales of
galaxies and the surrounding CGM we wish to simulate.  

This ideally needed dynamic range is beyond the reach of current
state-of-the-art simulations, requiring the representation of the effects of unresolved
processes by adopting so-called `sub-grid' schemes. For large scale
cosmological simulations, where the interstellar medium (ISM) is 
poorly resolved, these schemes must rely on dealing with stellar feedback at a
high level of abstraction. For example, such approaches
may use effective equations of state to approximate the effect of a multiphase
ISM pressurised by feedback energy \citep[e.g.][]{Springel2003, Teyssier2010}. 
Winds are often added with some predetermined mass loading, either temporarily
decoupling outflowing gas from the hydrodynamics, imposing some minimum
threshold temperature of the wind ejecta or switching off radiative cooling
losses for a given amount of time, to ensure sufficiently strong driving
\citep[e.g.][]{Springel2003, Oppenheimer2006,DallaVecchia2008,Sales2010}. Such
schemes are presently necessary to model large samples of galaxies but
lack predictive power on small scales. However, if the target of a simulation
is a single galaxy, either in an idealised, isolated setup or in a
cosmological `zoom-in', then the higher resolution available enables the
adoption of more explicit models of feedback, allowing investigations of how
feedback arises on comparatively smaller scales to be carried out.

Nevertheless, even in the individual galaxy simulations the resolution
requirements are still severe. In the case of SNe, one of the main obstacles
to physically consistent coupling of SN energy to the ISM is the ability to
resolve the Sedov-Taylor phase of a SN remnant. The expansion of
SN remnants has been well studied and can be broken down into several
distinct regimes \citep{Woltjer1972}. The SN explosion ejects material into
the ISM with typical kinetic energies of $10^{51}$ ergs. The SN ejecta expands
relatively unhindered into the ISM as long as the mass swept up in the forward
shock is smaller than the ejecta mass. Concurrently, the reverse shock
heats up the gas inside the remnants, leading to high temperatures and
pressures. Radiative losses are negligible so the expansion proceeds
adiabatically into the surrounding medium, which marks the Sedov-Taylor phase. During
this phase, the momentum of the remnant is boosted by up to an order of
magnitude \citep{Taylor1950, Sedov1959, Chevalier1974, Cioffi1988,
  Blondin1998, Kim2015, Martizzi2015}. Eventually, a thin, dense shell builds
up at the shock front and radiative losses become important, triggering the
transition from energy conserving to momentum conserving evolution. Because of
the large increase in momentum that occurs during the adiabatic expansion,
merely injecting energy (whether thermal or kinetic) into the surrounding gas
without properly resolving the length scales corresponding to the Sedov-Taylor phase
results in a severe underestimation of the amount of momentum imparted to the
ISM. \cite{Kim2015} found that the minimum requirements for correctly
modelling the evolution of SNe in this manner are that the shell forming
radius, $r_\mathrm{SF}$, is resolved by three resolution elements. For
evolution in an inhomogeneous medium, they quantified that
$r_\mathrm{SF}=30\ \mathrm{pc}\ (n/\mathrm{cm^{-3}})^{-0.46}$, meaning that at
a density of $100\ \mathrm{cm^{-3}}$ (typical mean density for a giant
molecular cloud) the resolution requirement is $\sim1\ \mathrm{pc}$. Failure
to meet these requirements when using a simple injection of SN energy will
result in `overcooling' as the energy is radiated away before it can do any
work.

Many strategies to circumvent this issue exist in the literature. One implicit solution
is to inject the energy of several SNe simultaneously resulting in more
energetic explosions. Often, this is achieved simply by
injecting a star particle's entire feedback energy budget at once, either
instantaneously or after some predetermined delay time. The strength of this
effect is therefore tied to the star particle mass. Alternatively, a stochastic
feedback approach, such as that proposed in \cite{DallaVecchia2012}, may be adopted, in which SN energy is
redistributed in time and space to produce fewer, more energetic events
guaranteeing the overcooling problem is avoided. Such schemes conserve the
total feedback energy in a globally averaged sense, but lose the connection to
individual SN events and are not spatially consistent. If the simulation is of
a coarse resolution and the structure of the ISM is not resolvable/of
interest, this may be an acceptable compromise. 

A different class of approaches involves switching off the radiative cooling
of gas that has received feedback energy, enforcing an adiabatic phase, for
some length of time \citep[see e.g.][]{Stinson2006, Governato2010, Agertz2011,
  Teyssier2013}. The length of time by which cooling is delayed is somewhat of
a tunable parameter, particularly in simulations with coarse resolution, but
physically motivated parameters can be arrived at by analytical arguments (see
for example the appendix of \citealt{Dubois2015}). A downside of `delayed
cooling' models is that the radiative cooling of the gas is physically
correct, even if the resolution effects responsible for the overcooling
phenomena are not. Thus it is possible for gas to occupy unphysical regions of
temperature-density phase diagrams when it should have cooled. 

In alternative to the `delayed cooling' schemes, it is possible to take
account of the momentum boost in the missed adiabatic phase rather than
enforcing such a phase. Some schemes skip the Sedov-Taylor phase entirely,
putting in a bubble at some fixed radius and adjusting the kinetic energy of the gas inside
to match the analytically determined values assuming some mass loading
\citep[see e.g.][]{Dubois2008}. Others determine the stage of a remnant's
evolution that can be resolved and boost the momentum by some appropriate
factor determined either analytically \citep[e.g.][]{Hopkins2014a,Kimm2014} or
by making use of fits to high resolution simulations of SN remnant evolution
(e.g. \citealt{Martizzi2015} as employed in \citealt{Martizzi2016}). These
schemes are often referred to as mechanical feedback. They feature few (if
any) explicitly tunable parameters, but rely on assumptions about the
structure of the ISM at small scales and how the remnant will interact with
it. For example, a porous ISM structure caused by turbulence may allow the
remnant to propagate preferentially down low density channels
\citep{Iffrig2015, Kim2015, Martizzi2015, Walch2015, Li2015, Haid2016} though
the net effect of this phenomenon is not well constrained and possibly
introduces further free parameters into the model. 

Of course, SNe are not the only form of stellar feedback. It is possible for
photoionisation to break up star forming clouds prior to the first SNe
occurring \citep{Vazquez-Semadeni2010, Walch2012, Dale2014, Sales2014}. Winds
from massive stars are unable to completely disrupt $10^4-10^5\ {\rm M_\odot}$
clouds, but can carve cavities of $\sim10$~pc which may enhance subsequent SNe
feedback \citep{Dale2014}. Radiation pressure can in principle supply as much momentum as
stellar winds (see e.g. \citealt{Leitherer1999}), though it is difficult to
assess the extent to which this can be coupled to the ISM. On the one hand,
H\textsc{\,ii} regions created by massive stars will blunt the impact of radiation
pressure, rendering the ISM transparent to Lyman-limit photons, but in the
presence of dust, multiple scattering of IR photons can boost the momentum
input to the ISM by up to a few orders of magnitude \citep{Murray2010}. Using
sub-grid models of radiation pressure feedback it has been
found that boost factors of $\sim10-100$ are necessary to drive strong
outflows \citep{Hopkins2011a,Hopkins2012b, Hopkins2012a, Agertz2013, Aumer2013, 
Roskar2014, Agertz2015}. However, using full radiative hydrodynamics (RHD),
\cite{Rosdahl2015} concluded that radiation pressure is unable to drive strong
outflows in their simulations, although they are unable to resolve gas at high
enough densities to become significantly optically thick to IR
photons. Nevertheless, a simple boosting of the IR optical depths resulted in
suppressing star formation and smoothing of the disk without generating
outflows. In reality, all of these stellar feedback mechanisms will interact
in a complex manner. For example, the FIRE project \citep{Hopkins2014a} has
produced encouraging results by including multiple stellar feedback processes
in sub-grid fashion, creating realistic looking galaxies relative to
observations. However, it is clear that before trying to unpick the
interaction of different processes and their impact on galaxy formation, it is
crucial to understand the numerical consequences of the individual feedback
schemes. 

To this end, in this work, we carry out a detailed study of various flavours
of SN feedback prescriptions commonly found in the literature. We perform
simulations of idealised, isolated galaxy models, in the absence of other
feedback prescriptions and with a simple star formation law, in order to
provide as clean a comparison as possible. The schemes tested are all chosen
to work with individually time resolved SN events, providing as direct a link
to the locations and timescales of star formation as possible (e.g. we do not
consider stochastic feedback such as \citealt{DallaVecchia2012}), and are
optimised for isolated or cosmological zoom-in simulations (rather than
cosmological boxes). We carry out our fiducial simulations of a
$10^{10}\ {\rm M_\odot}$ system at three resolutions (the highest of which is
chosen to largely eliminate the overcooling problem in a simple thermal dump
scheme) in order to test convergence properties, trialing six sub-grid
feedback schemes. Having presented our main findings with respect to resulting
galaxy morphologies, SFRs, and outflow properties as function
of feedback scheme, we briefly examine how these results depend on the mass of
the galaxy and simple changes of the star formation prescription.
\vspace{-4ex}
\section{Methodology}\label{Methodology}
\subsection{Basic Code Setup}\label{Setup}
We make use of the moving-mesh code \textsc{Arepo} \citep{Springel2010} with
our own novel implementation of star formation and SN feedback (described
below). \textsc{Arepo} uses a quasi-Lagrangian finite volume technique,
solving hydrodynamics on an unstructured mesh determined by a Voronoi
tessellation of discrete mesh-generating points. These points move with the
local gas velocity (with the addition of minor corrections to allow for cell
regularisation). By moving the mesh with the fluid and employing a smoothly
varying refinement and derefinement scheme, \textsc{Arepo} is able to keep
cell masses constant (to within a factor $\sim2$). \textsc{Arepo} benefits
from many of the advantages inherent to traditional Lagrangian approaches
(e.g. smoothed particle hydrodynamics (SPH)), such as continuously varying
resolution with density and Galilean invariance, while retaining advantages of
contemporary Eulerian codes (i.e. adaptive mesh refinement (AMR)) such as more
accurate resolution of shocks, contact discontinuities and fluid instabilities
\citep{Bauer2012, Keres2012, Sijacki2012, Torrey2012a, Vogelsberger2012}. We
include radiative cooling from both primordial species and metal-lines as
presented in \cite{Vogelsberger2013}: primordial heating and cooling rates are
calculated using cooling, recombination and collisional rates from
\cite{Cen1992} and \cite{Katz1996}, while lookup tables pre-calculated with
the photoionization code \textsc{Cloudy} are used to obtain the metal cooling
rates. Note that in this work we do not include a UV background.   
\vspace{-2ex}
\subsection{Non-thermal pressure floor} \label{subsec_floor}
Failing to sufficiently resolve the Jeans length can result in artificial
fragmentation \citep{Truelove1997}. To avoid this we include a non-thermal
pressure floor to ensure that the Jeans length is resolved by $N_\mathrm{J}$
cells, i.e.
\begin{equation}
P_\mathrm{min} = \frac{N_\mathrm{J}^2 \Delta x^2 G \rho^2}{\pi \gamma}, 
\label{pressure_floor}
\end{equation}
where $\Delta x$ is the cell diameter, $\rho$ is the gas density and 
$\gamma = 5/3$ is the adiabatic index. In principle, at sufficiently high
resolution if feedback is able to entirely prevent gas from entering a phase
where it is vulnerable to artificial fragmentation, it may be possible to
avoid the use of a pressure floor. This would ideally prevent the risk of
suppressing physical fragmentation which may occur when a pressure floor is in
place. Alternatively, the star formation prescription adopted could be formulated
to ensure that gas is turned into stars before artificial fragmentation occurs.
However, as we only include SNe feedback (note that there is a delay of
$\sim 3$ Myr before the first SNe go off) and wish to study the effects of the
feedback without a more involved  method of modelling star formation (see
below), we use a pressure floor to ensure numerically meaningful gas
conditions prior to feedback.   

Various values for $N_\mathrm{J}$ can be found in the literature. We find that
the value required varies depending on choice of code, cooling prescriptions,
initial conditions, resolution and included sub-grid physics. The choice is
therefore somewhat arbitrary and often ill defined. By performing an array of
numerical experiments, we find that $N_\mathrm{J}=8$ is a reasonable choice
for $1000\ \mathrm{M_\odot}$ cell resolution (see Fig.~\ref{J_comp_hr_proj} in
Appendix~\ref{appendix_floor}). It should be noted that in the
absence of feedback, the choice of $N_\mathrm{J}$ has a significant impact on 
the total stellar mass formed (see Fig.~\ref{J_comp_dens}). For the purposes of 
the comparison of feedback implementations in this work, our choice is therefore
motivated by our requirements to avoid the opposite extremes of artificial fragmentation
or total suppression of star formation by the pressure floor.

Using a fixed value of $N_\mathrm{J}$ with different resolutions ensures that 
the Jeans length is always resolved by the same number of cells. This means that, by design,
fragmentation is allowed to occur on smaller scales as simulations move to higher resolutions
and the minimum resolvable scale decreases.
While under most circumstances this is a desirable behaviour, the resulting lack of convergence
in the absence of feedback makes a meaningful study of the resolution dependence of SN feedback
schemes impossible. Thus, for this work, we adopt the scaling
$N_\mathrm{J}=8(m_\mathrm{cell}/1000\ \mathrm{M_\odot})^{-1/3}$ such that the pressure
floor corresponds to resolving the same length-scale across all
resolutions. This results in relatively similar gas morphologies,
temperature and density distributions and SFRs in the absence
of feedback at all numerical resolutions explored (see Fig.~\ref{J_comp_res_proj} in
Appendix~\ref{appendix_floor}). With this choice of
the pressure floor scaling, starting with relatively similar disk properties
in different resolution runs we can more readily isolate how feedback operates
at different resolutions. A much more detailed discussion of the use of the
pressure floor in this work and its effects on the simulations is presented in
Appendix~\ref{appendix_floor}. 
\vspace{-2ex}
\subsection{Star formation}\label{SF}
In our model, gas is marked as star forming if it is above some density
threshold $n_{\mathrm{SF}}$. We then compute a star formation rate for
the gas based on a simple Schmidt law, using the almost ubiquitous expression
\begin{equation}
\dot{\rho}_{*} = \epsilon_\mathrm{SF}\frac{\rho}{t_{\mathrm{ff}}}, 
\label{eq:schmidt}
\end{equation}
where $\rho$ is the gas density, $\epsilon_\mathrm{SF}$ is some efficiency and 
$t_{\mathrm{ff}}=\sqrt{3\pi/32G\rho}$ is the free-fall time. We use a fiducial 
value of $n_{\mathrm{SF}}=10\ \mathrm{cm}^{-3}$ and $\epsilon_\mathrm{SF}=1.5\%$ 
\citep[chosen to match observed efficiencies in dense gas, see e.g.][and references therein]{Krumholz2007}. 
These values are kept the same across all resolutions for our fiducial simulations 
(they are an appropriate choice for all resolutions explored), with the aim of removing the 
dependence the choice of star formation law prescription and allowing us to assess the convergence properties
of the SNe schemes alone\footnote{However, see Section~\ref{sf_law} where we
present results with a higher density threshold value of
$n_{\mathrm{SF}}= 100\ \mathrm{cm}^{-3}$ or a higher efficiency of 
$\epsilon_\mathrm{SF}=15\%$}. 
We then use these rates to stochastically convert gas cells to star particles 
(representing a single stellar population).
\vspace{-2ex}
\subsection{Supernova feedback}\label{SN}
Our implementation of SN feedback is directly related to individual star
particles and discretely resolves individual SNe in time. This is in contrast
to implementations which inject energy continuously at some rate related to
the SFR and to methods in which a fixed quantity of energy per stellar mass in
injected, possibly after some delay. Injecting the energy of multiple SNe at
once will help avoid the overcooling problem (the radius of the remnant at
the end of the Sedov-Taylor phase has a dependence on injected energy as
$E^{0.29}$ \citep{Kim2015}). However, the local evolution of the ISM with time
as it evolves prior to the first SNe and as SNe occur sequentially (for
example enhancing the strength of subsequent SNe) is non-trivial. Failing to
resolve individual SNe in time potentially misses important
physics. Therefore, each timestep, for each star particle, we tabulate SN
rates, $\dot{N}_\mathrm{SN}$, as a function of age and metallicity from
\textsc{Starburst99} \citep{Leitherer1999} assuming a \cite{Kroupa2002}
IMF. We then draw the number of SNe that occur from a Poisson distribution
with a mean $\bar{N}_\mathrm{SN}=\dot{N}_\mathrm{SN}\Delta t$, where $\Delta
t$ is the timestep. We further impose a timestep limiter for star particles
such that $\bar{N}_\mathrm{SN}\ll1$ to ensure that SNe are individually
resolved in time.  

When a SN occurs, mass, metals, energy and/or momentum (depending on the
feedback scheme, see below) are deposited into the gas cell hosting the star
particle and its immediate neighbours (i.e. all cells that share a face with
the host cell). The various quantities are distributed amongst these cells
using a weighting scheme that aims to guarantee an isotropic
distribution. This contrasts with the SPH-like (mass, volume etc.) weighting
schemes commonly used in Lagrangian codes. Because higher density regions will
contain more resolution elements, such a weighting scheme will preferentially
inject feedback quantities perpendicular to the local density gradient. In the
worst case scenario, we have found that this manifests itself in the unphysical
driving of strong feedback `rings' through the plane of a thin disk, similar
to those reported in \cite{Hopkins2017a,Hopkins2018} (see our Appendix~\ref{appendix_iso}
for more details). Our weighting scheme is based on the `vector weighting'
scheme from \cite{Hopkins2017a} \citep[see][for a full derivation of the scheme]{Hopkins2018}.
Essentially, the quantities are weighted both by the solid angle
subtended by the adjoining cell face and by projection operators to enforce
isotropy. To compute these quantities we use the mesh geometry used in the
hydrodynamic calculation. For reasons of numerical simplification we take the
centre of the SNe to be the mesh generating point of the cell hosting the star
particle, rather than the star particle itself. The effects of this are small,
since by definition the star particle is spatially unresolved in the context
of the hydrodynamic resolution. Further simplifications to the
\cite{Hopkins2017a} scheme arise since the Voronoi tessellation guarantees that
cell face norms are aligned with the position vector between two mesh generating
points and that cell faces lie exactly halfway between the two mesh generating
points in that direction. Note that the centre of the cell face is not
guaranteed to lie on the line between two mesh generating points. However, we
take this as an approximation. Due to the mesh regularisation schemes used in
\textsc{Arepo}, we find this to be reasonable.

We first find the cell that contains the star particle, hereafter referred to as the host cell. For each of the neighbour cells (cells that share a face with the host cell), $i$, we determine the vector weight $\bar{\mathbf{w}}_i$ defined as
\begin{equation}
\bar{\mathbf{w}}_i = \frac{\mathbf{w}_i}{\sum_{j} \left| \mathbf{w}_j \right|} \left(1 - f_\mathrm{host} \right),
\end{equation}
where the sum over $j$ is over all neighbour cells including $i$,
\begin{equation}
\mathbf{w}_i = \omega_i \sum_{+,-} \sum_{\alpha} \left( \mathbf{\hat{x}}^{\pm}_i \right)^{\alpha} f^{\alpha}_{\pm},
\end{equation}
\begin{equation}
f^{\alpha}_{\pm} = \left\{ \frac{1}{2} \left[ 1 + \left( \frac{ \sum_j \omega_j \left| \mathbf{\hat{x}}^{\mp}_j \right|^{\alpha}}{ \sum_j \omega_j \left| \mathbf{\hat{x}}^{\pm}_j \right|^{\alpha}} \right)^2 \right] \right\}^{1/2},
\end{equation}
\begin{equation}
\omega_i = \frac{1}{2} \left( 1 - \frac{1}{ \sqrt{ 1 + 4A_i / (\pi \left| \mathbf{x}_{i} \right|^2  )   }} \right),
\end{equation}
\begin{equation}
\left( \mathbf{\hat{x}^{+}}_i \right)^{\alpha} = \left| \mathbf{x}_{i} \right|^{-1} \mathrm{MAX}\left( \mathbf{x}_{i}^{\alpha}, 0 \right)\Bigr|_{\alpha=x,y,z},
\end{equation}
\begin{equation}
\left( \mathbf{\hat{x}^{-}}_i \right)^{\alpha} = \left| \mathbf{x}_{i} \right|^{-1} \mathrm{MIN}\left( \mathbf{x}_{i}^{\alpha}, 0 \right)\Bigr|_{\alpha=x,y,z},
\end{equation}
where $f_{\mathrm{host}}$ is the fraction of feedback quantities given to the
host cell, $A_i$ is the area of the face between the neighbour and the host
cell, $\mathbf{x}_{i}$ is the position vector between the mesh generating
points of the neighbour to the host, the superscript $\alpha$ denotes the
component in a given Cartesian direction, $x$, $y$ or $z$, while the $+$ and $-$ denote components with either a positive or negative value respectively. Given a total
ejecta mass $m_\mathrm{ej}$ and SN energy $E_{\mathrm{SN}}$, the total
momentum to be injected in the rest frame of the star particle is
\begin{equation} 
p_\mathrm{tot} = \sqrt{2 m_\mathrm{ej} f_\mathrm{kin} E_{\mathrm{SN}}}\,, 
\label{p_tot}
\end{equation}
where $f_\mathrm{kin}$ is the fraction of ejecta energy that is in
kinetic form (which we vary throughout this work). The portion of mass,
momentum and total energy each cell receives (in the rest frame of the star
particle) is then
\begin{equation}
\Delta m_i = \left| \bar{\mathbf{w}}_i \right| m_\mathrm{ej}
\end{equation}
\begin{equation}
\Delta \mathbf{p}_i = \bar{\mathbf{w}}_i p_\mathrm{tot}
\end{equation}
\begin{equation}
\Delta E_i = \left| \bar{\mathbf{w}}_i \right| E_\mathrm{SN}\,.
\end{equation}
Transforming back to the simulation frame (i.e. the rest frame of the simulated volume), the momentum and energy fluxes become
\begin{equation}
\Delta \mathbf{p}_i^{\prime} = \Delta \mathbf{p}_i + \Delta m_i \mathbf{v}_* \label{eq:momentum}
\end{equation}
\begin{equation}
\Delta E_i^{\prime} = \Delta E_\mathrm{i} + \frac{1}{2\Delta m_i} \left( \left| \Delta \mathbf{p}_i^{\prime} \right|^2 - \left| \Delta \mathbf{p}_i\right|^2 \right)\,,
\end{equation}
where $\mathbf{v}_*$ is the velocity of the star particle in the simulation
frame. Note that this implicitly deals with any momentum cancellation,
i.e. the `lost' kinetic energy becomes thermal energy. 

The host cell receives the following mass and energy
\begin{equation}
\Delta m_\mathrm{host} =  f_\mathrm{host} m_\mathrm{ej}
\end{equation}
\begin{equation}
\Delta E_\mathrm{host} = f_\mathrm{host} E_\mathrm{SN} + \frac{1}{2}
f_\mathrm{host} m_\mathrm{ej} \left| \mathbf{v}_* - \mathbf{v}_\mathrm{host}\right|^2\,. 
\label{eq:host_energy}
\end{equation}
The final term in equation~(\ref{eq:host_energy}) assumes complete
thermalisation of the kinetic energy carried by the star
particle. Empirically, we find that the mean number of neighbouring cells is
$\sim20$. We therefore adopt $f_\mathrm{host}=5\%$ to evenly distribute
feedback quantities. In practice, we find a very weak dependence on the value
of $f_\mathrm{host}$. In simulations described as containing no feedback, the
host cell and neighbours receive mass and metals as described above but their
energy and momentum are not altered. We adopt $m_\mathrm{ej} =
10\ \mathrm{M}_{\rm \odot}$, of which $2\ \mathrm{M}_{\rm \odot}$ is in metals (i.e. a metallicity of 0.2), and
$E_\mathrm{SN}=10^{51}\ \mathrm{ergs}$ throughout this work. 

\subsubsection{Classical feedback schemes}
For the purpose of this work, we refer to schemes that employ a simple dump of
thermal and/or kinetic energy as classical feedback schemes. These use the
methods outlined above with some value of $f_\mathrm{kin}$. For pure thermal
feedback, we use $f_\mathrm{kin}=0$. For pure kinetic feedback\footnote{This
  should not be confused with other schemes sometimes referred to as `kinetic'
  that boost the momentum input by some fixed mass loading factor
  \citep[see][]{Dubois2008, Kimm2015, Rosdahl2017}.}, we use
$f_\mathrm{kin}=1$. We also trial a mixed feedback scheme that uses
$f_\mathrm{kin}=0.28$, which distributes the energy into the ratio expected
during the Sedov-Taylor phase (e.g. see \cite{Ostriker1988,Cioffi1988} for
analytical arguments, also see \cite{Kim2015} for an example of this in a
numerical simulation).
\vspace{-2ex}
\subsubsection{Delayed Cooling}
Additionally to the classical feedback schemes we adopt a feedback
prescription based on the delayed cooling method of \cite{Teyssier2013}. This
method aims to take into account (sub-grid) non-thermal processes that might
store some of the feedback energy, such as, for example, unresolved
turbulence, magnetic fields and cosmic rays. The timescales on which
these processes dissipate energy is longer than the cooling time of the
thermal component, so energy may be stored for longer and released
gradually. We introduce a new variable $u_{\mathrm{FB}}$ which is used to
record the energy density from feedback that gas particles currently possess
and is advected with the gas flow, acting as a passive Lagrangian tracer
(which is to say it is not directly involved in the hydrodynamics). When a gas
cell is involved in a SN event, feedback energy is injected as described above
with $f_\mathrm{kin}=0$ (i.e. entirely thermally apart from the momentum
conserved from the star particle). The amount of energy received is also added
to $u_{\mathrm{FB}}$. This feedback energy store is allowed to dissipate as
\begin{equation}
\frac{\mathrm{d}u_{\mathrm{FB}}}{\mathrm{d}t} = - \frac{u_{\mathrm{FB}}}{t_{\mathrm{diss}}},
\end{equation}
where $t_{\mathrm{diss}}$ is some dissipation timescale as in \cite{Teyssier2013}. Note that $u_{\mathrm{FB}}$ can also be increased if the gas cell is involved in another SN event. We compute an effective velocity dispersion corresponding to the feedback energy,
\begin{equation}
\sigma_{\mathrm{FB}} = \sqrt{2u_{\mathrm{FB}}}\,,
\end{equation}
and the gas particle is not allowed to cool if this velocity dispersion is above some threshold. Following \cite{Teyssier2013} we use
\begin{equation}
\Lambda=0\ \mathrm{if}\ \sigma_{\mathrm{FB}}>10\ \mathrm{km\ s}^{-1}\,. 
\end{equation}
The motivation for switching off cooling when $\sigma_{\mathrm{FB}}$ is above
this threshold is to mimic a non-thermal contribution to the pressure. Once
the non-thermal contribution becomes comparable to the thermal contribution,
cooling is allowed to continue as normal. We also trial a larger threshold
value of $100\ \mathrm{km\ s^{-1}}$ in Appendix \ref{appendix_DFB}.

We use a fixed value for the dissipation time of 10 Myr, as in \cite{Teyssier2013}. We also trial a variable dissipation time, based on the effective crossing time for the turbulence within a cell,
\begin{equation}
t_\mathrm{diss}=\frac{\Delta x}{\sigma_\mathrm{FB}}, 
\label{eq:diss_time}
\end{equation}
where $\Delta x$ is the diameter of the cell.

\subsubsection{Mechanical feedback}
In this feedback scheme we aim to account for the $PdV$ work done during the
Sedov-Taylor phase of the SNe remnant expansion, where the momentum can be
boosted by around an order of magnitude. The correct momentum to couple to the
ISM therefore depends on the stage of the expansion (alternatively
parametrized in terms of swept up mass), limited by the final momentum at the
point when the remnant exists the Sedov-Taylor phase. Several such schemes
exist in the literature \citep[see e.g.][]{Hopkins2014a,Hopkins2017a,Hopkins2018,Kimm2014,Kimm2015,Martizzi2015}. In our
mechanical feedback scheme, momentum calculated in
equation~(\ref{eq:momentum}) is enhanced as follows
\begin{equation}
\Delta \mathbf{p}_i^{\prime\prime} = \Delta \mathbf{p}_i^{\prime} \mathrm{MIN} \left[ \sqrt{1 + \frac{m_i}{\Delta m_i}}, \frac{p_\mathrm{fin}}{p_\mathrm{tot}} \right]\,,
\end{equation}
where $p_\mathrm{fin}$ is the momentum as the remnant transitions to the
snowplough phase \citep{Blondin1998, Thornton1998, Geen2015, Kim2015,
  Martizzi2015}; following \cite{Kimm2015} we adopt
\begin{equation}
p_\mathrm{fin} = 3 \times 10^5\ \mathrm{km\ s^{-1}\ \mathrm{M}_{\rm \odot}}\,
E^{16/17}_{51} n^{-2/17}_\mathrm{H} Z^{\prime-0.14}, 
\label{p_fin}
\end{equation}
where $E_{51} = \left(E_\mathrm{SN} / 10^{51}\ \mathrm{ergs}\right) = N_\mathrm{SN}$, $n_\mathrm{H}$ is the hydrogen number density and $Z^\prime = \mathrm{MAX}\left( Z/Z_{\rm \odot}, 0.01\right)$ is the metallicity in solar units. Note that we calculate $\Delta \mathbf{p}_i^{\prime\prime}$ for each cell involved in the SN event {\it independently}.
\vspace{-4ex}
\section{Simulations}
\subsection{Initial conditions and simulation details}
We simulate isolated galaxies comprising of a stellar and gas disk, a stellar
bulge, a hot gaseous atmosphere and a static background potential representing
the dark matter 
component. The dark matter follows an NFW profile \citep{Navarro1997} with
concentration parameter $c = 10$ and spin parameter $\lambda = 0.04$ for all
galaxies simulated. The baryonic component is
generated using \textsc{MakeNewDisk} \citep{Springel2005a}. The disk density
profile is exponential in radius. The stellar disk has a Gaussian vertical density profile with a scale height
$0.1$ times the scale radius. The stellar bulge has a scale length $0.1$ times
the scale radius of the disk. The collisionless particles comprising the
stellar disk and bulge in the initial conditions do not contribute to stellar
feedback. The vertical structure of the gas disk is determined so as to obtain initial
hydrostatic equilibrium. 

We simulate three galaxies in this work with properties described in
Table~\ref{Table1}. The majority of simulations in this work are of a galaxy
with a total mass of $10^{10}\ {\rm M_\odot}$. We refer to this setup as the
fiducial galaxy. This setup is comparable to the G8 model in
\cite{Rosdahl2015}. We also simulate two additional systems, `Small' and
`Large', which are an order of magnitude lower and higher in mass,
respectively. For our fiducial model, we initialise the disk with a temperature of
$10^4\ \mathrm{K}$. We scale the initial disk temperature with the virial temperature of
the halo for the `Small' and `Large' models. This is to avoid an initially vertically diffuse disk for
the `Small' model, while also maintaining consistency between the models\footnote{While the disks will rapidly cool
once the simulation start and the vertical structure will settle into an equilibrium configuration, we find that
if the initial disk structure is too vertically diffuse in the `Small' model, the resulting collapse is too severe
and does not allow the disk to settle satisfactorily.}. The gas in the disk is initialised with a metallicity of
$Z=0.1\,\mathrm{Z_\odot}$. To roughly represent the CGM we include a
hot gas atmosphere of uniform density
$n_\mathrm{H}=10^{-6}\ \mathrm{cm}^{-3}$, uniform temperature $10^{6}\ \mathrm{K}$ and zero metallicity.

Gas cells and star particles (both those present in the initial conditions and newly created
stars) share the same mass, $2000\ \mathrm{M_\odot}$, $200\ \mathrm{M_\odot}$ and $20\ \mathrm{M_\odot}$ for
the low, intermediate and high resolution runs, respectively. At the highest
resolution, the mass of cells/star particles approaches that of the total ejecta mass per
SN ($10\ \mathrm{M_\odot}$). We have confirmed that the refinement/derefinement scheme in place in
\textsc{Arepo} is sufficiently effective such that star particles always have enough mass to provide
the full ejecta mass budget in all but a negligible fraction of SN events ($\lesssim1\%$).
Table~\ref{Table2} contains details of every simulation
presented in the main body of this work (i.e. not including simulations
presented in the appendices), including the galaxy model used, the resolution
and number of cells/particles, gravitational softenings, additional star
formation  and feedback parameters and the mass of new stars formed after 250
Myr. 

\begin{table}
\caption{Initial conditions of the three disk galaxies modelled in this work,
  referred to as `Small', `Fiducial' and `Large'. We list the total mass of
  the galaxy, $M_{\mathrm{tot}}$, (excluding the CGM, which is negligible),
  the halo virial radius, $R_{\mathrm{vir}}$, the mass in the disk component,
  $M_{\mathrm{disk}}$, the fraction of the disk component in gas,
  $f_{\mathrm{gas}}$, the scale radius of the disk, $r_{\mathrm{s}}$, the
  scale height of the stellar disk, $h_{\mathrm{s}}$, the mass of the stellar bulge,
  $M_{\mathrm{bulge}}$, the initial metallicity of the gas in the disk,
  $Z_{\mathrm{disk}}$, (the CGM initially contains no metals), 
  the initial temperature of the disk, $T_{\mathrm{disk}}$.}\label{Table1}
\bc
\begin{tabular}{llll}
\hline\hline
& Small & Fiducial & Large \\ \hline
$M_{\mathrm{tot}}$ & $10^{9} \mathrm{M}_{\rm \odot}$ & $10^{10} \mathrm{M}_{\rm \odot}$ & $10^{11} \mathrm{M}_{\rm \odot}$\\
$R_{\mathrm{vir}}$ & $16.3\ \mathrm{kpc}$ & $35.0\ \mathrm{kpc}$ & $75.5\ \mathrm{kpc}$\\
$M_{\mathrm{disk}}$ & $3.5\times10^{7} \mathrm{M}_{\rm \odot}$ & $3.5\times10^{8} \mathrm{M}_{\rm \odot}$ & $3.5\times10^{9} \mathrm{M}_{\rm \odot}$\\
$f_{\mathrm{gas}}$ & $0.5$ & $0.5$ & $0.5$\\
$r_{\mathrm{s}}$ & $0.33\ \mathrm{kpc}$ & $0.70\ \mathrm{kpc}$ & $1.52\ \mathrm{kpc}$\\
$h_{\mathrm{s}}$ & $33\ \mathrm{pc}$ & $70\ \mathrm{pc}$ & $152\ \mathrm{pc}$\\
$M_{\mathrm{bulge}}$ & $3.5\times10^{6}\ \mathrm{M}_{\rm \odot}$ & $3.5\times10^{7}\ \mathrm{M}_{\rm \odot}$ & $3.5\times10^{8}\ \mathrm{M}_{\rm \odot}$\\
$Z_{\mathrm{disk}}$ & $0.1\ \mathrm{Z_\odot}$ & $0.1\ \mathrm{Z_\odot}$ & $0.1\ \mathrm{Z_\odot}$\\
$T_{\mathrm{disk}}$ & $2.1\times10^3\ \mathrm{K}$ & $10^4\ \mathrm{K}$ & $4.6\times10^4\ \mathrm{K}$\\
\hline\hline
\end{tabular}
\ec
\vspace{-5ex}
\end{table}
\begin{table*}
\caption{Details of all simulations presented in this work. From left to right
  we list: galaxy model used (see Table~\ref{Table1}), feedback method used
  (and any additional information), target mass of gas cells (and star
  particles), number of gas cells (excluding CGM) and star particles in the
  initial conditions, cell diameter at the star formation density threshold of
  $10\ \mathrm{cm^{-3}}$ (note that due to our Lagrangian method, cells can
  become much smaller, with $\Delta x \propto \rho^{-1/3}$), minimum
  gravitational softening for gas cells (and fixed softening for star
  particles), feedback and star formation parameters, and total newly formed
  stellar mass present at 250 Myr (not including stellar mass returned to the
  ISM through feedback).} 
\label{Table2}
\bc
\setlength\tabcolsep{1.5pt}
\begin{tabular}{llcccclc}
\hline\hline
Galaxy & Feedback & $m_\mathrm{cell}\ [\mathrm{M}_{\rm \odot}]$ & $N_\mathrm{part}$ & $\Delta x_\mathrm{SF}\ [\mathrm{pc}]$ & $\epsilon_\mathrm{min}\ [\mathrm{pc}]$ & Parameters & $M_*\ [10^7\ \mathrm{M}_{\rm \odot}]$\\
\hline
Fiducial & None & 2000 & $1.925\times10^5$ & 22.7 & 8.1 & - & 7.64\\
Fiducial & Thermal & 2000 & $1.925\times10^5$ & 22.7 & 8.1 & $f_\mathrm{kin}=0.0$ & 6.82\\
Fiducial & Mixed & 2000 & $1.925\times10^5$ & 22.7 & 8.1 & $f_\mathrm{kin}=0.28$ & 6.63\\
Fiducial & Kinetic & 2000 & $1.925\times10^5$ & 22.7 & 8.1 & $f_\mathrm{kin}=1.0$ & 5.82\\
Fiducial & Delayed cooling & 2000 & $1.925\times10^5$ & 22.7 & 8.1 & $f_\mathrm{kin}=0.0, \sigma_\mathrm{thresh}=10\ \mathrm{km\ s^{-1}}$ & 0.18\\
         &                 &      &                   &      &     & $t_\mathrm{diss}=10\ \mathrm{Myr}$ & \\
Fiducial & Delayed cooling & 2000 & $1.925\times10^5$ & 22.7 & 8.1 & $f_\mathrm{kin}=0.0, \sigma_\mathrm{thresh}=10\ \mathrm{km\ s^{-1}}$ & 0.40\\
         & (variable $t_\mathrm{diss}$) &      &        &      &     & $t_\mathrm{diss}=\Delta x/\sigma_\mathrm{FB}$ & \\
Fiducial & Mechanical & 2000 & $1.925\times10^5$ & 22.7 & 8.1 & $f_\mathrm{kin}=1.0$ & 1.43\\
\hline
Fiducial & None & 200 & $1.925\times10^6$ & 10.6 & 3.8 & - & 9.56\\
Fiducial & Thermal & 200 & $1.925\times10^6$ & 10.6 & 3.8 & $f_\mathrm{kin}=0.0$ & 8.64\\
Fiducial & Mixed & 200 & $1.925\times10^6$ & 10.6 & 3.8 & $f_\mathrm{kin}=0.28$ & 8.39\\
Fiducial & Kinetic & 200 & $1.925\times10^6$ & 10.6 & 3.8 & $f_\mathrm{kin}=1.0$ & 7.93\\
Fiducial & Delayed cooling & 200 & $1.925\times10^6$ & 10.6 & 3.8 & $f_\mathrm{kin}=0.0, \sigma_\mathrm{thresh}=10\ \mathrm{km\ s^{-1}}$ & 0.21\\
         &                 &      &                   &      &     & $t_\mathrm{diss}=10\ \mathrm{Myr}$ & \\
Fiducial & Delayed cooling & 200 & $1.925\times10^6$ & 10.6 & 3.8 & $f_\mathrm{kin}=0.0, \sigma_\mathrm{thresh}=10\ \mathrm{km\ s^{-1}}$ & 0.59\\
         & (variable $t_\mathrm{diss}$) &      &        &      &     & $t_\mathrm{diss}=\Delta x/\sigma_\mathrm{FB}$ & \\
Fiducial & Mechanical & 200 & $1.925\times10^6$ & 10.6 & 3.8 & $f_\mathrm{kin}=1.0$ & 1.02\\
Small & None & 200 & $1.925\times10^5$ & 10.6 & 3.8 & - & 0.082\\
Small & Mechanical & 200 & $1.925\times10^5$ & 10.6 & 3.8 & $f_\mathrm{kin}=1.0$ & 0.014\\
Large & None & 200 & $1.925\times10^7$ & 10.6 & 3.8 & - & 124.1\\
Large & Mechanical & 200 & $1.925\times10^7$ & 10.6 & 3.8 & $f_\mathrm{kin}=1.0$ & 42.6\\
Fiducial & None & 200 & $1.925\times10^6$ & 10.6 & 3.8 & $\epsilon_\mathrm{SF} = 15\%$ & 11.9\\
& (High SF efficiency) & & & & & &\\
Fiducial & Mechanical & 200 & $1.925\times10^6$ & 10.6 & 3.8 & $\epsilon_\mathrm{SF} = 15\%$, $f_\mathrm{kin}=1.0$ & 1.33\\
& (High SF efficiency) & & & & & &\\
Fiducial & None & 200 & $1.925\times10^6$ & 10.6 & 3.8 & $n_\mathrm{SF} = 100\ \mathrm{cm^{-3}}$ & 10.1\\
& (High SF density threshold) & & & & & &\\
Fiducial & Mechanical & 200 & $1.925\times10^6$ & 10.6 & 3.8 & $n_\mathrm{SF} = 100\ \mathrm{cm^{-3}}$, $f_\mathrm{kin}=1.0$ & 1.43\\
& (High SF density threshold) & & & & & &\\
\hline
Fiducial & None & 20 & $1.925\times10^7$ & 4.9 & 1.8 & - & 9.23\\
Fiducial & Thermal & 20 & $1.925\times10^7$ & 4.9 & 1.8 & $f_\mathrm{kin}=0.0$ & 1.09\\
Fiducial & Mixed & 20 & $1.925\times10^7$ & 4.9 & 1.8 & $f_\mathrm{kin}=0.28$ & 0.94\\
Fiducial & Kinetic & 20 & $1.925\times10^7$ & 4.9 & 1.8 & $f_\mathrm{kin}=1.0$ & 0.92\\
Fiducial & Delayed cooling & 20 & $1.925\times10^7$ & 4.9 & 1.8 & $f_\mathrm{kin}=0.0, \sigma_\mathrm{thresh}=10\ \mathrm{km\ s^{-1}}$ & 0.27\\
         &                 &      &                   &      &     & $t_\mathrm{diss}=10\ \mathrm{Myr}$ & \\
Fiducial & Delayed cooling & 20 & $1.925\times10^7$ & 4.9 & 1.8 & $f_\mathrm{kin}=0.0, \sigma_\mathrm{thresh}=10\ \mathrm{km\ s^{-1}}$ & 0.79\\
         & (variable $t_\mathrm{diss}$) &      &        &      &     & $t_\mathrm{diss}=\Delta x/\sigma_\mathrm{FB}$ & \\
Fiducial & Mechanical & 20 & $1.925\times10^7$ & 4.9 & 1.8 & $f_\mathrm{kin}=1.0$ & 0.72\\
\hline\hline
\end{tabular}
\ec
\end{table*}
\vspace{-4ex}
\subsection{Disk morphologies and gas phases} \label{subsec_morph}
\begin{figure*}
\centering
\includegraphics[width=\textwidth]{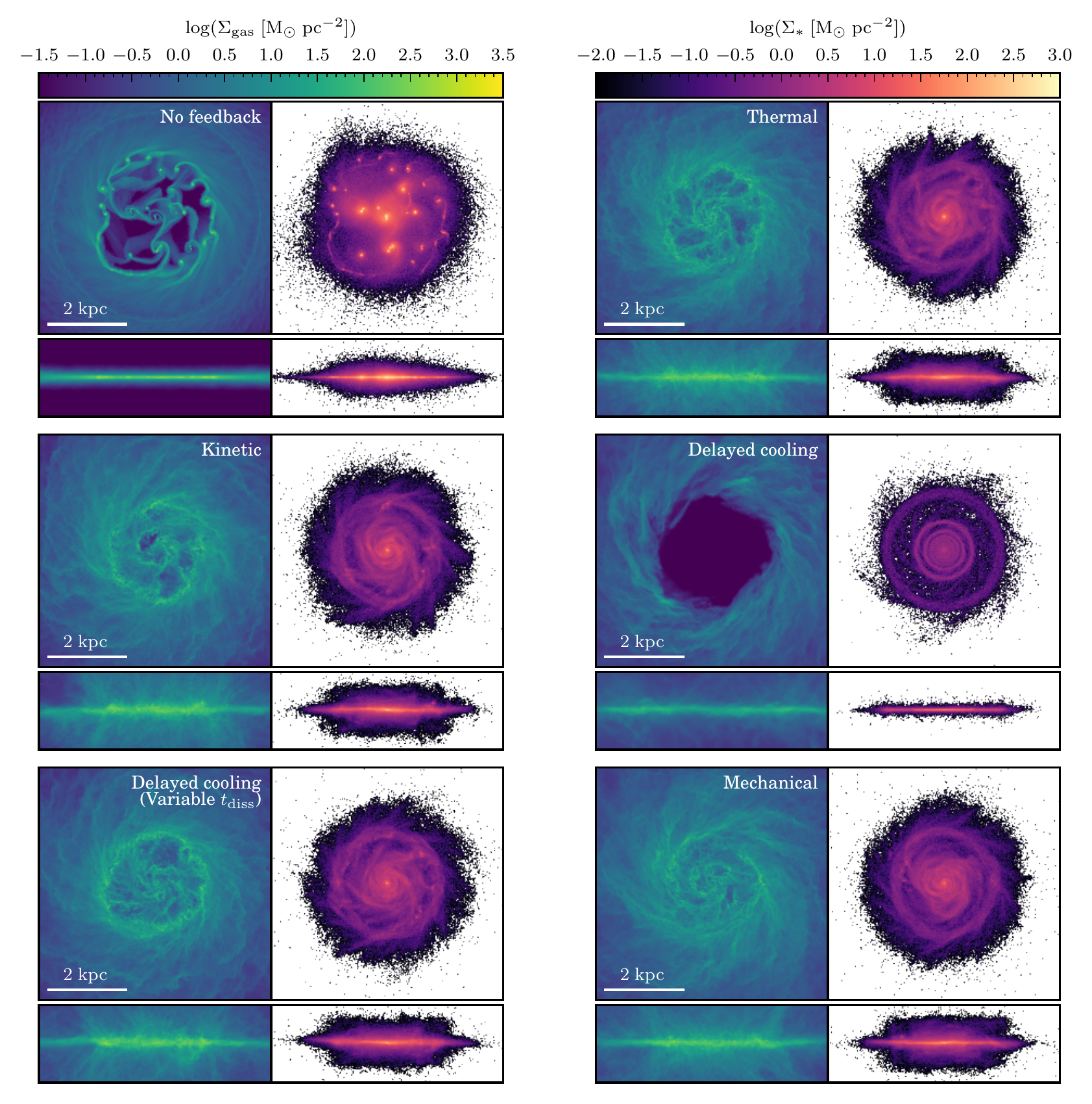}
\caption{Projections of gas and newly formed stars at 250 Myr for different
  feedback runs at $20\ \mathrm{M}_{\rm \odot}$ resolution, viewed both face-on and
  edge-on. The mixed feedback simulation is not shown as the results are similar to
  the thermal and kinetic feedback simulations. The simulation without feedback results in
  dense clumps of gas which produce stars at a high rate. The simulations with classical, delayed cooling 
  with variable $t_\mathrm{diss}$ and mechanical feedback schemes are able suppress the formation of
  dense clumps and reduce the mass of stars formed.
  They all show very similar disk morphologies
  with gas and stars exhibiting spiral patterns. The
  delayed cooling scheme is far too effective and blows up a large fraction
  of the gaseous disk leading to ring-like structures of newly formed
  stars. Equivalent plots for the lower
  resolution simulations can be found in Appendix~\ref{appendix_other},
  Figs.~\ref{disk_projections_lr} and \ref{disk_projections_hr}.}  
\label{disk_projections_uhr} 
\end{figure*}
\begin{figure*}
\centering
\includegraphics[width=\textwidth]{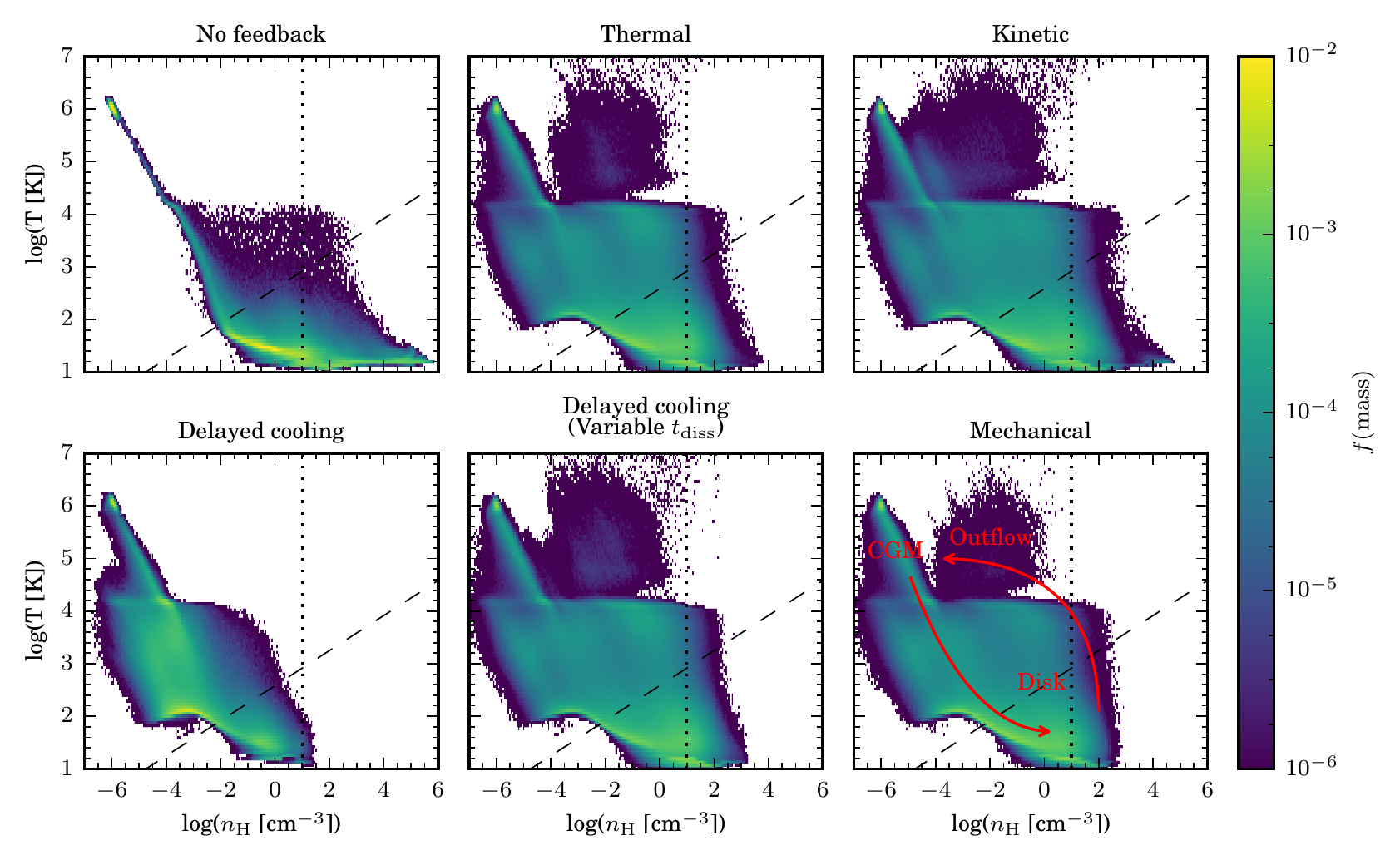}
\caption{Phase diagram for gas within the virial radius at 250 Myr for
  different feedback simulations at $20\mathrm{M_\odot}$ resolution. Colour coding is according to the fraction of mass in a given pixel. 
  The vertical dotted line shows the star formation density threshold, $n_\mathrm{SF}$. The region
  of the phase diagram below the diagonal dashed line is where the pressure is
  dominated by the non-thermal Jeans pressure floor, rather than conventional
  thermal pressure. The mixed feedback simulation is not shown as the results
  are similar to the thermal and kinetic feedback simulations. While the
  majority of gas resides at low temperature and high densities, i.e. within
  the disk, all feedback models are able to remove a fraction of gas from the
  ISM and to heat it to high temperatures above $10^4\,{\rm K}$ launching
  galaxy-scale outflow. The delayed cooling launches a large outflow at early times, 
  the majority of which is outside the virial radius at 250 Myr.
  The plot for the mechanical feedback simulation is labelled, showing the location of disk (gas denser than $10^{-4}\ \mathrm{cm^{-3}}$ is within 3 scale radii and heights), outflowing material (the region marked on the plot is all outflowing at more than $50\ \mathrm{km\ s^{-1}}$, but still within the disk region) and the CGM on the phase diagram. Also marked is the circulation of gas around the diagram due to the galactic fountain effect.}
\label{phase_uhr} 
\end{figure*}
\begin{figure*}
\centering
\includegraphics{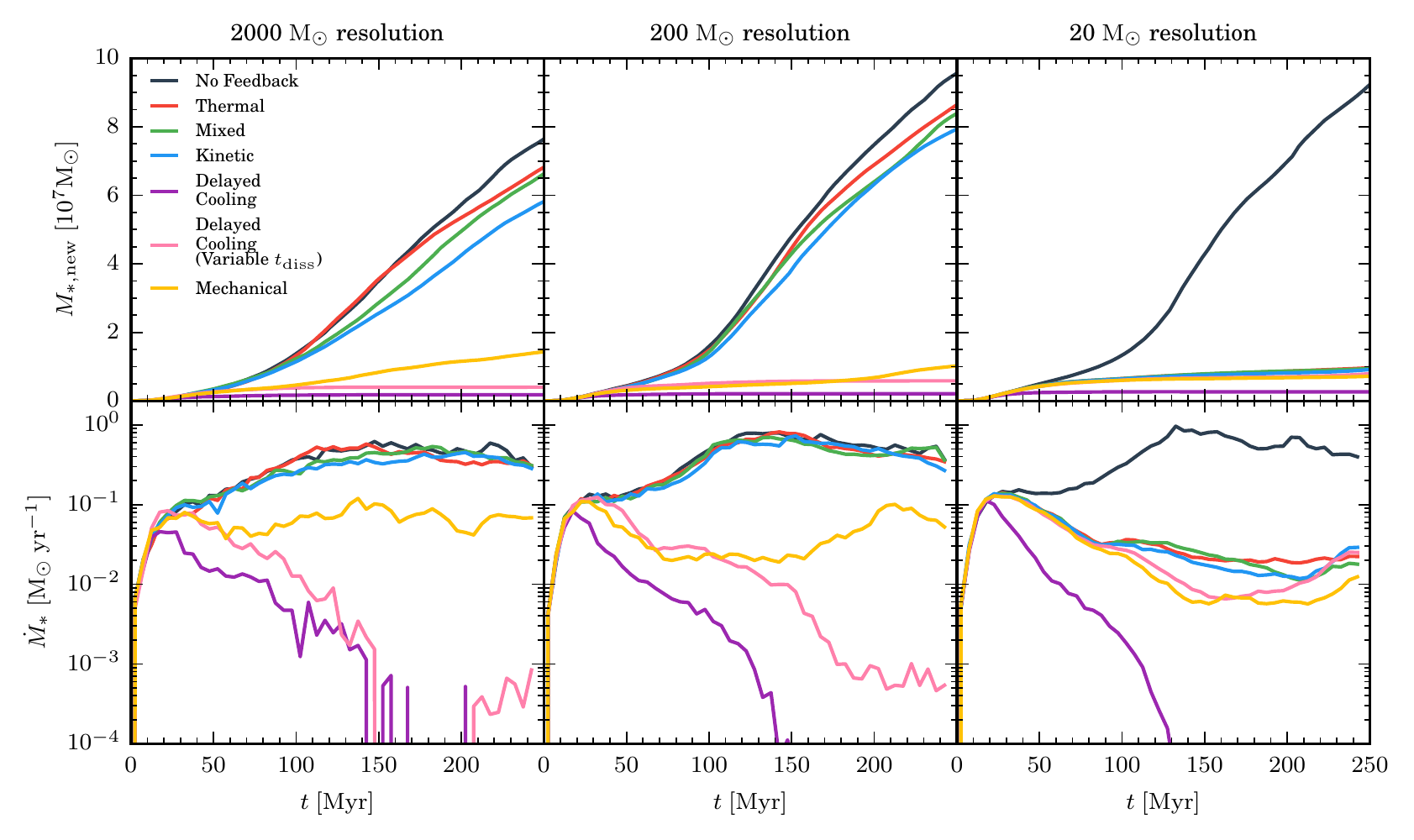}
\caption{Newly formed stellar mass (top) and SFRs (bottom) for
  our three resolutions. At the low and intermediate resolutions, the classical
  feedback schemes experience the overcooling problem and barely suppress star formation relative
  to the no feedback simulations. However, at the highest resolution they are able to
  suppress star formation. The delayed cooling schemes are in general too powerful, completely quenching
  star formation (with the exception of the highest resolution variable $t_\mathrm{diss}$ run which is
  barely delaying cooling in this regime). The mechanical feedback scheme suppresses star formation by a similar amount across
  all three resolutions, demonstrating reasonable convergence, while also being comparable to 
  the classical schemes at the highest resolution, suggesting it is converging onto the `correct'
  physical result.}
\label{disk_sfr} 
\end{figure*}
Fig.~\ref{disk_projections_uhr} shows face-on and edge-on projections of the gas
and newly formed stars in the highest resolution simulations after 250
Myr. Without feedback, the gas disk cools efficiently and adopts a highly
clumpy morphology on large scales. The gas in these clumps is extremely
efficient at forming stars. Hence, the distribution of newly formed stars also
follows this clumped morphology. There is no single dominant bulge component,
instead there are multiple large clumps near the centre of the disk. Seen edge-on, the gas disk is very thin as, having cooled, it lacks vertical pressure
support. The morphology for the simulations without feedback is similar to
those carried out at lower resolutions (see Appendix~\ref{appendix_other}). 

The thermal, mixed (not shown) and kinetic feedback schemes are able to
prevent the formation of gas clumps, instead forming complex structures of
dense gas and spiral arms. This structure is also reflected in the disk of
newly formed stars. The multiple clumps of stars seen in the no feedback
case are not present, though there is a definite overdensity of new stars in
the centre of the disk. The global surface density of newly formed stars is
greatly reduced (see Section~\ref{subsec_sfr_out}). Seen edge-on, a complex
vertical gas structure is evident with outflows present. At lower resolutions,
this morphology is not evident (see Appendix~\ref{appendix_other},
Figs. \ref{disk_projections_lr} and \ref{disk_projections_hr} for equivalent
projections). Instead, the thermal, mixed and kinetic feedback schemes are
unable to prevent the formation of dense clumps of gas. The subsequent
evolution of the disk is then broadly similar to that of the runs without
feedback. This clearly indicates that a mass resolution of at least $\sim20 \,
\mathrm{M}_{\rm \odot}$ is needed for these feedback schemes to become effective.

The simulation with delayed cooling using a fixed dissipation time results in
a completely disrupted disk. When the first SNe occur, they are able to eject
most of the gas from the centre, leaving behind a central, low density region
at 250 Myr, as evident in Fig.~\ref{disk_projections_uhr}. The projection of
newly formed stars shows an unusual ring-like structure. This is caused by the
violent ejection of gas from the centre, forming stars in areas of compression
as the resulting shock is transmitted through the disk plane, essentially
leading to a `positive' feedback. This behaviour is also apparent in the lower
resolution simulations but can largely be regarded as a numerical
artifact. The strength of the feedback, and resulting gas and stellar
morphologies, indicate that the choice of parameters for this scheme is not
appropriate and further 
tuning is necessary.
Further discussion of the issues
with the delayed cooling in our simulation can be found later and
simulations carried out with different parameters can be found in
Appendix~\ref{appendix_DFB}.\newline
\indent The delayed cooling run with variable dissipation time
($t_\mathrm{diss}=\Delta x/\sigma_\mathrm{FB}$) is not as strong. The disk
morphology is similar to the thermal, mixed and kinetic feedback schemes at
this resolution, with suppression of large scale clumping without destruction
of the disk. However, perhaps counter-intuitively, this feedback scheme
becomes stronger at lower resolutions (as evidenced by
Figs. \ref{disk_projections_lr} and \ref{disk_projections_hr}), disrupting the
disk in the $2000\ \mathrm{M}_{\rm \odot}$ simulation. This is because the dependence
on the cell diameter results in very short dissipation times at high
resolution. At our highest resolution of $20\ \mathrm{M}_{\rm \odot}$, the cooling is
essentially not delayed at all, resulting in a straight thermal dump and hence
similarity to the classic thermal feedback scheme. Qualitatively, this is the
desired behaviour, with the delayed cooling being reduced at resolutions high
enough to resolve the Sedov-Taylor phase, but increased at lower
resolution. However, the lack of convergence with resolution (and disk
destruction at low resolution) indicates some form of tunable parameter might
need to be introduced to refine this scheme. 

Finally, the mechanical feedback scheme results in morphologies similar to the classical feedback schemes. Uniquely among the schemes tested, the mechanical feedback is able to produce these morphologies across two orders of magnitude in resolution. While the classical feedback schemes overcool at lower resolutions, the mechanical feedback scheme is still able suppress large scale clumping and the formation of high density gas, without destroying the disk. Unlike the variable dissipation time delayed cooling scheme, the implicit modulation of small scale feedback strength as a function of resolution in the mechanical scheme is able to produce convergent disk morphologies at our three resolutions.

Fig.~\ref{phase_uhr} shows phase diagrams for the gas in the highest
resolution simulations at 250 Myr (similar plots for the lower resolution runs
may be found in Appendix~\ref{appendix_other}). In the no feedback simulation,
the majority of the gas in the disk has cooled well below $10^2\ \mathrm{K}$
and there is a substantial quantity of gas at high density, as far as $\sim
10^6\ \mathrm{cm^{-3}}$. Once the gas enters this cold, dense phase, the
resulting evolution is regulated by the non-thermal pressure floor and is
highly dependent on the choice of parameters (for more details see
Appendix~\ref{appendix_floor}), though our results are qualitatively similar
to comparable simulations in the literature \citep[e.g.][]{Rosdahl2015,
  Rosdahl2017, Hu2016, Hu2017} . At lower resolution, the results are similar, though gas does
not reach quite such high densities, an expected consequence of lower
resolution. At the highest resolution, the classical feedback schemes are able
to maintain a warm phase in the disk. While cold, dense gas is still present
it does not reach the high densities seen in the no feedback simulation. An
additional component of gas is apparent on the phase diagrams, between $T \sim
10^4 - 10^7\ \mathrm{K}$ and $n \sim 10^{-5} - 10\ \mathrm{cm^{-3}}$. This is
gas that has received feedback energy and is expanding up out of the plane of
the disk. When viewed as a function of time, a cyclical pattern anticlockwise
around the phase diagram may be observed with gas cooling and contracting to
star forming densities (moving down and right), being heated by feedback
(moving up), expanding (moving left). Gas which rains back on the disk (the
so-called galactic fountain, see Section~\ref{subsec_sfr_out}) cools and drops
back down the phase diagram and may enter the cycle again.
\setlength{\parskip}{0pt plus 0pt}

\vspace{2ex}
The phase diagram for the delayed cooling with fixed dissipation time feedback
at 250 Myr shows a complete absence of dense gas. In addition, because the
feedback has efficiently quenched star formation (see
Section~\ref{subsec_sfr_out}), there are no further SNe after the initial
budget has been exhausted. This results in the lack of gas above
$10^4\ \mathrm{K}$ (with the exception of that in the CGM). When the delayed
cooling scheme is used with a variable dissipation time, at the highest
resolution the phase diagrams are similar to those for the classical
schemes. However, at lower resolution as the feedback becomes stronger, they
become similar to the delayed cooling with fixed dissipation time. In the
highest resolution simulation, the mechanical feedback scheme produces phase
diagrams similar to the classical schemes, although the highest density gas
has been curtailed. The phase diagrams look similar across all resolutions.
\setlength{\parskip}{0pt plus 0.1pt}

\vspace{-4ex}
\subsection{Star formation rates and outflows} \label{subsec_sfr_out}
\begin{figure*}
\centering
\includegraphics{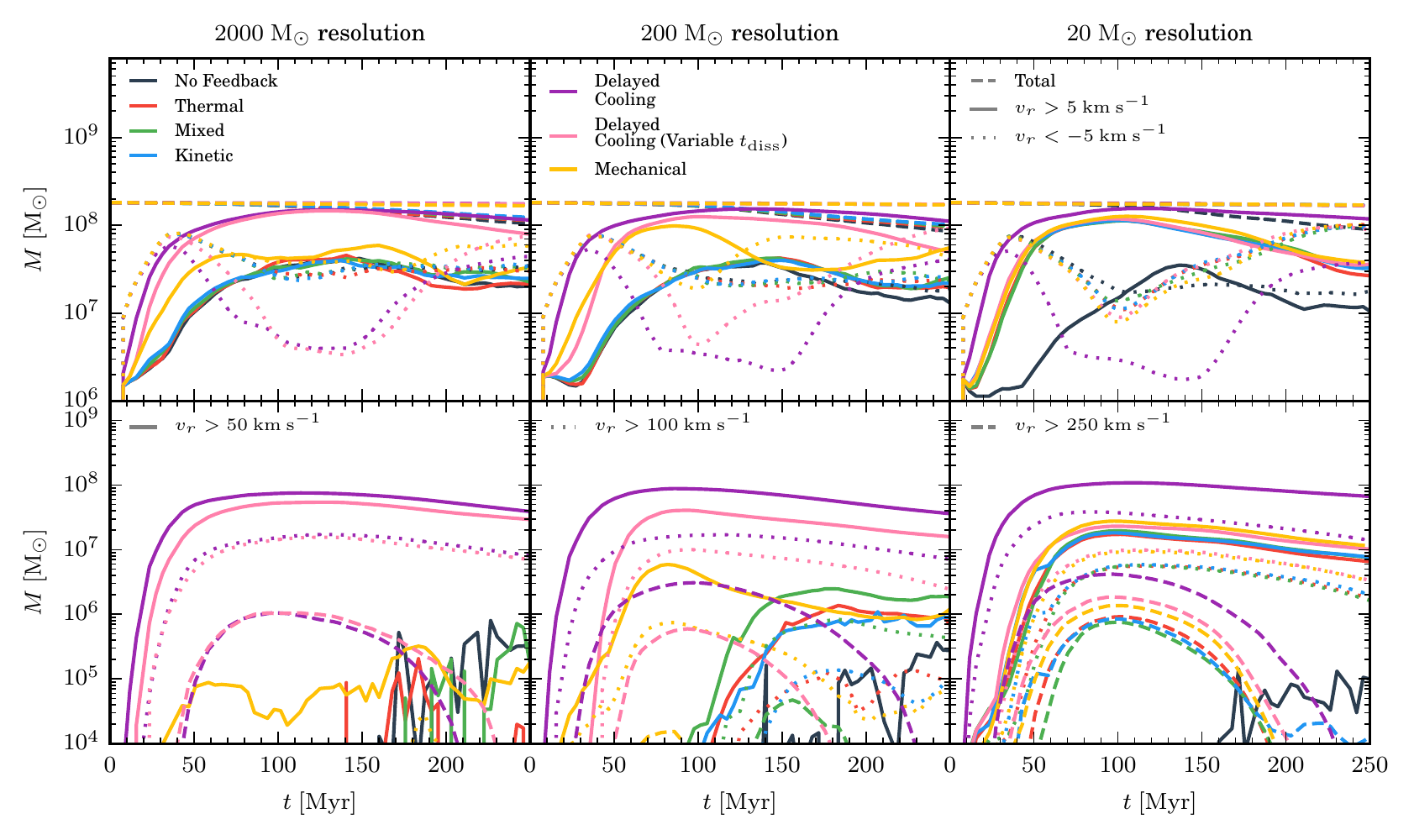}
\caption{Mass of gas moving at various radial velocities within the virial
  radius as a function of time for our different feedback schemes at all three
  resolutions. \textit{Top:} Total gas mass within the virial radius (dashed
  curves), mass of gas radially outflowing and inflowing over
  $5\ \mathrm{km\ s^{-1}}$ (solid and dotted curves,
  respectively). \textit{Bottom:} mass of gas radially outflowing at more than
  $50$, $100$ and $250\ \mathrm{km\ s^{-1}}$ (solid, dotted and dashed curves,
  respectively). Mass of the outflowing gas is very sensitive to the
  resolution of simulations and only in the highest resolution runs do
  feedback schemes launch significant outflows.} 
\label{bulk_outflow} 
\end{figure*}
Fig.~\ref{disk_sfr} shows the mass of stars formed and the star formation
rates as a function of time for all feedback schemes at all three
resolutions. The no feedback simulations are similar across all three
resolutions. As gas cools and reaches star forming densities, star formation
begins. After a sudden jump in star formation at the beginning of the
simulation, the SFR rises gradually to a roughly constant rate $\sim 0.2 -
1\ \mathrm{M}_{\rm \odot}\ \mathrm{yr^{-1}}$. The higher resolutions result in denser
clump formations leading to slightly higher SFR on average, but the total
stellar mass formed is similar between the three resolutions
($7.64\times10^7\, {\rm M_\odot}$, $9.56\times10^7\, {\rm M_\odot}$ and
$9.23\times10^7\,{\rm M_\odot}$ for the $2000\,{\rm M_\odot}$, $200\,{\rm
  M_\odot}$, and $20\,{\rm M_\odot}$ resolutions, respectively). It should be
noted that in the absence of effective feedback, the SFR and its time
dependence become regulated by the choice of non-thermal pressure floor since
that impacts the scale of fragmentation and the densities reached. However,
our scaling of the pressure floor with resolution (as described in
Section~\ref{subsec_floor}) results by construction in reasonably convergent
behaviour with resolution, though it is not convergent with choice of pressure
floor parameter (see Appendix~\ref{appendix_floor}). 

In the lower resolution simulations, the classical feedback schemes are unable
to suppress star formation by more than $\sim20\%$ in the best case as they
catastrophically overcool and follow the same behaviour as the no feedback
simulations. A slight trend of increased effectiveness with increased
$f_\mathrm{kin}$ is apparent but the total impact on SFRs is weak. However, at
our highest resolution of $20\,{\rm M_\odot}$, the classical feedback schemes become effective, reducing SFR and total stellar mass by around an order of magnitude with respect to the no feedback simulations.

The delayed cooling scheme with a fixed dissipation time efficiently quenches
star formation with a single period of SNe activity, expelling all star
forming gas from the centre of the system. The results are well converged with
resolution, forming almost exactly the same stellar mass. The total stellar
mass formed is only a few percent of the no feedback case. As described in the
previous section, the delayed cooling with variable dissipation time results
in stronger feedback at lower resolution. This can be seen in
Fig.~\ref{disk_sfr} where the SFR is similar to the fixed dissipation time
simulation at $2000\, {\rm M_\odot}$, higher at $200\, {\rm M_\odot}$ and
close to the classical schemes at $20\,{\rm M_\odot}$.

At the lowest resolution, the mechanical feedback scheme results in a steady
star formation rate below $10^{-1}\ M_{\rm \odot}\ \mathrm{yr^{-1}}$,
suppressing total stars formed by approximately a factor of $5$. With
increasingly higher resolution, the SFRs are suppressed slightly more, but
encouragingly the total stellar mass formed is within a factor of $2$ from the
lowest to the highest resolution simulations. The SFR exhibits a slight dip in
the $200\,{\rm M_\odot}$ resolution simulations, increasing slightly towards
$200$~Myr. This is due to a `galactic fountain' effect, with gas launched from
the disk returning and forming new stars, which we discuss in greater detail
below. At the highest resolution, the mechanical feedback scheme is reasonably
similar to the classical schemes because the Sedov-Taylor phase is resolved in
the majority of SNe events, so the momentum boost factor is close to
unity. The mechanical scheme is slightly stronger than the classical schemes
due to the few SNe at this resolution that are not sufficiently resolved by
the classical schemes because they occur at high densities (for further
details see Section~\ref{host}).

\begin{figure*}
\centering
\includegraphics{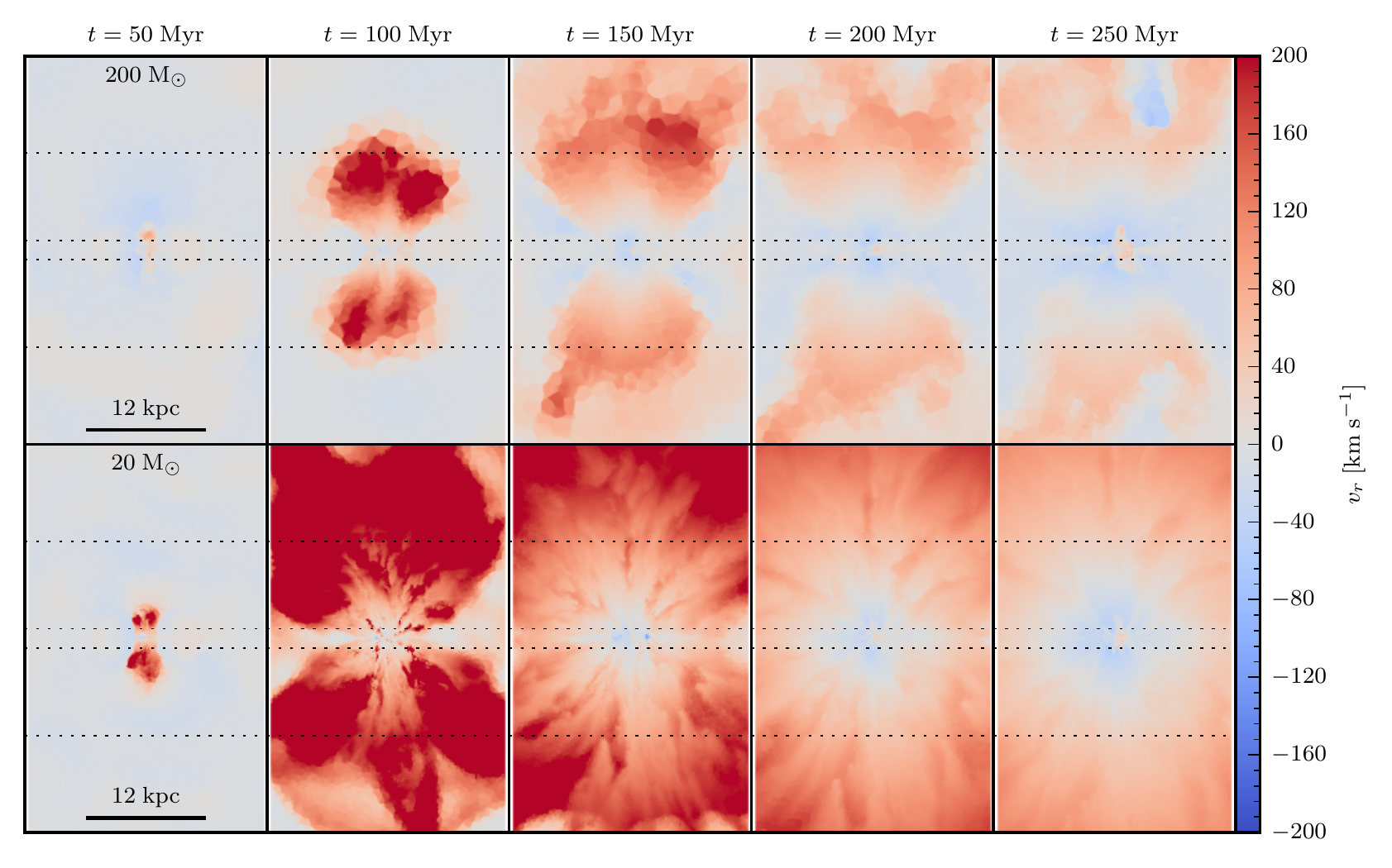}
\caption{Slices through the centre of the mechanical feedback simulation at
  $200\,{\rm \Msun}$ (top) and $20\,{\rm \Msun}$ (bottom) resolutions showing
  radial velocity at various times (outflowing gas is in read, while the
  inflowing gas is in blue). The horizontal dotted lines show the
  planes at 1 and 10 kpc away from the midplane of the disk, used examine the
  outflows in Fig.~\ref{massload}. Not only is the outflow stronger in the
  higher resolution run but the spatial structure of inflowing and outflowing
  gas changes as well with resolution, where the galactic fountain effect is
  more pronounced in the lower resolution simulation.} 
\label{vel_slice} 
\end{figure*}
\begin{figure*}
\centering
\includegraphics{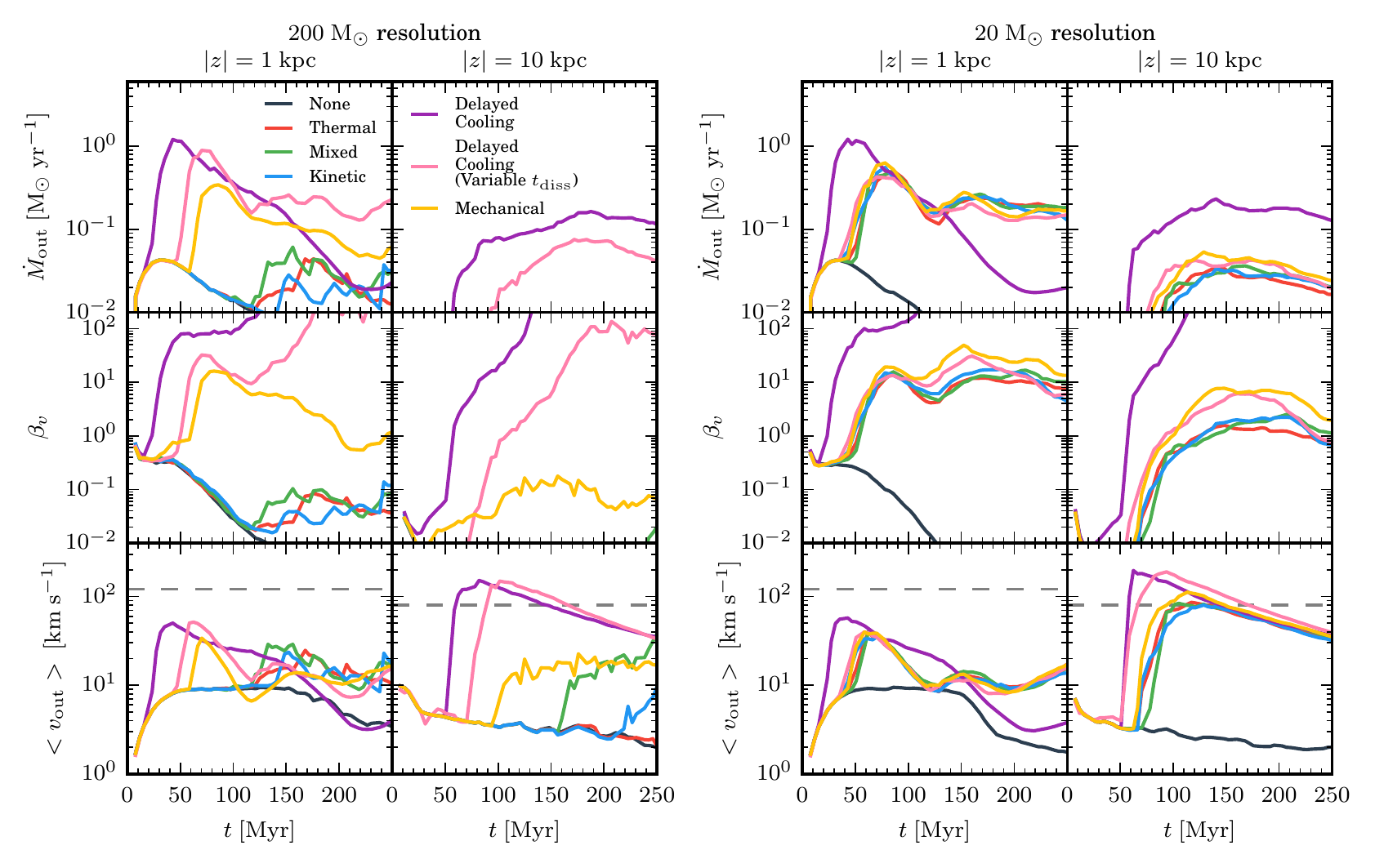}
\caption{Mass outflow rates (top), mass loading factors (middle) and
  mass-weighted average outflow velocities (bottom) for our different feedback
  models across planes at $1$ and
  $10$~kpc from the disk midplane for our two highest resolution simulations
  (left and right panels). The dashed grey lines indicate the escape velocity
  at the relative disk height. The $2000\ \mathrm{M}_{\rm \odot}$ simulations are not
  shown as outflows for all except delayed cooling are negligible (see
  Figs.~\ref{bulk_outflow} and \ref{disk_slice_lr}). Outflow velocities
  comparable to the escape velocity and mass loading factors of a few are only
reached at the highest resolution simulations for all feedback runs (except
for delayed cooling runs which are over-efficient).} 
\label{massload} 
\end{figure*}
\begin{figure*}
\centering
\includegraphics{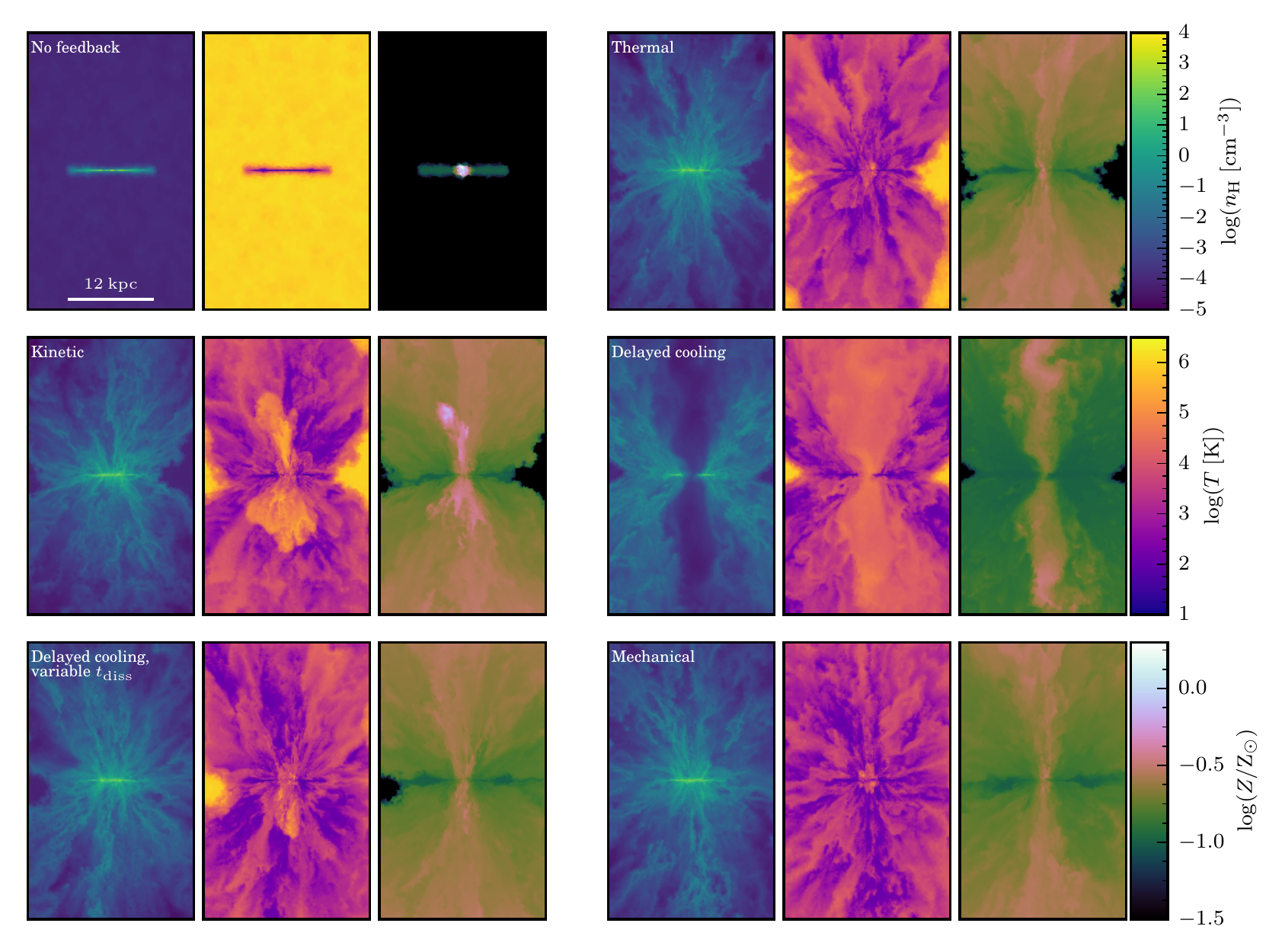}
\caption{Density, temperature and metallicity slices at 250 Myr for $20\ \mathrm{M}_{\rm
    \odot}$ resolution runs. The no feedback simulation unsurprisingly
    launches no outflows, leaving a high concentration of metals in the centre
    of the system. The classical, delayed cooling with variable $t_\mathrm{diss}$
    and mechanical schemes all produce comparable outflows (differences are largely
    due to variability with time). The outflows have a complex filamentary structure
    and are multiphase with temperatures spanning $\sim6$ orders of magnitude. The 
    delayed cooling scheme evacuates gas from the centre of the system, leaving a 
    column of very hot, low density gas. Equivalent plots for the lower resolution
    simulations, which are unable to launch such strong outflows, can be found in
    Appendix~\ref{appendix_other}, Fig. \ref{disk_slice_lr}.}
\label{disk_slice_uhr} 
\end{figure*}

While the general features of SFRs are converged for feedback schemes that do
not overcool across different resolutions, such as the `Mechanical' run,
interestingly the
same is not true for outflows (with the exception of delayed cooling runs,
which drive strong outflows at all resolutions). Fig.~\ref{bulk_outflow} shows
the total gas mass within the virial radius moving at various radial
velocities for our various feedback schemes at three resolutions. In the no
feedback run, the behaviour is simple and is essentially the same across all
resolutions. Initially, there is a rise in gas inflow as the disk collapses
vertically. Gas tagged as outflowing (moving more than $5\ \mathrm{km\ s^{-1}}$ 
radially outwards) is apparent despite the lack of feedback, but this is caused by 
motions in the disk rather than a true outflow. After $\sim100$ Myr, the inflow 
and outflow rates are approximately equal. The gas
settles in the disk plane with small net motions due to the movements of the
clumps and the gas reservoir is rapidly converted to stars. Note that in a
cosmological context, there would be a constant net inflow from well outside
the disk due to cosmic accretion. In our setup, the initially inflowing gas is
simply the disk settling into an equilibrium configuration as it cools, our
background uniform CGM has long cooling times, largely stays in place and
does not accrete onto the disk. 

The same behaviour is apparent for the classical feedback schemes at the lower
two resolutions. Due to overcooling, the feedback is unable to suppress the
inflowing gas, i.e. neither to stabilise the disk with a larger scale height
nor to drive any appreciable outflows. At the lowest resolution, there is no
gas outflowing faster than $100\ \mathrm{km\ s^{-1}}$. There is a small
fraction ($\sim 0.1\ \%$ of the total gas mass) moving faster than
$50\ \mathrm{km\ s^{-1}}$ at late times, but these are merely motions of the
clumpy disk and are present in the no feedback run. At the $200\, {\rm
  M_\odot}$ resolution, the feedback is able to generate some outflows faster
than $50\ \mathrm{km\ s^{-1}}$ ($\sim1\ \%$ of the total gas mass) and a small
amount moving faster than $100\ \mathrm{km\ s^{-1}}$ once the density in the
disk has dropped slightly due to conversion of gas to stars, reducing the
overcooling effect. However, this little mass moving at relatively low
velocities is not able to make it far from the disk plane and the net inflow
and outflow rates are still comparable. Despite significantly suppressing star
formation in the lowest resolution simulation, mechanical feedback struggles
to launch outflows. It generates marginally larger outflow rates than the
classical feedback schemes, but the inflow and outflow rates match after 50
Myr. In the $200\, {\rm M_\odot}$ resolution simulation, the feedback is able
to drive a much stronger outflow at early times, suppressing inflow and
launching a significant mass of material faster than $50\ \mathrm{km\ s^{-1}}$
(and a smaller amount faster than $100\ \mathrm{km\ s^{-1}}$). However, this
material is not moving fast enough to escape the galaxy, so it returns back to
the disk in a galactic fountain, with the inflow rates overtaking the outflow
rates at around $\sim125$~Myr. 

At the highest resolution, the classical and mechanical feedback schemes are
relatively similar and are able to launch strong outflows. From around
$50$~Myr onwards, over $10^7\ {\rm M_\odot}$ of gas is moving faster than
$50\ \mathrm{km\ s^{-1}}$, the majority of which is moving faster than
$100\ \mathrm{km\ s^{-1}}$ and a non-negligible fraction is moving faster than
$250\ \mathrm{km\ s^{-1}}$. The net outflow rates are approximately an order
of magnitude larger than the inflow rates at 100 Myr. However, after this
point, inflow rates rise and outflow rates drop slightly as the lower velocity
gas begins to stall and flow inwards. At around 180 Myr, the inflow rates
exceed the outflow rates. This galactic fountain effect can be seen in
Fig.~\ref{vel_slice}, which shows gas in a vertical slice through the system
colour coded by radial velocity at several times for the mechanical feedback
scheme at the higher two resolutions. In the top panel, in the $200\,{\rm
  M_\odot}$ simulation, an outflow is launched at $50$~Myr. As time
progresses, the outflowing gas moves up away from the centre of the system,
but gas begins to flow inwards, segregated by velocity. By $200$~Myr, the
centre is dominated by returning gas while only the fastest moving gas has
continued to outflow. At 250 Myr, a new outflow has just been launched,
resulting in a complex, interleaved pattern of gas outflowing, inflowing and
outflowing with increasing height from the disk. At the higher resolution, the
initial outflow is much faster. At $100$~Myr and $150$~Myr, the covering
fraction of outflowing gas is much larger that in the lower resolution run
with all but the very central regions dominated by outflows. A complex
structure of outflowing gas is apparent, with a `finger' like pattern of
various outflow velocities. Again, at the final snapshot at 250 Myr, the
central regions contain largely inflowing gas, but far more gas continues to
outflow as compared to the lower resolution. 

As previously discussed, the delayed cooling schemes launch very strong
outflows. As can be seen in Fig.~\ref{bulk_outflow}, once again the delayed
cooling with fixed dissipation time is extremely well converged across the
three resolutions, drastically suppressing inflows and launching large
quantities of gas at high velocities from essentially a single period of SNe
activity (although, inflow rates begin to pick up at late times once SNe have
been shut off due to star formation quenching). However, as previously
mentioned, this feedback seems unphysically strong. The delayed cooling with
variable dissipation time follows similar behaviour to the fixed dissipation
time scheme, but once again converges to the classical schemes with higher
resolution.

Having examined the bulk mass in outflows, it is also instructive to consider
the properties of the outflow at certain heights above the disk
plane. Fig.~\ref{massload} shows the outflow rates (i.e. only considering
outflowing gas, not inflowing), mass loading factors and mass-weighted average
outflow velocity at $1$~kpc and $10$~kpc above the disk as a function of time
for the highest two resolutions. We calculate the mass outflow as 
\begin{equation}
\dot{M}_\mathrm{out} = \frac{\sum_i m_i v_{\mathrm{out},i}}{\Delta z} \,,
\end{equation}
where the sum is over all cells within a slice (parallel to the disk plane) of
thickness $\Delta z$ centred on the target height (i.e. $1$ or $10$~kpc) that
have a positive outflow velocity $v_\mathrm{out}$ (vertically away from the
disk plane, rather than radially as in the previous figures). We adopt $\Delta
z=200\ \mathrm{pc}$. The mass loading factor, $\beta_v$ is the ratio of the
mass outflow rate to the star formation rate and is essentially a measure of
the efficiency of stellar feedback to drive outflows. There is of course a
delay between the formation of a given stellar population and the outflow its
feedback eventually drives reaching a given height.  However, here we use the
instantaneous ratio between $\dot{M}_\mathrm{out}$ and the SFR rather than a
more complex binning scheme as our SFRs are, on the whole, steady over long
periods (with the exception of the delayed cooling schemes, for which the mass
loading must be interpreted with some caution). Finally, we plot the
mass-weighted mean outflow velocity (i.e. ignoring inflowing gas) alongside
the escape velocity at that height\footnote{We calculate the escape velocity
  at the relevant height directly above the centre of the disk from the
  initial conditions. Deviations from the initial conditions during the course
  of the simulation have a negligible impact on $v_\mathrm{esc}$ at $1$~kpc
  and are insignificant at $10$~kpc as the dominant component is the static
  halo potential.}.

As demonstrated in Fig.~\ref{bulk_outflow}, with $200\ {\rm M_\odot}$
resolution, the classical feedback schemes are unable to drive much of an
outflow. After some initial gas flow over $1$~kpc as the system settles from
the initial conditions (also apparent in the no feedback simulations; this is
a `spurious' mass loading amplified by low SFRs), the mass outflow drops
off. There is a small increase after $100$~Myr as the feedback becomes more
efficient and the gas reservoir is used up, but very little of this outflow
reaches $10$~kpc. At $1$~kpc, this outflow has a mass loading factor below
$0.1$ i.e. significantly lower than the $\beta_v \gtrsim 1$ required by observations and models (for
a more detailed discussion see introduction). The
mean outflow velocities (admittedly dominated by the slower moving gas) are
well below the escape velocity at $1$~kpc. The mixed feedback has a higher
mean outflow velocity at $10$~kpc, however its seemingly increased
effectiveness over the other methods can be put down to stochasticity
amplified by the exceedingly small mass of gas that is actually outflowing at
that distance. The mechanical feedback is able to generate a slightly more
vigorous outflow, with a mass loading factor between $1 - 10$ at
$1$~kpc. However, it is unable to sustain the outflow as previously discussed,
with gas returning in a galactic fountain. Again, very little of the outflow
reaches $10$~kpc.

At the higher resolution of $20\ {\rm M_\odot}$, the classical and mechanical
feedback schemes are able to launch much stronger, sustained outflows. Once
the outflow has reached the heights we are investigating, mass loading is
around $10$ at $1$~kpc and over unity at $10$~kpc. The mean velocities are
below the escape velocity at $1$~kpc, but a significant quantity of gas (see
Fig.~\ref{bulk_outflow}) is moving much faster. By $10$~kpc, the slower moving
gas having begun to drop back to the disk, the mean velocities are comparable
to the escape velocity. 

The delayed cooling simulations with fixed dissipation time are able to launch
strong, but short lived outflows. Having completely quenched star formation,
there is no source for additional driving of outflows beyond the initial
burst. The instantaneous mass loading factor becomes an unreliable metric in
such conditions, since it must naturally tend to infinity as SFR tends to
zero, however we plot it here for reference. Again, using the variable
dissipation time results in strong outflows at lower resolutions but similar
results to the classical and mechanical feedback schemes at the highest
resolution.

Fig.~\ref{disk_slice_uhr} contains vertical slices through the disk at
$250$~Myr for the highest resolution simulations, showing gas density,
temperature and metallicity. Generating no outflows, the no feedback
simulation shows a cold, thin, dense disk. The central regions have very high
metallicities since the ejecta from SNe stay within the star forming
regions. The thermal, mixed (not shown), kinetic, mechanical and variable
dissipation time delayed cooling feedback schemes have qualitatively similar
outflows. The outflows are multiphase (with temperatures in the range $\sim
10^2 - 10^7\ \mathrm{K}$) and have a complex structure, with many individual
filaments apparent. Observations of galactic outflows reveal them
to be comprised of multiphase gas: molecular gas at $\sim10-10^3\ \mathrm{K}$
observed at radio wavelengths \citep[e.g.][]{Walter2002,Walter2017,Bolatto2013},
material around $\sim10^4\ \mathrm{K}$ observed in the optical and near-UV
\citep[e.g.][]{Pettini2001,Martin2012,Soto2012} and
$\sim10^7 - 10^8\ \mathrm{K}$ plasma seen in X-rays \citep[e.g.][]{Martin1999,Strickland2007,
Strickland2009}. At approximately the peak of the outflow (150~Myr), for the mechanical
feedback simulation, in material
moving radially outwards at more than $100\ \mathrm{km\ s^{-1}}$, the proportions of cold
($<2000\ \mathrm{K}$), warm ($2000 - 4\times10^5\ \mathrm{K}$) and hot ($>4\times10^5\ \mathrm{K}$)
material are 3.1\%, 78.9\% and 17.9\% by mass, respectively, or 0.1\%, 56.8\% and 43.1\% by volume.
Thus, while the dominant fast moving wind component is warm, a cold component is present in the outflow.
The cold component dominates by mass in material moving outwards between $5-100\ \mathrm{km\ s^{-1}}$,
with the proportions of cold, warm and hot components being 56.4\%, 39\% and 4.6\% respectively (by
volume, 0.6\%, 4.2\% and 95.2\%, the very slow moving CGM material dominating the hot component
here). As the outflow progresses and the galactic fountain effect becomes apparent, the proportions
remain similar. The cold component dominates the returning gas, but warm gas also returns. It appears
that the returning cold component contains both initially cold outflowing gas as well as material
from the warm phase that has cooled. In summary, we find that the cold gas mainly traces the lower
velocity outflows while the warm and hot medium probe the faster moving outflows; if this effect
is present in real galaxies, observing one component alone will give a biased measurement of the
outflow velocity.

The results are similar for the thermal, mixed, kinetic and variable
dissipation time delayed cooling feedback schemes at this resolution.
Small differences between schemes apparent in the figure
are largely due to stochasticity. The outflowing gas is enriched with metals
and there is a dependence of metallicity on opening angle. The most metal
enriched regions of the outflow are in the centre ($Z \gtrsim 0.25Z_{\rm
  \odot}$), containing the highest concentration of SNe ejecta, whereas
towards the edges of the outflow the metallicity is closer to the initial disk
gas metallicity of $0.1Z_{\rm \odot}$. The kinetic feedback simulation shows a
high metallicity outflow of $\sim10^6\ \mathrm{K}$ gas, having had an outflow
event shortly before 250 Myr. The delayed cooling with fixed dissipation time
simulation exhibits a column of low density, high temperature gas that extends
all the way though the centre of the disk, with cold, denser gas building up
on the fringes of the outflow. At lower resolutions (see
Fig.~\ref{disk_slice_lr}), the outflows are much
weaker (as described above) for all simulations except the delayed cooling
schemes. At $200\ \mathrm{M_\odot}$ resolution, the classical feedback schemes
launch warm $10^4\ \mathrm{K}$ outflows with dense, cold edges at the
interface with the CGM. The outflows are highly enriched because the mass
loading is so low i.e. a substantial amount of SNe occurred to launch the
outflows. 
\vspace{-4ex}
\subsection{The Kennicutt-Schmidt Relation}
The link between gas surface density, $\Sigma_\mathrm{gas}$, and SFR surface density,
$\Sigma_\mathrm{SFR}$, is an important diagnostic of star formation in galaxies. Specifically, the 
Kennicutt-Schmidt relation, $\Sigma_\mathrm{SFR}\propto\Sigma^{1.4}_\mathrm{gas}$ \citep{Kennicutt1998},
has been well established by observations of galaxies in the local Universe. Thus,
in addition to suppressing absolute SFRs, it is necessary for simulations to 
simultaneously reproduce this relation. It is possible to have very different values of $\Sigma_\mathrm{SFR}$ for the same global
$\Sigma_\mathrm{gas}$, dependent on the small scale star formation and the degree of clustering in star formation.
Our choice of small scale star formation law to some extent impacts the resulting global KS relation. For example the
choice of $\dot{\rho}_{*} \propto \rho/t_\mathrm{ff} \propto \rho^{3/2}$ generally leads to the correct slope,
but this does not guarantee the correct normalization, as shown below.

Fig.~\ref{ks} shows the global star formation rate surface density,
$\Sigma_\mathrm{SFR}$, as a function of global gas surface density,
$\Sigma_\mathrm{gas}$ for our simulations, each point representing one of the
simulations at a particular time (points are evenly spaced by 25 Myr between
25 and 250 Myr). We define the surface densities as 
\begin{equation}
\Sigma_\mathrm{SFR}=\frac{\dot{M}_{*}\left(<R_\mathrm{SFR,90\%}\right)}{\pi R_\mathrm{SFR,90\%}^2}\,,
\end{equation}  
and
\begin{equation}
\Sigma_\mathrm{gas}=\frac{M_\mathrm{gas}\left(<R_\mathrm{SFR,90\%}\right)}{\pi R_\mathrm{SFR,90\%}^2}\,,
\end{equation}
where $R_\mathrm{SFR,90\%}$ is the disk radius enclosing 90\% of the total SFR\footnote{Results are fairly insensitive to the choice of the fraction of the SFR enclosed, merely sliding points up and down the Kennicutt-Schmidt relation. We only include gas within 2 kpc of the disk plane, although our results are insensitive to removing this constraint because the gas surface density is completely dominated by mass near the disk plane.}.
For comparison, we plot global measurements form 61 normal spirals \citep{Kennicutt1998},
similar global measurements from 19 low surface brightness galaxies \citep{Wyder2009}
and sub-kpc observations of 18 nearby galaxies \citep{Bigiel2008}. For reference, we also plot
the power law fit with a slope of 1.4 from \cite{Kennicutt1998}, however it is worth noting
that the slope is possibly too shallow for this range of measurements. This fit was made simultaneously to 
the 61 spirals plotted as well as 36 higher surface density starburst galaxies (not plotted). At lower
surface densities, the relation appears to steepen, possibly due to some form of star formation threshold
\citep[e.g.][]{Kennicutt1989, Martin2001, Bigiel2008}. Thus, it makes more sense to compare our results
to the data points rather than the fit plotted.

Except at the highest resolution, the no feedback and classical feedback simulations all lie well
above the observed relation, although once the system has finished clumping
after $\sim100$ Myr, the points have approximately the correct slope. This is mainly due to
the small scale star formation law adopted forcing $\dot{\rho}_{*} \propto
\rho^{3/2}$. The simulations then progress to lower SFR and gas surface
densities as the gas reservoir is consumed. At the highest resolution, the classical
schemes are able to quench star formation efficiently and so drop into agreement with observations.
Relative to the classical
feedback schemes, the other three feedback mechanisms produce an order of
magnitude lower SFR surface densities for the same gas surface density, lying
close to the observed relation at all three resolutions. The delayed cooling
with fixed dissipation time efficiently destroys the disk, so the majority of
the snapshots lie outside the range of the plot. The same is true of the
variable dissipation time run at the lower resolution, though the high
resolution run is well within the observed points. The mechanical feedback runs
at all resolutions agree well with the observations.
The simulations track up and down the relation with time as the gas surface density changes, partly due
to gas consumption, but mainly due to outflows. For example, the cluster of $20\ \mathrm{M_\odot}$ resolution
mechanical feedback points (yellow squares) near the bottom left of the relation correspond to the period after
 the peak of the outflow at about 100 - 200~Myr, but returning gas from the galactic fountain causes the disk to have
moved back up the relation by 250~Myr (open yellow square). In addition to variation over time, this effect
also causes the differences between the three resolutions: the lower resolutions tend to lie higher up the 
relation because their weaker outflows do not drop the disk surface density as much. Despite this, the mechanical
feedback points all lie close to the observations even though their exact position on the relation varies with resolution, this difference caused by the failure of resolution convergence with respect to outflows.
\begin{figure}
\centering
\includegraphics{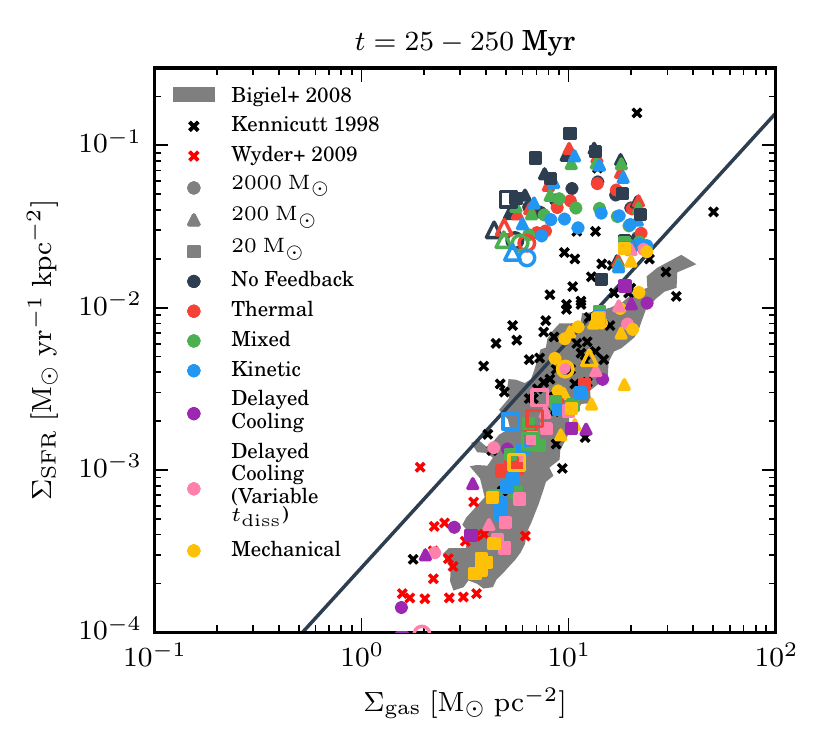}
\caption{SFR surface density plotted as a function of gas surface density for
  different feedback runs at all three resolutions (as indicated by different
  symbols). Each symbol represents the entire galaxy at one time between 25 -
  250 Myr, the open 
  symbols corresponding to
  the final snapshot at 250 Myr. The black crosses are 
  global measurements of normal spirals from \protect\cite{Kennicutt1998}, while red
  crosses are similar measurements for low surface brightness galaxies from \protect\cite{Wyder2009}.
  The contour is derived from multiple sub-kpc measurements of 18 galaxies, including spirals and dwarfs
  from \protect\cite{Bigiel2008}. We plot here the contour corresponding to more than 5 data points per 0.05 dex-wide
  cell. For reference, we also plot the power law with a slope of 1.4 from \protect\cite{Kennicutt1998},
  fitted to both the data plotted here and higher surface density starburst galaxies.
  While classical
  feedback schemes agree with the observed Kennicutt-Schmidt relation only at
  the highest resolution, the mechanical feedback produces realistic SFR and
  gas surface densities at all three resolutions.}
\label{ks} 
\end{figure}
\vspace{-4ex}
\subsection{Host sites of star formation and supernovae} \label{host}
\begin{figure*}
\centering
\includegraphics{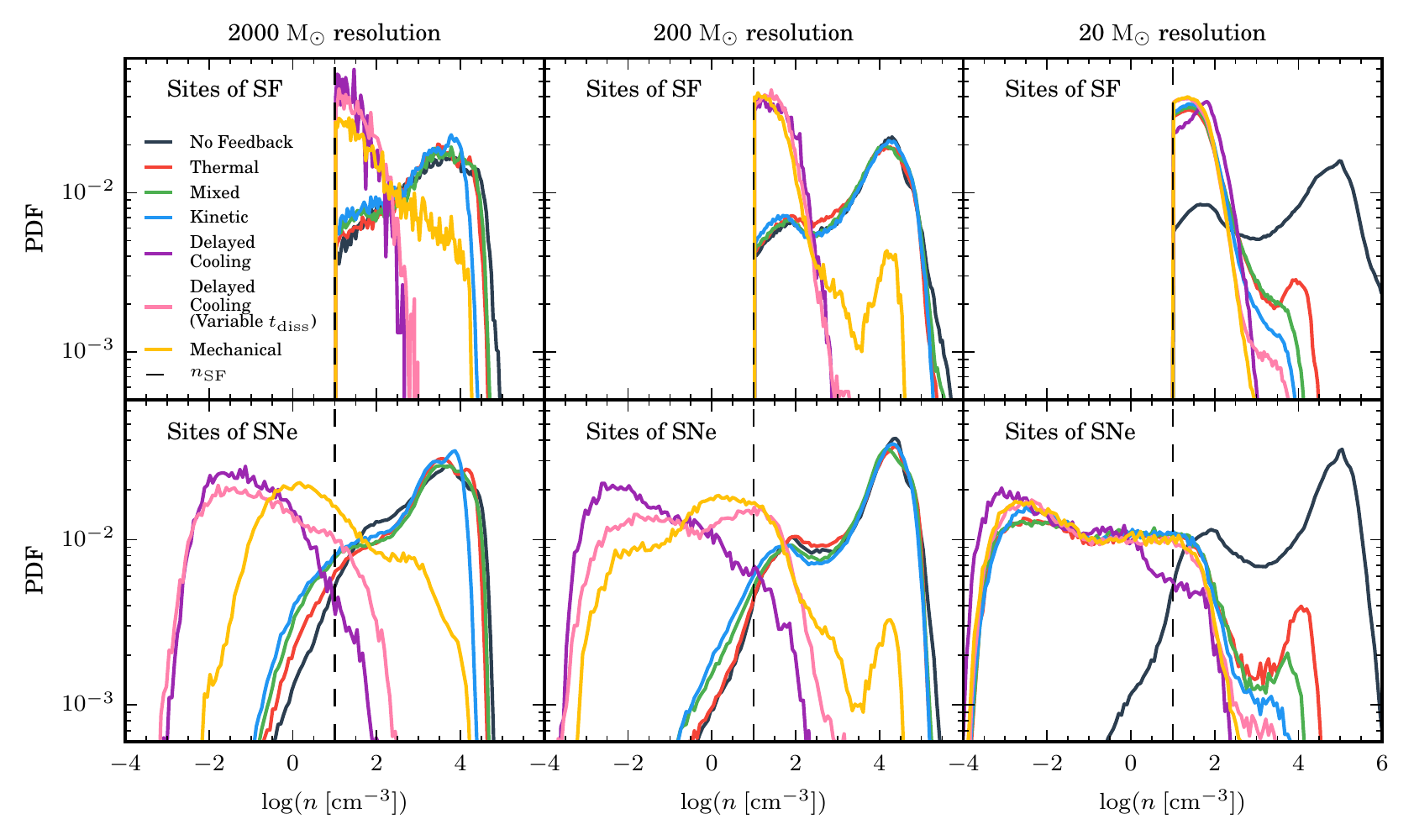}
\caption{PDFs of the densities of the sites where stars are formed (top)
  and where SNe occur (bottom) throughout the entire simulation. In addition,
  the star formation density threshold is marked with a vertical dashed line. Without
  efficient feedback, the majority of stars form at high densities and SNe occur
  in these regions. If the feedback is able to disrupt the dense birth
  clouds, then subsequent SNe occur at much lower densities, leading to a tail
  in the PDF well below the star formation density threshold. This also prevents star forming
  gas from reaching such high densities.}
\label{dens_pdf_lr} 
\end{figure*}
\begin{figure*}
\centering
\includegraphics{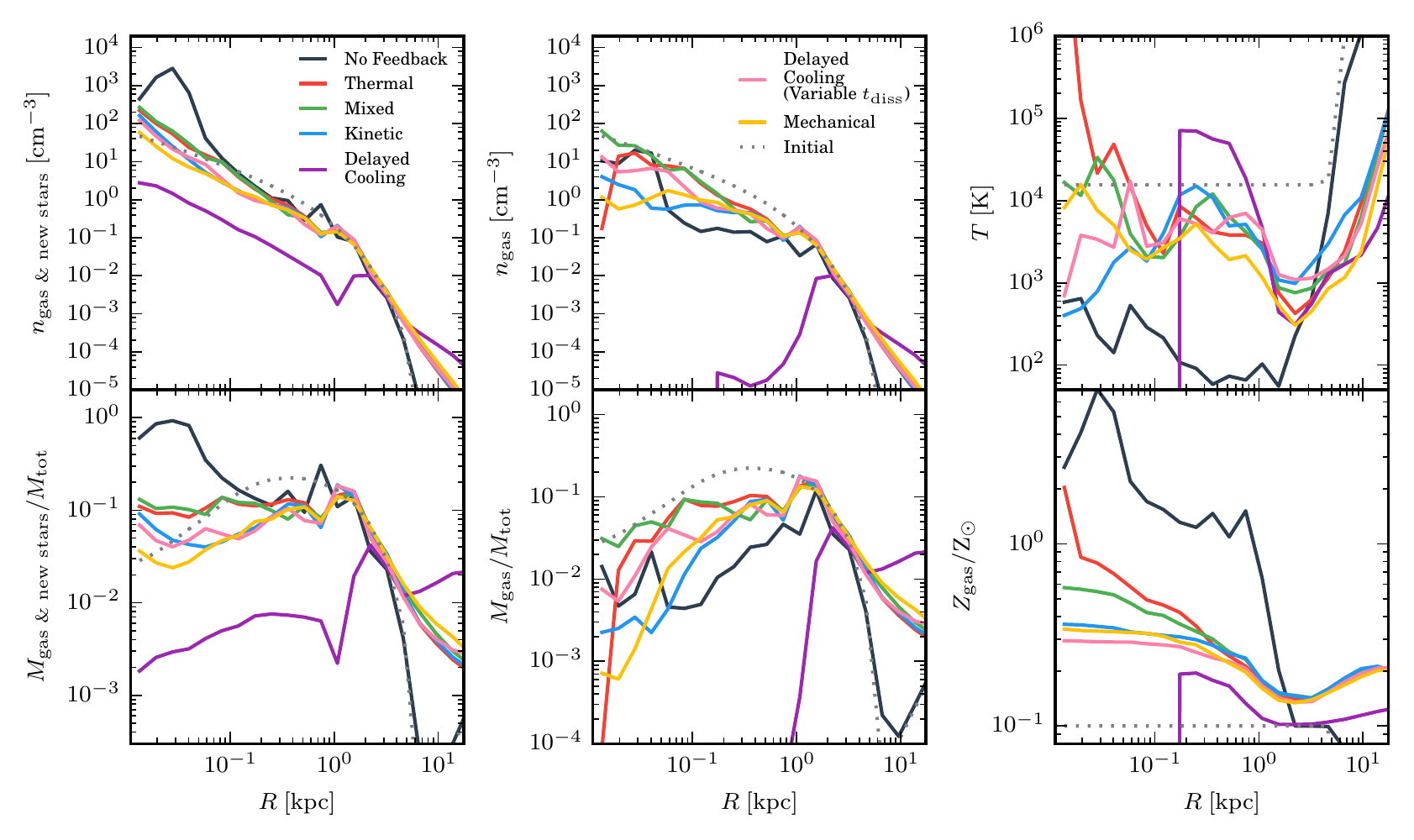}
\caption{Spherically averaged radial profiles of number density of gas and newly formed stars, gas number density, gas
  temperature (top panels) and mass fraction of gas and newly formed stars, gas mass fraction and gas
  metallicity (bottom panels) for our different feedback runs at $250$~Myr for
  simulations with $20\,{\rm M_\odot}$ resolution. The profiles from the initial
conditions are shown with gray dotted curves.}
\label{profiles_uhr} 
\end{figure*}

Fig.~\ref{dens_pdf_lr} shows PDFs of the local densities where stars are
formed (top panels) and SNe explode (bottom panels) for our different feedback
runs and at all three resolutions. Looking at the sites of star formation in
the simulations without feedback, a double peak form is apparent (though in
the lowest resolution, the lower density peak is suppressed into more of a
tail). This shape is a consequence of using a star formation threshold
density (indicated by a vertical dashed line in Fig.~\ref{dens_pdf_lr}). At
the beginning of the simulation, as gas densities increase and cross this
threshold, the first burst of star formation occurs, building a peak in the
PDF just above the threshold density. The gas continues to clump until it
reaches the maximum density the resolution and pressure floor allows. The
majority of star formation then occurs at this density, building a high
density peak in the PDF, enhanced by the fact that the SFR is higher in denser
regions (i.e. $\dot{\rho}_*\propto\rho^{3/2}$). The sites where SNe occur (in
this case, where mass and metals are returned but no feedback energy is
deposited) are therefore an almost direct mapping from the star formation PDF
because the local ISM is essentially unchanged from the star particle being
born to its SNe events occurring (although continued star formation will act to
drop the local density by transferring gas mass into star particles,
while gas may continue to collapse to higher densities before the first SNe occur). 

\setlength{\parskip}{0pt plus 0pt}
Inefficient feedback (i.e. the classical schemes at the lower two resolutions)
follow the same behaviour as they are unable to disrupt the dense clumps of
gas where the star particles are formed. In contrast, efficient feedback is
able to prevent the gas from clumping to high densities, therefore increasing
the fractional contribution of lower density star formation. In the case of
very strong feedback that disrupts the disk (i.e. delayed cooling) the PDF of
star formation is entirely dominated by the first burst of star formation,
before the SNe that those star particles produce completely quench star
formation. For more moderate feedback runs, the effect of a cycle of varying
SFRs, caused by the return of previously ejected mass (the galactic fountain),
can allow a building up of a small peak at high density. For example, in the
$200\ {\rm M_\odot}$ resolution mechanical feedback run, after a period of low
SFR between $50$ and $200$~Myr (see Fig.~\ref{disk_sfr}), the small peak at
$\sim10^4\ \mathrm{cm^{-3}}$ is able to form before the feedback is able to start
destroying clumps again. The degree to which this effect occurs is an
indication of how effective a given feedback scheme is at dispersing dense gas
at a low local SNe rate, since this is directly
linked to the local SFR (with an offset arising from a delay between massive
star formation and SNe occurrence). In other words, once the SFR has been
reduced by feedback, if the resulting lower SNe rate is unable to efficiently
prevent the return of gas, a high density peak in the SF PDF will
occur. Specifically, in the highest resolution simulations,
the thermal feedback shows a small peak at $\sim10^4\ \mathrm{cm^{-3}}$ while
the mechanical feedback does not (with mixed, kinetic and delayed cooling with
variable dissipation time lying between the two in order of their
effectiveness). Note that the shape of the PDF above the star formation threshold
will also be dependent upon the details of the small scale star formation prescription
adopted; this is discussed in Section~\ref{sf_law}.

\vspace{2ex}
With efficient feedback, the shape of the PDF of densities of SNe sites above
the star formation threshold corresponds closely to the star formation
PDFs, as in the inefficient feedback case. However, the PDF extends well below
the star formation density threshold. Since by definition the star particles
from which the SNe are occurring cannot have been formed at these densities,
these SNe are occurring after previous SNe have disrupted the star forming
regions of their birth cloud.\footnote{An alternative mechanism by which SNe
  can occur outside of dense star forming regions requires the SNe progenitors
  to have moved out of their birth clouds \citep[i.e. OB runaways, see e.g.][]{Conroy2012}, most likely as a result of
  interactions with other stars. Because we do not resolve the dynamics of
  individual stars in their clusters, this effect is not present in our
  simulations. We could adopt an additional sub-grid recipe to replicate this
  \citep[e.g.][]{Ceverino2009,Kimm2014,Kim2016a}, but this is beyond the scope of this work.} Of course, more
efficient feedback results in more SNe occurring in low density
environments. These subsequent SNe are themselves more efficient because
momentum input into the ISM is higher at lower ambient densities (but also
numerically, in the case of the classical schemes, because it is easier to
resolve the Sedov-Taylor phase). Thus, particularly when other faster acting
stellar feedback effects are not included (as in our work), a major
requirement of an efficient SNe feedback scheme is that the first SNe to occur
in a star forming region are able to disperse the dense gas to allow the
efficiency of later SNe to be increased. The classical feedback schemes are
only able to achieve this at the highest resolution probed. The mechanical
feedback scheme is more successful at all resolutions, but with increasing
resolution more SNe go off in lower density environments. At the highest
resolution, the shape of the PDF below the star formation density threshold is
similar for all feedback schemes. Because the delayed cooling schemes result
in a rapid clearing of gas from the centre of the system, most SNe occur in
low density gas at all resolutions.
\setlength{\parskip}{0pt plus 0.1pt} 

\subsection{Structure and kinematics}
Fig.~\ref{profiles_uhr} shows the spherically averaged radial profiles of
number density and mass fraction of gas together with new stars, and of gas
separately, as well as mass-weighted gas temperature and metallicity, for the
highest resolution simulations at $250$~Myr. The profiles from the initial
conditions are also plotted for comparison. The no feedback shows an
enhancement in baryon number density and mass fraction, due to a large centrally positioned clump (see
Fig.~\ref{disk_projections_uhr}). A smaller peak caused by another clump
further out is also apparent. However, the density and mass fraction of the
gas component taken on its own are significantly reduced from the initial
conditions, indicative of a major conversion of gas to stars in situ. The
profiles are unchanged at large radii as there has been no outflow of
material. The temperature within the disk gas has dropped by several orders of
magnitude due to metal cooling. The metallicity of disk gas has increased by a
factor of $10-100$ because of the high SFR in conjunction with the lack of
outflows, resulting in very short cooling times.

\begin{figure*}
\centering
\includegraphics{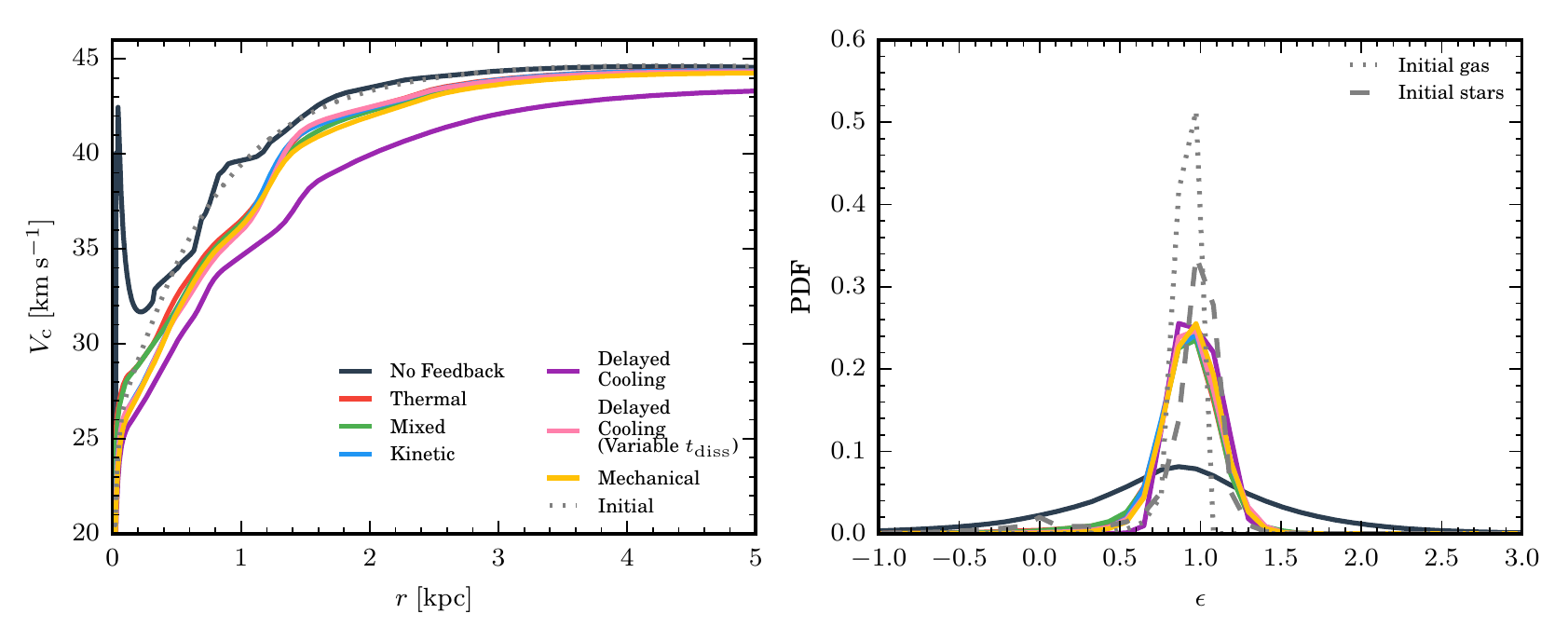}
\caption{Circular velocity profiles (left) and circularities, $\epsilon =
  j_z/j_c$, (right) for newly formed stars at $250$~Myr for our different
  feedback simulations at $20\,{\rm M_\odot}$ resolution. We also plot the
  initial circular velocity profile (left) and distribution of circularities
  for stars present in the initial conditions and of gas (within 3 scale radii and 3 scale heights) (right).
  The circular velocity profiles are largely unchanged from the initial conditions (due to the largely
  unchanged initial stellar disk and the static halo potential). The no feedback simulation is peaked
  close to the centre due to presence of a clump of gas and stars. The simulations with feedback reduce the circular
  velocity slightly by transporting mass outwards. All simulations have circularity distributions centred
  at $\sim1$, indicative of a disk. There is no signature of a bulge component. The no feedback simulation
  has a broader distribution of circularities due to the highly clumped disk structure.}
\label{circ_uhr} 
\end{figure*}

The classical, mechanical and delayed cooling with variable dissipation time
feedback schemes are similar to the initial conditions with respect to the
baryon number density and mass fraction in the central regions. The gas mass
fraction has been reduced, as have the central densities, partly due to
conversion of gas to stars (as in the no feedback case) but mainly due to
outflowing material. This outflow material can be seen at larger radii, where
the gas mass fraction outside $\sim4\ \mathrm{kpc}$ has been significantly
increased. The gas within the disk has been prevented from runaway cooling and
the average temperature within the is mostly between
$10^3-10^4\ \mathrm{K}$. Variations in temperature between the different
feedback schemes are largely transient and stochastic, particularly at small
radii (where the average is over less gas mass). Temperatures are significantly
reduced from the initial conditions outside a few kpc, as colder outflowing
gas displaces the hot CGM. Central metallicities are increased from the
initial conditions by a factor of $\sim3 - 8$ (with the exception of a
metallicity spike from the recent feedback event in the thermal simulation
which has not yet dispersed). Metals have been transported into the region
initially occupied by the CGM. As previously described, the delayed cooling
with fixed dissipation time evacuates gas from the central regions extremely
efficiently, resulting in a large drop in the central gas density and mass
fraction, a spike in temperature of the remaining gas but a very small
increase in metallicity (because there have been very few SNe). Despite the
explosive nature of delayed cooling feedback, the mass-averaged temperature at
the outer radii ($1-10\ \mathrm{kpc}$) is still much lower than the original
CGM temperature. The lower resolutions simulations (not shown) have very
similar profiles to those described above, dependent on whether how effective
the feedback is (i.e. the classical schemes overcool so adopt the same
behaviour as the no feedback case). We have also examined surface 
density profiles (not shown) and find that the results are similar
to the radial profiles. Simulations with feedback preserve the initial exponential
density profile of the disk in terms of total baryons. The gas profiles are centrally
cored relative to the initial profile and densities are enhanced at outer radii due to
outflows.

Fig.~\ref{circ_uhr} shows the circular velocity profiles and distribution of
circularity parameter, $\epsilon=j_z/j_\mathrm{circ}$, for newly formed stars
for our different feedback runs at the highest resolution. The circular
velocities are generally similar to initial conditions, because the distribution 
of the initial stellar disk and (small) bulge of old stars is largely unchanged 
(these components making up over half of the initial baryonic mass), while at larger radii the circular
velocity is dominated by the static halo potential. The no
feedback case shows a peak at small radii due to the centrally positioned
clump remarked upon earlier. Note, as described above, this clump cannot be
taken to be indicative of bulge formation. As seen in
Fig.~\ref{disk_projections_uhr}, there are multiple clumps present in the
disk. The position of the clump is somewhat stochastic. We have found other
simulations at a variety of resolutions to produce the same degree of
clumping, yet not necessarily with a clump positioned close enough to the
centre to produce a central peak in the circular velocity profile. However, as
seen in Fig.~\ref{circ_uhr}, efficient feedback results in a reduction in
circular velocity, particularly at small radii, due to the transport of gas
mass out to further radii.

The circularity parameter gives an indication of the degree of rotational
support in a system by comparing the specific angular momentum in the
$z$-direction (i.e. out of the disk plane) to that required for a circular
orbit at the same radius.  Thus, stars belonging to a disk will have
$\epsilon\sim1$, whereas a non-rotating spheroid would have a symmetric
distribution about $\epsilon=0$.  All simulations have a peak around
$\epsilon\sim1$, indicative that the newly formed stars form a disk, while 
there is no clear signature of a bulge component (such as the small enhancement
at $\epsilon=0$ present in the initial stars). This is not surprising because
the stars have formed directly from the gas disk. The distribution is peaked
very slightly below $\epsilon=1$; this is inherited from the initial gas distribution
which has some degree of pressure support in addition to the rotational support (as
can be seen by comparing the initial distribution of gas circularities to those of
the stellar disk). The difference between no feedback and feedback simulations
is apparent in the width of the distribution. With feedback, the stellar circularities are relatively
narrowly distributed around 1 (irrespective of the feedback scheme adopted), whereas in the no feedback gas the distribution is considerably broader.
This is caused by the highly clumpy disk formed in this run, with stars acquiring significant offsets
from the circular velocity due to local interactions with clumps.

We conclude this section by remarking that a proper study into the effects of
the feedback schemes on galaxy structure and kinematics should make use of
galaxies formed self-consistently in a cosmological context, rather than in a
disk set up `by hand' as in this work. However, we find it informative to
examine the extent to which an ideal system is maintained in the presence of
our feedback schemes, ranging from a strongly clumped distribution in the no
feedback case to total disk destruction due to overstrong feedback in the
delayed cooling case. 
\vspace{-4ex}
\subsection{Varying galaxy mass}
\begin{figure*}
\centering
\includegraphics{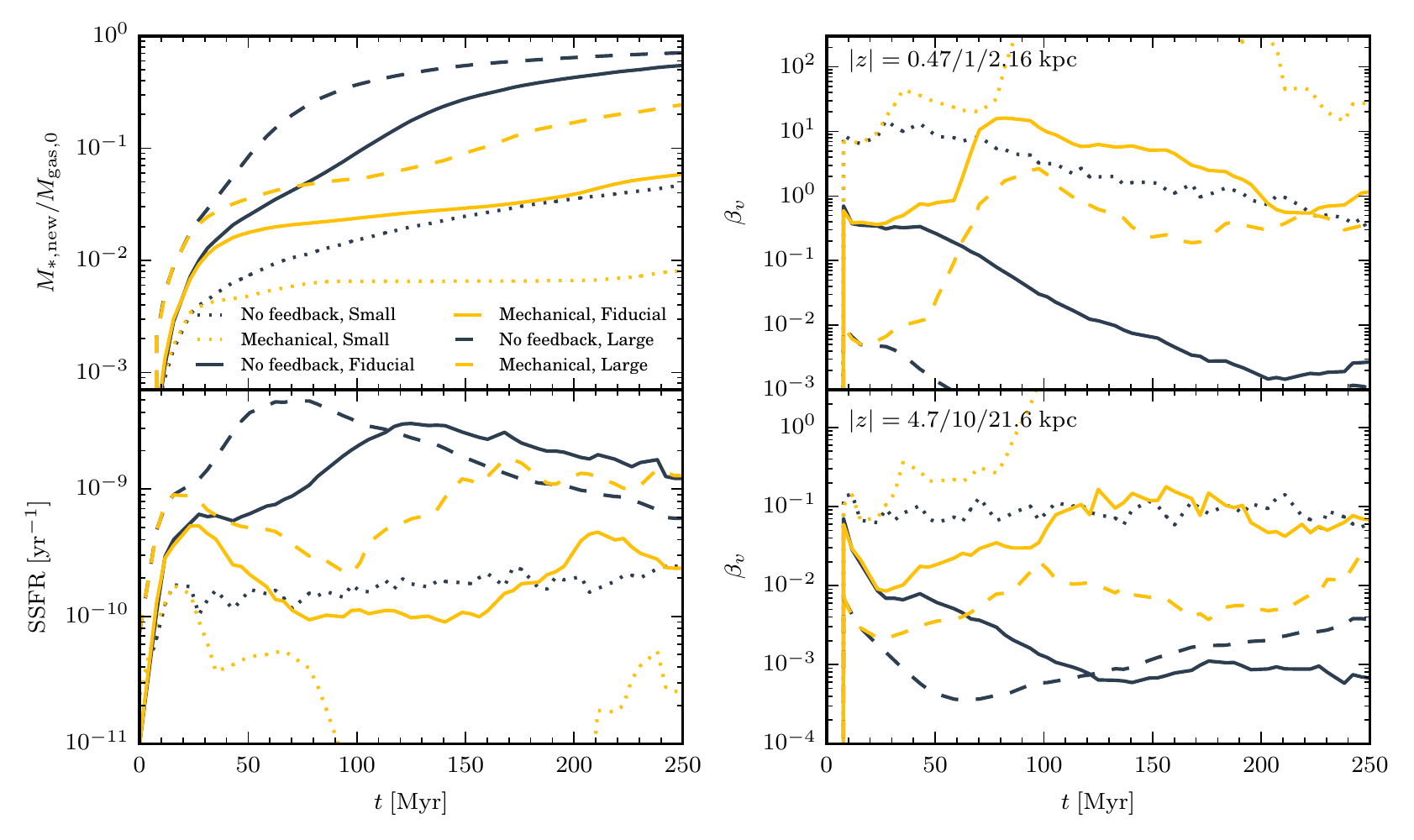}
\caption{A comparison of simulations with no feedback and with mechanical
  feedback for our three galaxy masses (see Table~\ref{Table1}) at $200\ {\rm
    M_\odot}$ resolution. \textit{Top left}: Newly formed stellar mass
  expressed as a fraction of the initial gas disk mass. \textit{Bottom left}:
  specific star formation rates ($\dot{M}_*/M_*$ where $M_*$ includes both old and new stellar mass). \textit{Right}: mass loading factor across two planes
  at different distances from the disk midplane. For the fiducial galaxy,
  these are 1 and 10 kpc as in Fig.~\ref{massload}. For the small and large
  galaxies, the planes are at the same distance relative to the virial radius
  as in the fiducial case (0.47 and 4.7, 2.16 and 21.6 kpc,
  respectively). Global star formation efficiency increases with increasing system mass (a trend in line
  with abundance matching), though this appears to be independent of feedback in our setup. Feedback suppresses
  star formation by a similar factor in all three systems. Outflows become weaker with increasing system mass.}
\label{disk_multiG} 
\end{figure*}
In addition to our fiducial $10^{10}\ \mathrm{M}_{\rm \odot}$ galaxy, we have also run
simulations of smaller ($10^{9}\ \mathrm{M}_{\rm \odot}$) and larger ($10^{11}\ \mathrm{M}_{\rm
  \odot}$) systems without feedback and with mechanical feedback at our
intermediate resolution of $200\ {\rm M_\odot}$. The results of these
simulations are summarised in Fig.~\ref{disk_multiG} alongside the equivalent
simulations of our fiducial system, showing mass of newly formed stars
(expressed as a fraction of the initial disk gas budget), specific star formation rates (SSFR) 
and mass loading factors at two distances from the midplane of the disk. For a comparison 
between the systems, the mass loading is measured across planes the same distance away
from the centre scaled by the virial radius of the systems. For our fiducial
galaxy mass this is $1$~kpc and $10$~kpc, so we use $0.47$~kpc and $4.7$~kpc
for the smaller system\footnote{We use a thickness $\Delta z=100\ \mathrm{pc}$
  for determining the mass outflow rates for the smaller mass system.} and
$2.16$~kpc and $21.6$~kpc for the larger system.

In the smaller system, the simulation without feedback quickly starts
forming stars from the beginning of the simulation, though it quickly
establishes a steady SSFR of $\sim2\times10^{-10}\ \mathrm{yr}^{-1}$ as
the smaller surface densities prevent
gas from clumping to high densities. 
This SSFR is approximately an order of magnitude smaller than the fiducial
simulation. The mechanical feedback
initially follows the same evolution as the no feedback case. However,
once the SFR has reached its peak, the SNe are able to unbind the gas from
system due to the shallow potential well, quenching star formation. At
$250$~Myr, the ratio of newly formed stellar mass to the initial gas disk mass
is approximately an order of magnitude lower than the fiducial simulation. This 
trend agrees with the general results from abundance matching that lower
mass galaxies are less efficient at forming stars (below $\sim10^{12}\ \mathrm{M_\odot}$) \citep[see e.g.][]{Moster2013,Behroozi2013}. However,
the factor by which feedback has suppressed star formation relative to the
no feedback simulation is similar to the fiducial case. This suggests that
the lower star formation efficiency relative to the fiducial simulation is inherent
to our particular setup, rather than being caused by more efficient feedback as is
commonly posited to explain the phenomenon. After the spuriously high mass loading due to
the low SFR at the beginning of the simulation has reduced (note that the no feedback
simulation continues to have a relatively high apparent mass loading for the same reason), the mass loading
at $0.47$~kpc and $4.6$~kpc rises dramatically as the gas is expelled from the
system. The mass loading then tends to infinity because the SFR has dropped to
zero (the instantaneous mass loading factor is not a good metric in a highly bursty
regime).

Without feedback, the larger system follows an evolution similar to that of
the fiducial system. However, it forms stars more than proportionally faster
than the fiducial case, resulting in $\sim3$ times more stellar mass formed
than a simple scaling with the system mass. With mechanical feedback, the
result is similar relatively speaking, with star formation suppressed by a
similar factor. Once again, the trend of a greater star formation efficiency
seems to be qualitatively in line with abundance matching, but, as with the
low mass system, this effect seems to be inherent to the set-up rather than
caused by less efficient feedback.
The mass loading factor is mostly within a factor of a few of the fiducial
simulation at both distances, though is always lower. The mass loading factor
is mostly below unity at $2.16$~kpc and reaches a maximum of only $0.1$ at
$21.6$~kpc, i.e. the outflows are even more inefficient than in our fiducial
galaxy model, which is not unsurprising given the deeper potential well
outflows need to overcome.
\vspace{-4ex}
\subsection{Varying star formation law parameters} \label{sf_law}
\begin{figure*}
\centering
\includegraphics{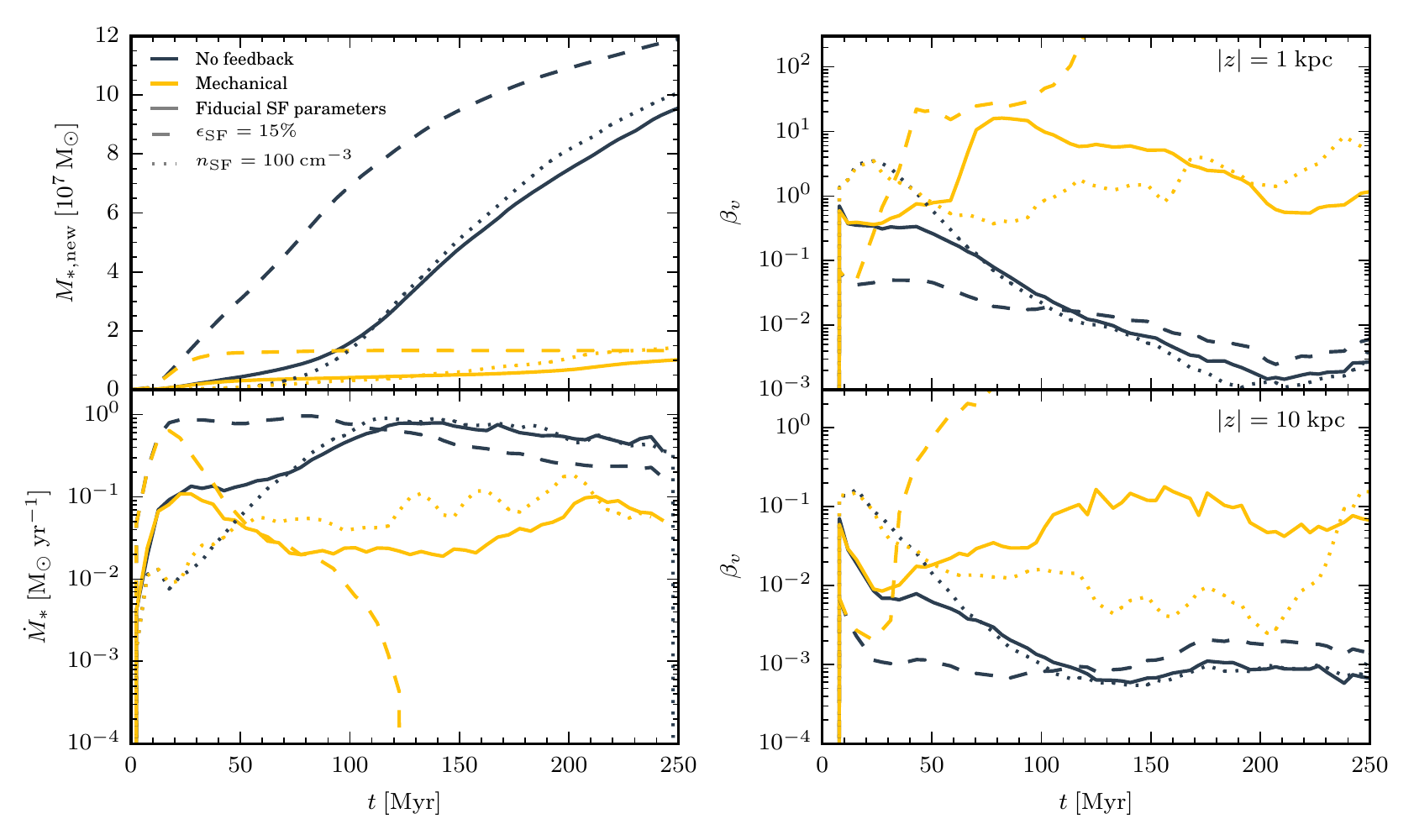}
\caption{Simulations with no feedback and mechanical feedback with varying
  star formation criteria at $200\ {\rm M_\odot}$ resolution for our fiducial
  galaxy. We compare out fiducial values for star formation
  ($\epsilon_\mathrm{SF}=1.5\%$, $n_\mathrm{SF}=10\ \mathrm{cm^{-3}}$) with an
  increased star formation efficiency ($\epsilon_\mathrm{SF}=15\%$) or an
  increased star formation threshold density
  ($n_\mathrm{SF}=100\ \mathrm{cm^{-3}}$). \textit{Top left}: Newly formed
  stellar mass. \textit{Bottom left}: SFRs. \textit{Right}: mass loading
  factor across two planes at different distances from the disk
  midplane. These are at $1$~kpc and $10$~kpc as in Fig.~\ref{massload}.
  Increasing $\epsilon_\mathrm{SF}$ results in faster star formation, leading to
  much a much stronger burst of feedback which quenches subsequent star formation. Increasing
  $n_\mathrm{SF}$ results in a similar evolution of stellar mass to the fiducial case,
  but produces weaker outflows.}
\label{disk_multi_sf} 
\end{figure*}

While the focus of this work is on the difference between different feedback
schemes, we also briefly examine here the effect of varying the parameters
used with our adopted star formation law (see
equation~(\ref{eq:schmidt})). We rerun our $200\ {\rm M_\odot}$ resolution
simulations of the fiducial galaxy without feedback and with mechanical
feedback but with an increased star formation efficiency parameter,
$\epsilon_\mathrm{SF}=15\%$ rather then the fiducial $1.5\%$ and also with an
order of magnitude higher density threshold,
$n_\mathrm{SF}=100\ \mathrm{cm^{-3}}$. The results are summarised in 
Fig.~\ref{disk_multi_sf} alongside the fiducial simulations. 

Increasing the star formation efficiency parameter by an order of magnitude
results in the initial SFR being an order of magnitude higher than the
fiducial case both with and without feedback. The simulation without feedback
maintains this high SFR ($\sim1\ \mathrm{M}_{\rm \odot}\ \mathrm{yr^{-1}}$), dipping
slightly below the fiducial simulation's SFR, which has risen to this value,
at $\sim110\ \mathrm{Myr}$ as the gas reservoir is consumed. The final newly
formed stellar mass is approximately 1.25 times larger than the fiducial
run. In the mechanical feedback case, the high SFR leads to a burst of strong
feedback at 20 Myr which expels the gas from the centre of the system and
quenches star formation. The mass in newly formed stars at 250 Myr is similar
to the fiducial simulation, but the majority of stars have formed in the first
50 Myr. Once again, the instantaneous mass loading factor for a regime that quenches star
formation is an unreliable metric (as it tends to infinity as the SFR
plummets). However, there is a brief period between $20-100$~Myr where
the SFR is non-zero, leading to a mass loading factor between $1-50$ at
$1$~kpc. The outflow also easily reaches the $10$~kpc plane as can be seen by
the high mass loading factor.

Increasing the star formation density threshold by an order of magnitude
results in an initially lower SFR in the simulation without feedback as it
takes slightly longer for the gas to reach the higher star forming
densities. However, by $70$~Myr, the SFR has reached the levels of the
fiducial simulation and subsequent evolution is similar, resulting in the
almost the same mass in new stars at $250$~Myr. The simulation with mechanical
feedback is similar until $50$~Myr, when the SNe are able to halt further
rising of the SFR. A stable SFR is established, a factor of a few higher than
the fiducial simulation. The stellar mass at $250$~Myr is only $1.4$ times
that of the fiducial simulation. The outflow is weaker at $1$~kpc by a factor
of a few than the fiducial case for most of the simulation, stable at around
unity. However, at $10$~kpc the outflow is very weak, with mass loading factor
between $10^{-3}-10^{-2}$ (an order of magnitude lower than the fiducial
simulation). 

\begin{figure}
\centering
\includegraphics{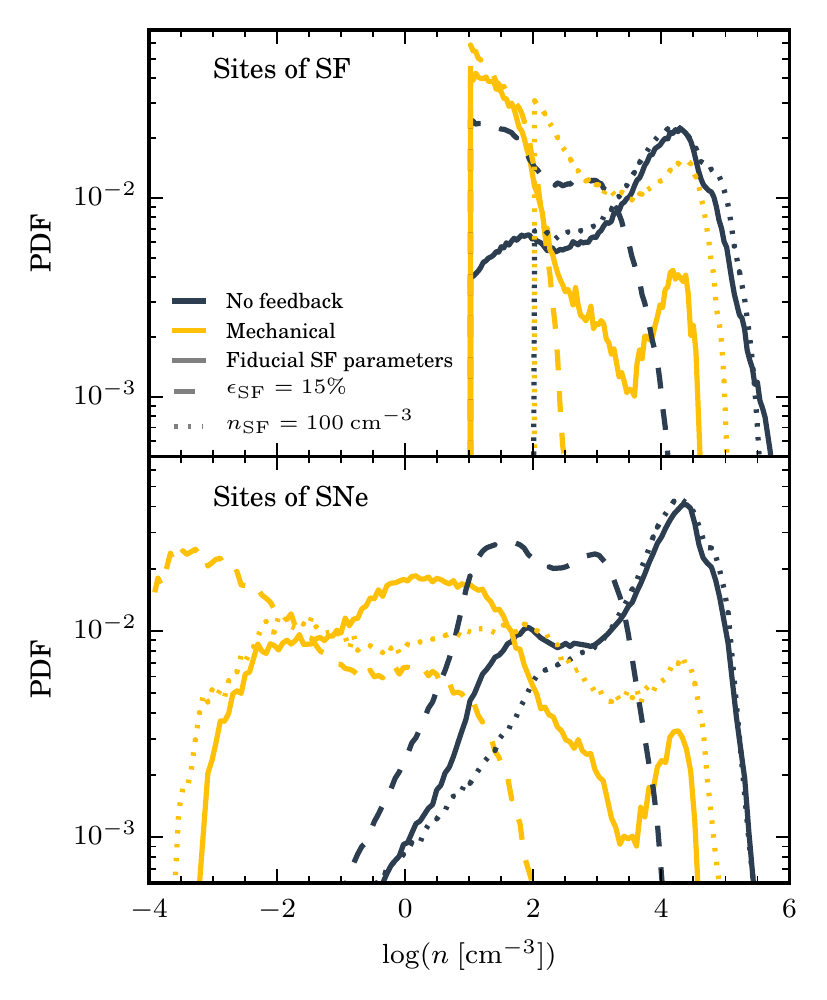}
\caption{PDFs of the densities of the sites where stars are formed (top)
  and where SNe occur (bottom) for the simulations in
  Fig.~\ref{disk_multi_sf}, comparing the effects of changing the star
  formation law parameters using our fiducial galaxy mass and $200\ \mathrm{M}_{\rm
    \odot}$ resolution. Increasing $\epsilon_\mathrm{SF}$ results in gas
    being rapidly converted to stars before it can reach high densities. A strong
    burst of initial feedback disrupts star forming clouds, so the majority of SNe
    occur at very low densities. Increasing $n_\mathrm{SF}$ has very little impact
    on the PDFs, though slightly more SNe occur at high densities.}
\label{dens_pdf_multi_SF} 
\end{figure}

Fig.~\ref{dens_pdf_multi_SF} shows PDFs of the gas densities at the sites of
star formation and SNe explosions, comparing the models with altered star
formation law parameters to the fiducial simulation. The simulation with
increased star formation efficiency produces most of its stars in gas that is
several orders of magnitude less dense than the fiducial case, because gas is
rapidly converted to stars before it can collapse to higher densities. This
means that SNe occur in lower density gas, enhancing their momentum input into
the ISM. This, coupled with the increased SNe rate relative to the fiducial
run gives rise to the quenching of star formation and the generation of
stronger outflows.

In the simulations with a higher star formation threshold density, without
feedback, the PDFs of star formation and SNe site densities are similar,
because most star formation occurs in gas at $n=10^4\ \mathrm{cm^{-3}}$, well
above the thresholds. However, in the runs with mechanical feedback, the
situation is different. In the fiducial simulation, feedback shifts the peak
star formation density down to the threshold. In the simulation with a higher
threshold density, stars are born at much higher densities. The result is that
SNe occur in gas of higher density which reduces their momentum input to the
ISM. The feedback is still strong enough to disrupt star forming regions,
which is why the SFR is close to the fiducial case, but the reduction in
momentum results in weaker outflows. Note that the reduction in momentum input
is not due to overcooling, but is physical (see equation~(\ref{p_fin})),
although the subsequent development of outflows is subject to resolution
effects, as discussed in Section~\ref{subsec_sfr_out}. 

What we have demonstrated in this section is that changing the star formation
prescription (to not unreasonable values) can have a non-negligible effect on
SFR and outflows, although a more comprehensive study of these effects is
beyond the scope of this work. We note that a low mass system such as our
fiducial model is likely to be less robust to changes in the star formation
prescription because it is easy to unbind gas if the feedback increases in
strength, leaving little margin for self-regulation, although we find our
results to be broadly in agreement with similar tests in
\cite{Rosdahl2017}. It is also likely that the inclusion of other feedback
processes that act in the time between star formation and the resulting SNe
might possibly help mitigate the dependence on the star formation prescription
by preventing further collapse of gas (see \cite{Hopkins2011a,Hopkins2013} for
examples of self-regulating systems that are somewhat robust to the star
formation prescription in terms of the global SFR). 

\section{Discussion} \label{Discussion}
\vspace{-1ex}
\subsection{Comparison of SN feedback implementations}
We find that our `classical' schemes (simple injection of thermal energy, kinetic energy or
a mixture of the two) all give very similar results. There is a slight trend for an injection
of kinetic energy to result in stronger feedback but the effect is minor. At all but the highest
resolution of $20\ \mathrm{M_\odot}$ these schemes suffer from the overcooling problem, barely
suppressing star formation relative to the no feedback case and producing similar clumpy
morphologies. This is not unexpected. 

One can alleviate the problem slightly by injecting
the energy of several SNe at once, but only up to a point. For example, \cite{Rosdahl2017} inject
the energy of 40 SN simultaneously which allows a thermal dump to be efficient in their $10^{11}\ \mathrm{M_\odot}$ system but not in their $10^{12}\ \mathrm{M_\odot}$ system, due to a combination of the deeper potential,
stronger metal line cooling, higher densities and lower resolution.
 In addition, such an approach requires the adoption of
an artificial delay time between the birth of a star particle and the triggering of a SN event.
\cite{Kimm2015} find that allowing for a realistic delay time, with individual SNe distributed between
$\sim3-40\ \mathrm{Myr}$, prevents the build up of dense gas prior to SNe occurring relative to a fixed
delay time of 10~Myr, while also allowing later SNe to explode in low density environments produced by
earlier events. The caveat, of course, is that individual SNe are more susceptible to overcooling.

Our trial of delayed cooling schemes is unsatisfactory. Unlike the other schemes explored, these schemes
have adjustable parameters, which is something we wish to avoid if possible. In addition, delayed cooling
schemes circumvent unphysical results caused by lack of resolution by enforcing an equally unphysical
adiabatic phase on large scales. The scheme with a fixed
dissipation time of 10~Myr produces far too violent feedback at all resolutions, completely destroying the disk and giving rise to an unphysical pattern of star formation as gas is ejected from the system. This suggests
our choice of parameters is incorrect, though we test a higher effective velocity dispersion 
threshold ($100~\mathrm{km\ s^{-1}}$ instead
of our fiducial $10~\mathrm{km\ s^{-1}}$) without a drastic change in results (see Appendix \ref{appendix_DFB})
and also note that our choice of parameters is not wildly different from others used in the literature at
similar resolutions (see e.g. \citealt{Teyssier2013,Rosdahl2017}, though \citealt{Dubois2015} determine lower
values for the dissipation time). Our attempts to modulate the dissipation time with resolution ($t_\mathrm{diss}
=\Delta x /\sigma_\mathrm{FB}$), suggested in \cite{Teyssier2013} as an alternative parametrization, do not converge with resolution, being similar to the fixed dissipation time at low resolution while essentially
acting as a simple thermal dump at high resolution. No doubt both these schemes could be improved were we to
spend more time tuning the parameters, but this would not achieve our goal of finding a physically motivated
model ideally free from adjustable parameters. We also note the concerns of \cite{Rosdahl2017} that their delayed cooling
scheme trialled does not converge with their thermal dump when the adiabatic phase is resolved (as we find), suggesting that the scheme does not necessarily converge to the correct answer.

The most successful scheme explored is the mechanical feedback scheme. It suppresses star formation by similar
factors across two orders of magnitude in mass resolution (though is slightly stronger at higher resolution), prevents the formation of highly dense clumps
of gas, preserves the disk structure and agrees with observations of the Kennicutt-Schmidt relation (though
the exact position on the relation has a resolution dependence caused by non-convergent outflow properties,
discussed below). It also
gives similar results to the classical schemes at the highest resolution, suggesting it is converging onto the correct answer. This latter feature was also noted in \cite{Rosdahl2017} and demonstrates that the ability of the
scheme to converge on the final momentum input to the ISM per SN \citep[as shown in][]{Kimm2014} translates into convergent behaviour for global properties. 
The mechanical feedback is slightly stronger than the classical feedback schemes because, even at this resolution, they are likely to still experience some overcooling at the highest density SN sites. 
The one area where the mechanical scheme does not converge is in outflows, which
we discuss next.
\vspace{-4ex}
\subsection{Difficulties in outflow generation and the possible effects of missing physics}
Having concluded in the previous section that the mechanical feedback scheme is the best amongst those explored, we choose to focus on it for this discussion. At the highest resolution, the scheme produces well developed multiphase outflows with appropriate mass loadings compared to observations and theory
($\beta_v \approx 1-10$), which is very encouraging
(similar results are obtained by the classical schemes and the variable $t_\mathrm{diss}$ delayed cooling at this
resolution). What is not so encouraging, however, is that the outflows are considerably weaker at a mass resolution of $200\ \mathrm{M_\odot}$ and practically non-existent with $2000\ \mathrm{M_\odot}$, despite similar
results with other galaxy properties. This also has the effect of moving the disks up the Kennicutt-Schmidt
relation with decreasing resolution because the disk surface density is increased (though it should be noticed
that these still match observations). \cite{Rosdahl2017} also report difficulties in driving outflows with
mechanical feedback with a resolution similar to our lower resolutions simulations.

Such inefficient outflows could be caused by an oversimplified model of SN expansion. The mechanical feedback
scheme treats the unresolved evolution of the SN remnant as expanding through a uniform medium. In reality, the
ISM is likely to be porous due to a turbulent structure, containing low density channels through which gas
accelerated by the SN can escape, leading to higher velocities. \cite{Haid2016} model this effect by considering
the ISM surrounding the SN as a set of cones of different densities (randomly drawn from a log-normal
distribution appropriate for the level of turbulence assumed) and use the results for a uniform medium within
each cone. They find that momentum can be boosted by up to a factor of 2 in a low density environment. Our scheme
already approximates this approach because it calculates the boost factor for each neighbour cell independently,
a point that is argued by \cite{Hopkins2018}. However, we would caution that this assumes that the turbulent
structure of the ISM is well resolved in the simulation, which is very unlikely to be the case, and will introduce a resolution dependence. However,
the momentum boost measured in \cite{Haid2016} is weak; similar results are found in other studies with full
3D simulations \citep[e.g.][]{Iffrig2015,Martizzi2015,Kim2015,Li2015,Walch2015}, some of which find a slight
negative impact on final momentum input versus the uniform case. However, while the final momentum input into
the ISM may be only weakly effected by a turbulent medium, the amount of mass involved in the expansion and
therefore the wind velocities reached can be altered. \cite{Kimm2015} trial a modification of their version of
mechanical feedback in AMR simulations where they reduce the mass entrained from the host cell to 10\% to
replicate this effect, resulting in a greater suppression of star formation and higher mass loading factors. They
state, however, that this fraction was somewhat arbitrarily chosen. Unless a physically motivated method of
determining the fraction to be entrained based on unresolved structure was used, this could easily become just
another tunable parameter. A final problem is that with a constant mass, as by definition imposed with a Lagrangian code, there is a minimum mass
that can be momentum boosted. Even if we inject the correct momentum, the effects can be diluted if it is
injected into too much mass resulting in lower velocity winds, effectively imposing a minimum resolution
requirement.

Another potential cause of inefficient outflows experienced here could be the lack of other stellar feedback
mechanisms. In particular, the ability of other feedback mechanisms to disrupt GMCs will enhance subsequent
SN feedback. An obvious mechanism is that of photoionisation \citep[see e.g.][]{Vazquez-Semadeni2010, Walch2012, Dale2014, Sales2014}. \cite{Geen2015} found that the final momentum input to the ISM by SN is increased when
the surrounding medium has been preprocessed by photoionisation feedback by forming an over-pressurised and
lower density region in which the SN occurs. \cite{Kimm2017} modify their mechanical feedback prescription
to include this momentum boost when they under-resolve the Str\"{o}mgren sphere in their RHD simulations.
\cite{Hopkins2012b,Hopkins2014a} find that if they turn off radiative feedback (radiation pressure, photoionisation and photoelectric heating), outflow mass loadings are reduced because GMCs are no longer
efficiently disrupted prior to SN occurring (though the strength of the effect is dependent on the mass of the
system). Simulating an isolated system similar to our fiducial model, \cite{Hu2017} find that while SNe are the
dominant feedback mechanism, the inclusion of photoionisation increases outflow rates by reducing the ambient
density at SN sites. However, in their simulations, the inclusion of photoelectric heating reduces outflows because it reduces the
SFR and therefore the number of SN occurring, while being unable to drive outflows itself.

As noted in Section \ref{sf_law}, the choice of star formation prescription can also impact the effectiveness
of feedback. We found that increasing the star formation efficiency parameter by a factor of 10 led to stronger
outflows (and the destruction of the disk) because the SFR was initially higher and gas could not
reach high densities before SN occurred, leading to a sudden, strong burst of efficient feedback. Instead,
increasing the threshold density had only a marginal impact on the SFR, but produced weaker outflows because
SN occurred in slightly denser environments. The adoption of other feedback mechanisms would probably
mitigate this effect. It is also worth noting that we have only tested changes of parameters to our simple star
formation prescription. More complex prescriptions may rely on the selection criteria
of star forming gas rather than a efficiency parameter \citep[see e.g.][and subsequent
papers, which use
an efficiency of 100\%, but require gas to be self-gravitating, self-shielding and very dense]{Hopkins2013,Hopkins2014a}. Alternatively, it has been suggested that while the globally averaged
star formation efficiency may be on the order of a few percent, small scale efficiencies vary
based on the local properties of the ISM \citep[see e.g.][]{Krumholz2005,Padoan2011,Hennebelle2011,Federrath2012}. In line with this, a star formation prescription could adopt a variable efficiency \citep[see e.g.][]{Kimm2017}. Such schemes are likely to impact the distribution of gas densities by allowing high density non-star forming
gas to exist, as well as impacting SN feedback effectiveness by altering the clustering properties of stars in
both space and time.
\vspace{-6ex}
\section{Conclusion}
Using an isolated disk galaxy setup and a new implementation of star formation
and SN feedback in the moving mesh code \textsc{Arepo}, we tested several
SN feedback prescriptions commonly found in the literature and assessed their
impacts on a variety of galaxy metrics, paying particular attention to how
well they converge as a function of resolution. The bulk of our simulations
were of a $10^{10}\ \mathrm{M}_{\rm \odot}$ system, although simulations were carried
out of systems an order of magnitude lower and higher in mass. In order to
test the convergent properties of the feedback schemes with resolution,
simulations were carried out with resolutions of $2000\ {\rm M_\odot}$,
$200\ {\rm M_\odot}$ and $20\ {\rm M_\odot}$. The schemes tested were designed
to be used in isolated galaxy or cosmological zoom-in simulations, using
individually time resolved SN events. Specifically, we investigated
`classical' dumps of thermal 
and/or kinetic energy, two parametrizations of delayed cooling and a
mechanical feedback scheme which injects the correct amount of momentum
relative to the stage of the SN remnant evolution resolved.

Without feedback, our simulations produce a highly clumpy disk and overproduce the mass of newly formed stars.
As expected, the `classical' feedback schemes overcool at all but the highest resolution.
The delayed cooling schemes tested are far too strong, unphysically destroying the
disk. We note that we could tune these simulations more carefully to avoid this effect,
but because we wish to avoid adjustable parameters as much as possible we do not consider
these schemes to be suitable for our purpose. Our mechanical scheme is the best tested, suppressing star formation 
by similar factors at all three resolutions, preventing the formation of highly dense clumps
of gas, agreeing with observations of the Kennicutt-Schmidt relation while also preserving
the disk structure. It also produces similar results to the `classical' schemes at $20\ \mathrm{M_\odot}$
resolution, suggesting it is converging onto the physically correct results.

At the highest resolution our mechanical scheme produces multiphase outflows with reasonable mass loading
factors relative to observations and theory ($\beta_v \approx 1-10$), as do the `classical' schemes at the highest
resolution. However, we struggle to produce outflows at lower resolution.
This may be due to an oversimplification of the way in which we model SN remnant evolution, for example failing to
adequately account for the unresolved porous structure of the ISM. The situation may also be improved by the inclusion of other
forms of stellar feedback that are able to preprocess the ISM and enhance the ability of the SN feedback to drive
galactic winds. In addition, alternative star formation prescriptions that aim to better capture small scale star
formation physics will impact the effectiveness of feedback by altering the clustering properties of SNe (both in space
and time). Finally, it is worth noting that there exists some minimum resolution requirement for the driving
of outflows with individually time resolved SN because the injection of momentum (even if it is the physically 
correct amount) into too much mass will result in unphysically slow gas velocities. 

Finally, it is worth pointing out that the resolution requirements for the mechanical feedback scheme to work well in
terms of outflows is within the reach of next generation cosmological zoom-in simulations (at least for low mass
systems). This will allow us to explore realistic SN feedback in a full cosmological environment, self-consistently
taking into account the circulation of complex gas flows all the way from the cosmic web to the ISM. The adoption
of a more accurate star formation prescription in concert with the inclusion of other forms of stellar feedback in
such simulations may ultimately help us unveil what shapes star formation in low mass systems.
\vspace{-4ex}
\section{Acknowledgements}
MCS is supported by the Science and Technology Facilities Council (STFC). DS and SS acknowledge support by the STFC and
the ERC Starting Grant 638707 ``Black holes and their host galaxies:
co-evolution across cosmic time''. 
SS also acknowledges support from ERC Advanced Grant 320596 ``The Emergence of Structure During the Epoch of Reionization''.
This work
was performed on the following: DiRAC Darwin
Supercomputer hosted by the University of Cambridge
High Performance Computing Service
(http://www.hpc.cam.ac.uk/), provided by Dell Inc.
using Strategic Research Infrastructure Funding from
the Higher Education Funding Council for England and
funding from the Science and Technology Facilities Council; DiRAC Complexity
system, operated by the University of Leicester IT Services. This equipment is
funded by BIS 
National E-Infrastructure capital grant ST/K000373/1 and
STFC DiRAC Operations grant ST/K0003259/1; COSMA Data Centric system at Durham
University, operated by the Institute for Computational Cosmology on behalf of
the STFC DiRAC HPC Facility. This equipment was funded by a BIS National
E-infrastructure capital grant ST/K00042X/1, STFC capital grant ST/K00087X/1,
DiRAC Operations grant ST/K003267/1 and Durham University. DiRAC is
part of the National E-Infrastructure.
\vspace{-4ex}
\bibliographystyle{mn2e} 
\bibliography{mcs2017}
\appendix
\vspace{-4ex}
\section{Non-thermal pressure floor} \label{appendix_floor}
\begin{figure*}
\centering
\includegraphics{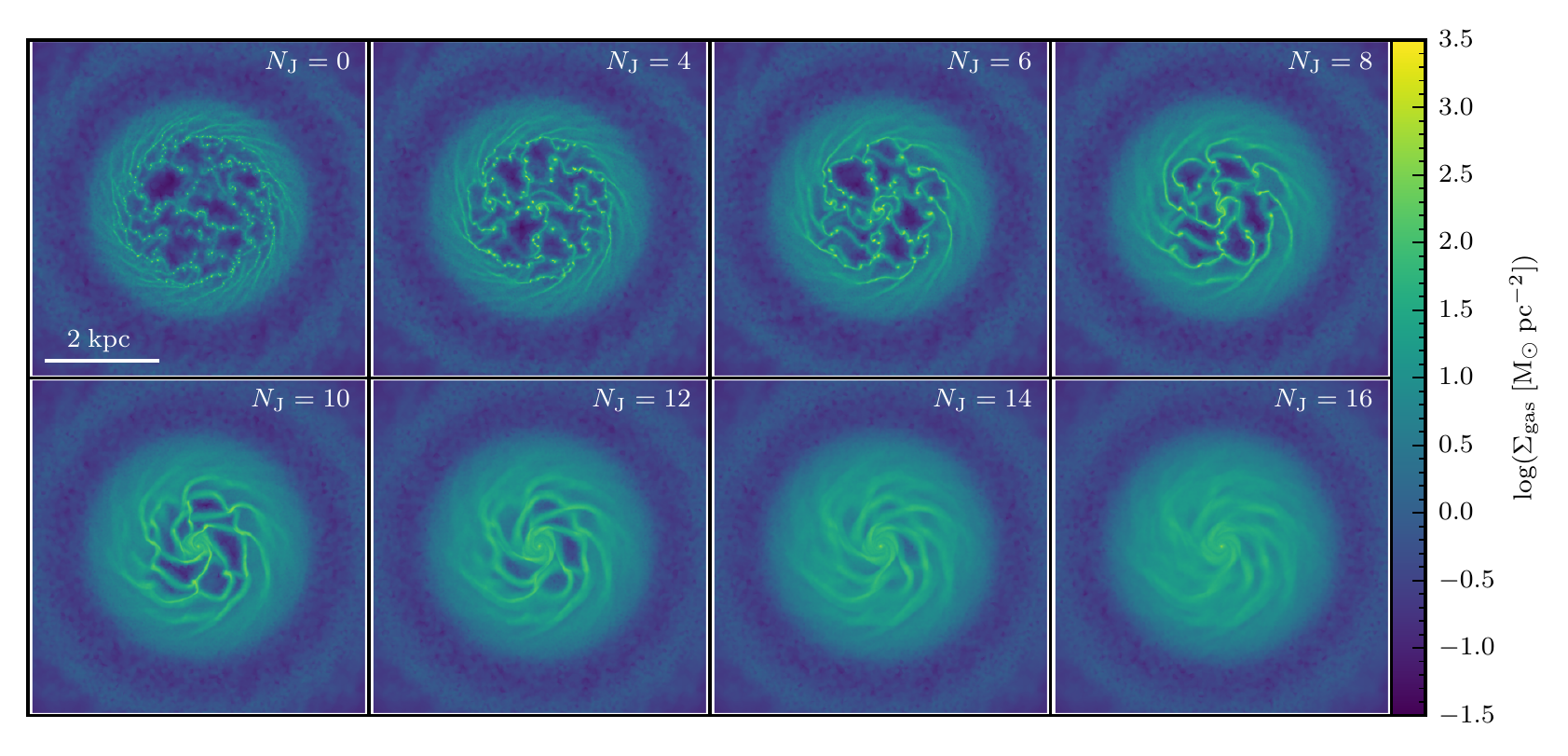}
\caption{Face-on gas density projections of simulations after $100$~Myr with
  varying values of $N_\mathrm{J}$ used to determine the artificial pressure
  floor, where 
  $N_\mathrm{J}$ is the number of cells by which the Jeans length must be
  resolved. Each simulation is carried out with no feedback and at
  $1000\ {\rm M_\odot}$ resolution using our fiducial galaxy model. While
  imposing no floor results in artificial fragmentation on the smallest
  scales (top left panel), the $N_\mathrm{J} = 16$ run (bottom right) washes out
  the physical structures present in the disk.} 
\label{J_comp_hr_proj} 
\end{figure*}

As described in Section~\ref{subsec_floor} we impose a non-thermal pressure
floor to avoid artificial fragmentation that may occur when the Jeans length
is not properly resolved. Setting a minimum pressure using equation~(\ref{pressure_floor})
ensures that the resulting Jeans length is always
resolved by at least $N_\mathrm{J}$ cells. \cite{Truelove1997} suggests that,
at a minimum, the Jeans length must be resolved by at least 4 cells. This
criteria is widely adopted in gravitational hydrodynamic simulations, but a
variety of values can be found in the literature. The choice of $N_\mathrm{J}$
is non-trivial: too low and artificial fragmentation will occur, too high and
the formation of small (physical) structures that would otherwise be resolved
is suppressed. Fig.~\ref{J_comp_hr_proj} shows the effect of various choices
of $N_\mathrm{J}$ on the morphology of our fiducial galaxy model at $100$~Myr
with no feedback with $1000\ \mathrm{M_\odot}$ resolution. It can be seen that with no pressure floor
($N_\mathrm{J}=0$) the disk fragments into multiple, small, high density
clumps. At the other extreme, enforcing resolution of the Jeans length by 16
cells ($N_\mathrm{J}=16$) results in the washing out of all small structure
save some weakly defined spiral arms. Other choices of $N_\mathrm{J}$ in
between these values results in a corresponding sliding scale of
morphologies. We can reasonably confidently assume that the adoption of
$N_\mathrm{J}=16$ results in a morphology that is
over-smoothed. Unfortunately, it is not so easy to say what the correct lower
limit of $N_\mathrm{J}$ is. Determining exactly when the onset of artificial
fragmentation occurs is a non-trivial problem that is beyond the scope of this
work.

\begin{figure}
\centering
\includegraphics{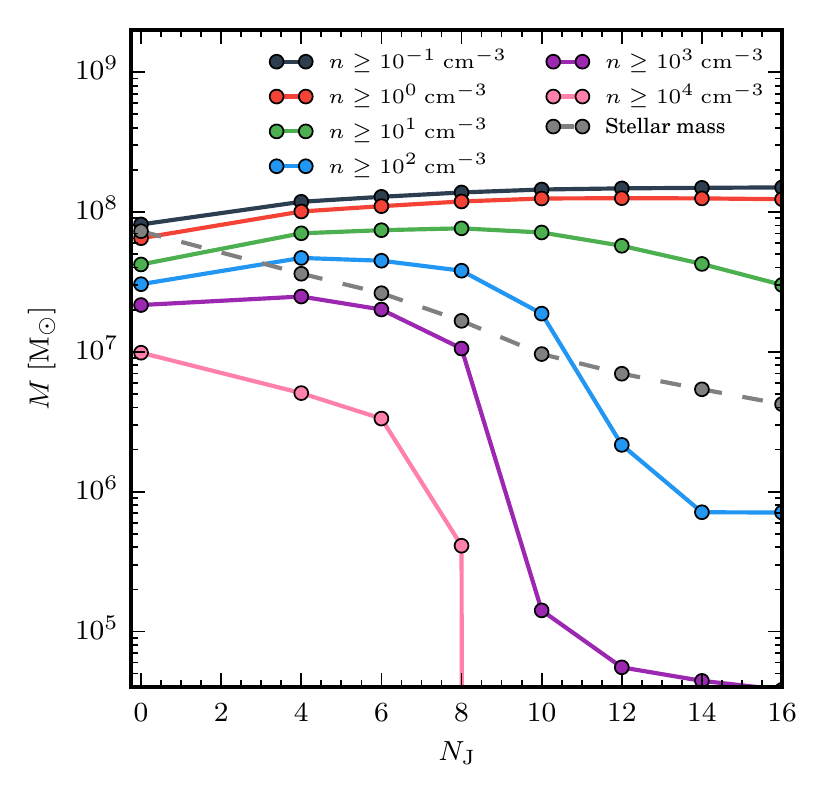}
\caption{The mass above a given density (or stellar mass, grey dashed curve)
  as function of $N_\mathrm{J}$ at $100$~Myr. Each value is measured from a
  simulation with no feedback at $1000\ {\rm M_\odot}$ resolution using our
  fiducial galaxy model. Note that there is a clear suppression of gas mass
  above a given density for densities above $100\ \mathrm{cm^{-3}}$ which
  motivates our choice of the reasonable $N_\mathrm{J}$ value to adopt, but
  for the stellar mass no such trend exists.} 
\label{J_comp_dens} 
\end{figure}

We attempt to crudely quantify the degree of fragmentation in
Fig.~\ref{J_comp_dens} by plotting the mass of gas above some density and the
mass of newly formed stars for our various choices of $N_\mathrm{J}$. Mass of
gas above $100\ \mathrm{cm^{-3}}$ decreases by well over an order of magnitude
across the range of $N_\mathrm{J}$ probed, with even more dramatic increases
when examining higher densities and, perhaps more worryingly, mass in new
stars. It can be seen that a definite transition occurs from a regime of
suppression of high density gas to a regime where gas may reach these
densities, though this transition happens at different values of
$N_\mathrm{J}$ depending on the density examined. For example, for densities
above $100\ \mathrm{cm^{-3}}$ the transition occurs at $N_\mathrm{J} \sim 10$,
while for densities above $10^3\ \mathrm{cm^{-3}}$ and
$10^4\ \mathrm{cm^{-3}}$ the transitions occurs at $N_\mathrm{J} \sim 8$ and $\sim 6$, respectively. This transition could
represent the transition between an artificially fragmenting regime to a
stabilised regime, but without a more careful determination of what
`artificial fragmentation' is, one could just as easily state that it merely
marks the transition from over-smoothing of structure to a properly resolved
regime. Hence, the choice of $N_\mathrm{J}$ becomes somewhat arbitrary, which
is certainly not ideal given the impact the choice has on the subsequent
evolution of the galaxy. With this in mind, we choice a fiducial value of
$N_\mathrm{J}=8$ for $1000\ {\rm M_\odot}$ resolution simulations. This lies
between the two extremes of small scale fragmentation and total suppression of
high density gas. It also produces a galaxy morphology similar to that found
in other works containing simulations of a similar type that use pressure
floors \citep[e.g.][]{Rosdahl2015, Rosdahl2017} though it should be noted that
these works adopt different values of $N_\mathrm{J}$ from us and from each
other. This is not a particularly satisfactory way of
choosing the strength of the pressure floor but it is necessary to allow the
comparison of our feedback models with those in other works.

\begin{figure*}
\centering
\includegraphics{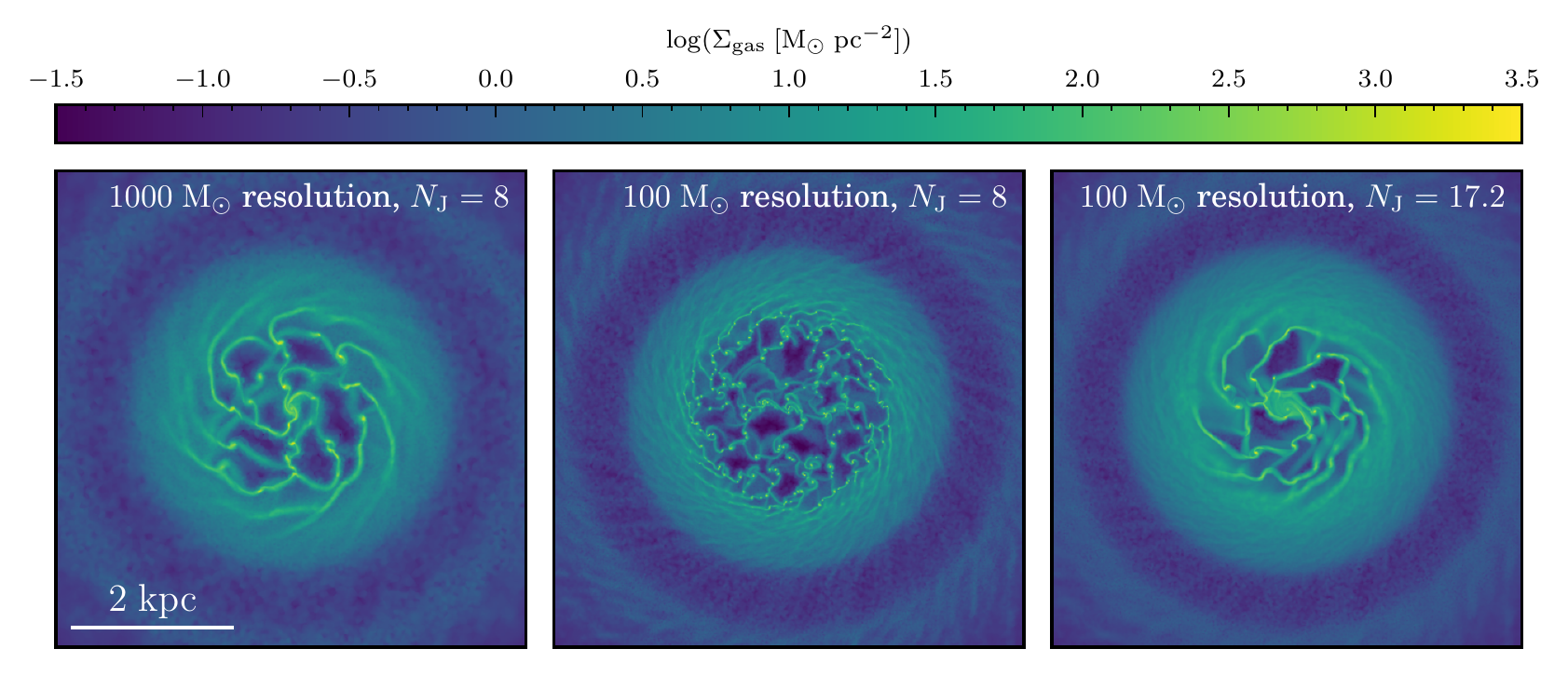}
\caption{Face-on gas density projections after 100 Myr comparing our fiducial
  choice of $N_\mathrm{J}=8$ for a $1000\ \mathrm{M}_{\rm \odot}$ resolution simulation
  with the same value for a $100\ {\rm M_\odot}$ resolution simulation (i.e. assuming that the `correct' choice for $N_\mathrm{J}$ is resolution independent) and $N_\mathrm{J}=17.2$ (i.e. such that the Jeans length is resolved by the same physical scale between resolutions). Each simulation is with no feedback using our fiducial galaxy model.}
\label{J_comp_res_proj}
\vspace{-3ex} 
\end{figure*}

If one assumes that there is a single `correct' value of $N_\mathrm{J}$ that
should be used to avoid artificial fragmentation, then it follows that
this value is resolution independent. In other words, there exists some
minimum required number of cells to correctly resolve fragmentation. Using a fixed
value of $N_\mathrm{J}$ then allows fragmentation to occur on smaller scales as the
minimum resolvable length decreases as resolution is increased. While this is often a desirable
behaviour, particularly when concerned with ISM properties on the edge of the resolution limit,
it necessarily leads to divergent galaxy properties.  Instead, scaling $N_\mathrm{J}$ such that the minimum
resolved Jeans length corresponds to the same physical scale at all resolutions
results in convergent morphologies. In Fig.~\ref{J_comp_res_proj}, we
compare the $1000\ {\rm M_\odot}$ resolution simulation with $N_\mathrm{J}=8$
(as in Fig.~\ref{J_comp_hr_proj}) with two simulations with a resolution of
$100\  {\rm M_\odot}$. One uses a value of $N_\mathrm{J}=8$, the other uses a
value of $N_\mathrm{J}=17.2$ (i.e. scaled with mass resolution such that the
minimum Jeans length is the same at both resolutions). It is clear that using
the same value of $N_\mathrm{J}$ results in very different morphologies,
whereas the adoption of a higher value with better resolution results in a
similar morphology.
This is not to say that the $1000\ {\rm M_\odot}$
resolution, $N_\mathrm{J}=8$ and $100\ {\rm M_\odot}$ resolution,
$N_\mathrm{J}=17.2$ results are more `correct' than the other, since as
previously mentioned, it is difficult to determine when artificial
fragmentation occurs. However, for the purposes of this paper where we are
primarily concerned about the effects of differing SNe feedback
implementations, particularly {\it examining the role of the resolution
  adopted on the feedback}, we find that it is advantageous to enforce approximately similar galaxy
evolutions in the no feedback case across our resolutions by scaling the value
of $N_\mathrm{J}$ with resolution as described in
Section~\ref{subsec_floor}. This we broadly achieve, with the no feedback
simulations at different resolutions producing same amount of stars within a
factor of a few and having comparable morphologies (see
Figs.~\ref{disk_projections_uhr}, \ref{disk_projections_lr} and
\ref{disk_projections_hr}). It is worth highlighting
that we are not unique in our choice to scale $N_\mathrm{J}$ with resolution \cite[see for
example][]{Rosdahl2015,Rosdahl2017}.

When we try to run simulations without a pressure floor but with feedback, the
effect is similar to increasing the star formation efficiency parameter (see
Fig.~\ref{disk_multi_sf}), with a sudden burst of high SFR followed by
extremely strong feedback that largely destroys the disk and quenches star
formation. It is possible that if we included other stellar feedback
mechanisms (stellar winds, radiation pressure, photoionisation) that are
active in the intervening time between star formation and SN occurrence it
might be possible to keep gas from entering the regime where it is vulnerable
to artificial fragmentation, thus removing the need for a pressure
floor. Alternatively, a more complex star formation criteria that identifies
fragmenting gas could also circumvent the issue (e.g. \cite{Hopkins2017a} argue
that the Jeans unstable gas should be turned into stars rather than using a
pressure floor), but if the Jeans length is significantly under-resolved this
could result in the spurious boosting of SFRs. We conclude
this section by remarking that, on the whole, when artificial pressure floors
are adopted, the motivation behind the choice of parameters is often not
clear. Given the strong dependence of results on this choice, we suggest that
this is an issue that needs to be addressed in more detail in future work.
\vspace{-12pt}
\section{SPH-like kernel weighting vs. explicitly isotropic weighting scheme for SNe feedback} \label{appendix_iso}
\begin{figure}
\centering
\includegraphics{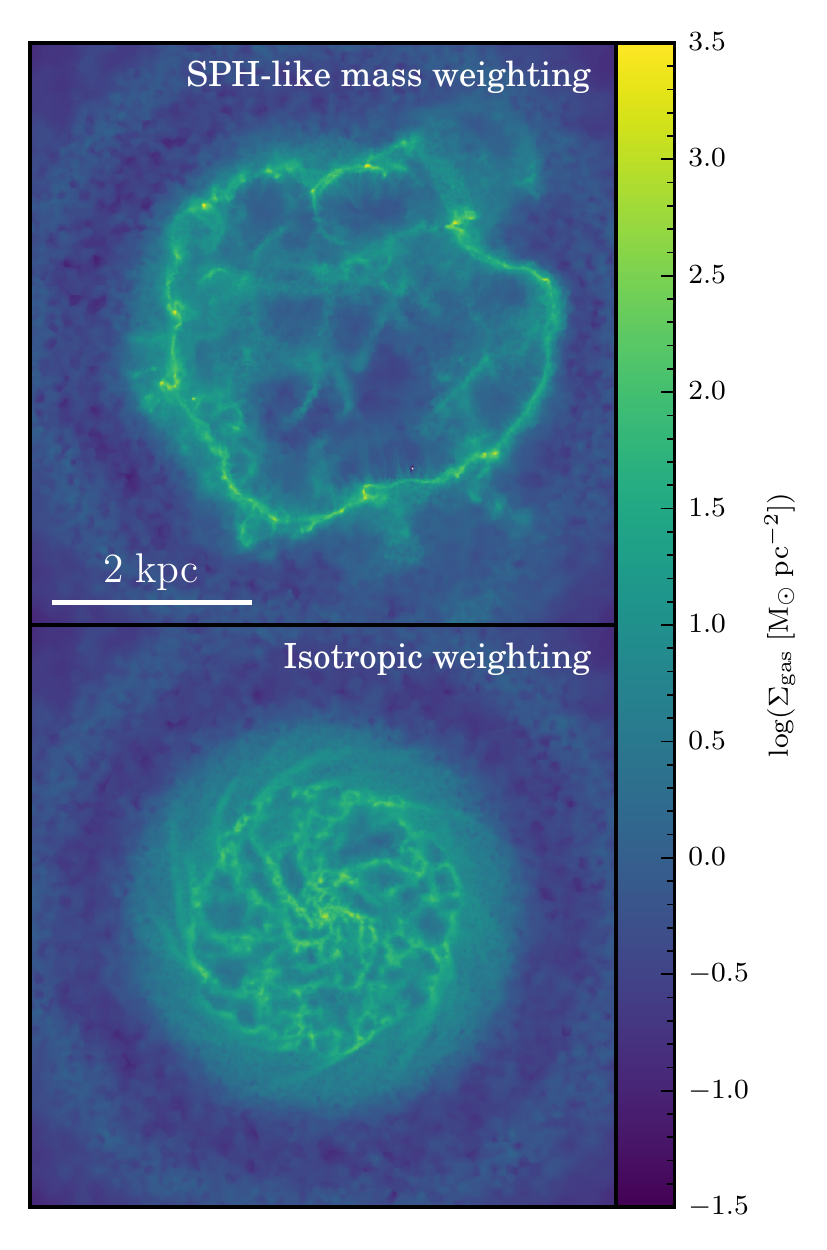}
\caption{Face-on gas density projections of simulations at $100$~Myr comparing
  the use of an SPH-like mass weighting scheme for distributing SNe mass,
  energy and momentum with our explicitly isotropic weighting scheme described
  in Section~\ref{SN}. The simulations have a resolution of $2000\ {\rm
    M_\odot}$, use our fiducial galaxy model with a modified form of our
  mechanical feedback (see main text). SPH-like weighting leads to
  unphysical shells propagating through the disk, sweeping up most of its mass.
Our isotropic weighting scheme avoids this numerical issue correctly coupling
the SN ejecta to the surrounding gas regardless of its density.}
\label{inject_comp}
\end{figure}
\renewcommand{\thefigure}{C\arabic{figure}} 
\setcounter{figure}{0} 
\begin{figure}
\centering
\includegraphics{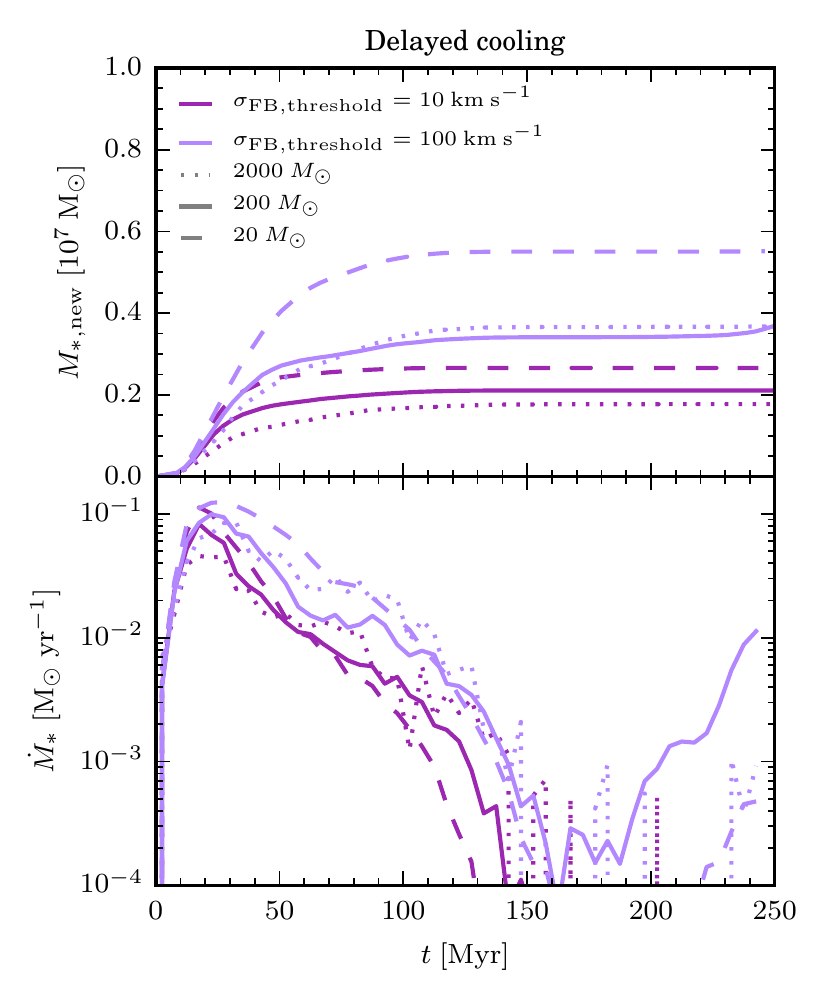}
\caption{Newly formed stellar mass (top) and SFRs (bottom) for
  simulations at all three resolutions for runs with delayed cooling using our
  fiducial threshold of $10\ \mathrm{km\ s^{-1}}$ and a higher value
  of $100\ \mathrm{km\ s^{-1}}$. The results are not very sensitive to
  the change of $\sigma_\mathrm{FB,threshold}$ and while the SFRs are suppressed somewhat
  less with the choice of the higher value, the disk is still largely
  disrupted after the first peak in SFR.} 
\label{disk_DFB_sfr}
\vspace{-3ex}
\end{figure}
As mentioned in Section~\ref{SN}, we have found that under certain conditions,
the use of a simple SPH-like kernel-based weighting scheme for distributing
feedback quantities (mass, metals, energy and momentum) into the gas local to
the SNe can result in significant violations of the desired isotropic
distribution. Because such a weighting scheme preferentially injects into
denser regions where there are more cells, if there is a strong density
gradient in a particular direction injection of feedback quantities will be
injected perpendicular to the gradient. For example, in the case of a SN
occurring in a thin disk, more resolution elements lie in the disk plane than
lie above and below it so feedback quantities will be preferentially injected
into the disk plane. The situation is exacerbated by poor resolution and by
the use of efficient momentum-based feedback schemes (thermal injection based
schemes can mitigate the situation slightly since heated cells will tend to
expand along the path of least resistance). This can result in the unphysical
driving of expanding shells through the disk plane, with little ejecta going
in the vertical direction \citep[see also][]{Hopkins2017a,Hopkins2018}.

Fig.~\ref{inject_comp} demonstrates this effect. We compare a simulation using
a standard SPH-kernel based mass weighting scheme for distributing feedback
quantities with our explicitly isotropic weighting scheme as described in
Section~\ref{SN}. For numerical reasons, we cannot use our full mechanical feedback scheme with an SPH-like scheme, so we use a hybrid of our kinetic and mechanical feedback schemes for this comparison; we inject $2.41\times10^5\ \mathrm{km\ s^{-1}}\ {\rm M_\odot}$ of
momentum per SNe, corresponding to the final momentum of a SN occurring in gas
of density $100\ \mathrm{cm^{-3}}$ and metallicity $0.1\ Z_{\rm \odot}$ as
calculated using equation~(\ref{p_fin}). The simulations are of our fiducial
galaxy at $2000\ {\rm M_\odot}$ resolution and the projections shown in
Fig.~\ref{inject_comp} are at $100$~Myr. The SPH-like weighting scheme sweeps
the disk mass into a thin expanding ring, whereas our isotropic scheme
prevents this occurring. This scenario is where the effect is most noticeable,
but it is still present to some extent at higher resolutions.  

It should be noticed that switching to a volume weighting scheme rather than
mass weighting does not have much of an effect. For a reasonable neighbour
number (32 - 64, as used with a cubic spline kernel) most identified
neighbours will be in the disk plane. If cells are found within the smoothing
length containing the neighbours that lie above the plane of the disk, the
extra weighting they will receive for being of a larger volume (because they
are less dense) is likely to be subdominant compared to the `penalty' they
receive for being furthest away from the star particle.

\section{Other delayed cooling parameters} \label{appendix_DFB}
As our fiducial parameters for the delayed cooling with fixed dissipation time
we have adopted $t_\mathrm{diss}=10\ \mathrm{Myr}$ and
$\sigma_\mathrm{FB}=10\ \mathrm{km\ s^{-1}}$, as used in
\cite{Teyssier2013}. As noted above, with our galaxy models at all resolutions
explored, this feedback scheme appears to be very strong relative to our other
schemes and produces unphysical results. We therefore tried a higher threshold
velocity dispersion, $\sigma_\mathrm{FB,threshold}=100\ \mathrm{km\ s^{-1}}$, as used in
\cite{Rosdahl2017}. Fig.~\ref{disk_DFB_sfr} shows the effect of using these
parameters on the star formation rate and stellar masses with our fiducial
galaxy model at all three of our resolutions. The SFRs are not suppressed to
the same degree with this weaker feedback, with final new stellar mass being
approximately a factor of $2$ larger at all resolutions. However, the feedback
still destroys the disk in the same manner as our fiducial simulations (though
gas returns to the centre, resulting in a second burst of star formation). As
mentioned above, with a more careful approach to tuning these parameters, we
could perhaps arrive at a less aggressive scheme.
\renewcommand{\thefigure}{D\arabic{figure}} 
\setcounter{figure}{0} 
\begin{figure*}
\centering
\includegraphics[width=\textwidth]{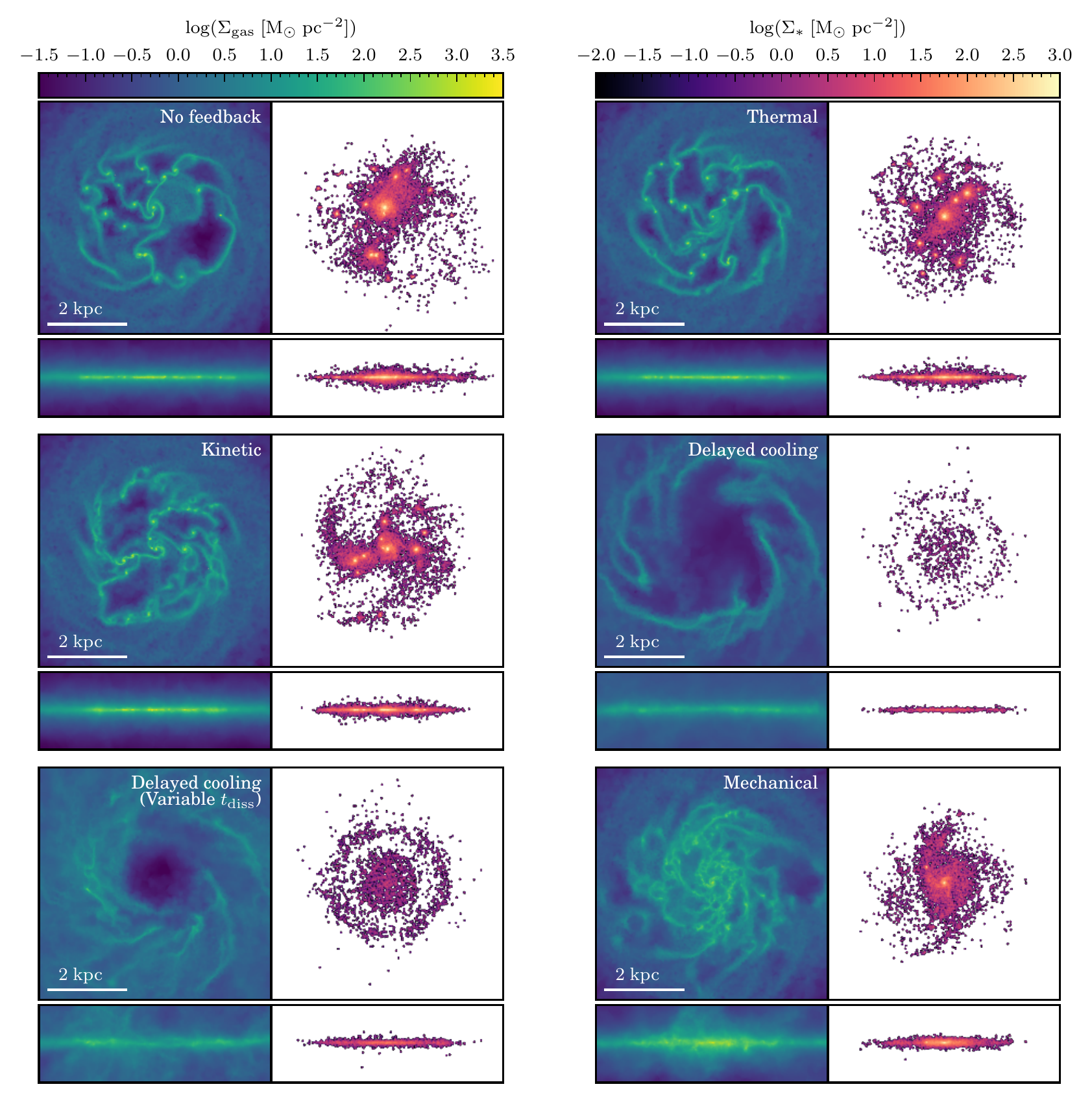}
\caption{Projections of gas and new stars formed for our different feedback
  simulation at $250$~Myr for $2000\, {\rm M_\odot}$ resolution runs. The mixed feedback simulation is not shown as the results
  are similar to the thermal and kinetic feedback simulations. The morphology
  of the no feedback simulation disk is similar to the highest resolution case (see
  Fig.~\ref{disk_projections_uhr}), with a highly clumpy distribution of gas and newly formed
  stars, although the structure is less well defined at this resolution (particularly in
  the stellar component). At this resolution, the classical feedback schemes overcool, so they
  produce similar morphologies to the no feedback simulation. Both delayed feedback schemes are
  too powerful, disrupting the gas disk and producing ring-like structures of newly formed stars.
  The mechanical feedback scheme is able to suppress the formation of dense clumps without destroying
  the disk. The resulting morphology is similar to the high resolution simulation, although it is not as
  well defined.}
\label{disk_projections_lr}
\vspace{-4ex} 
\end{figure*} 
\begin{figure*}
\centering
\includegraphics[width=\textwidth]{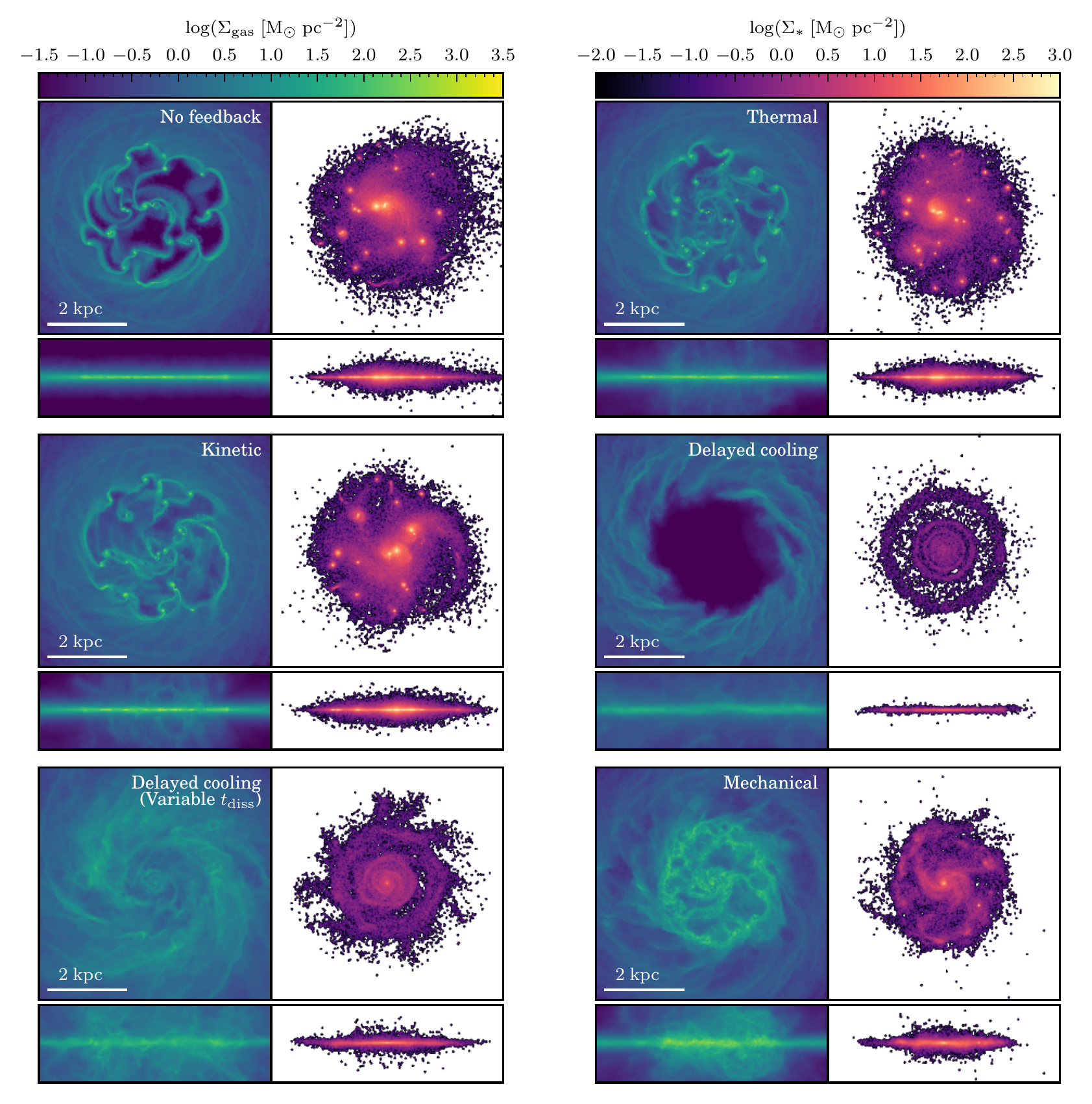}
\caption{Projections of gas and new stars formed for our different feedback
  simulation at $250$~Myr for $200\, {\rm M_\odot}$ resolution runs. The mixed feedback simulation is not shown as the results
  are similar to the thermal and kinetic feedback simulations. The morphology
  of the no feedback simulation disk is similar to the highest and lowest resolution cases (see
  Figs.~\ref{disk_projections_uhr} and~\ref{disk_projections_lr}). As in the lowest resolution simulation,
  the classical feedback schemes overcool and produce a similar clumped morphology to the no feedback simulation. The delayed feedback schemes remain
  too powerful, disrupting the gas disk, although the scheme with variable $t_\mathrm{diss}$ is weaker.
  The mechanical feedback scheme produces a similar morphology to the lower and higher resolutions simulations.}
\label{disk_projections_hr} 
\vspace{-4ex}
\end{figure*}
\begin{figure*}
\centering
$2000\ \mathrm{M_\odot}$ resolution\\
\includegraphics[width=0.8\textwidth]{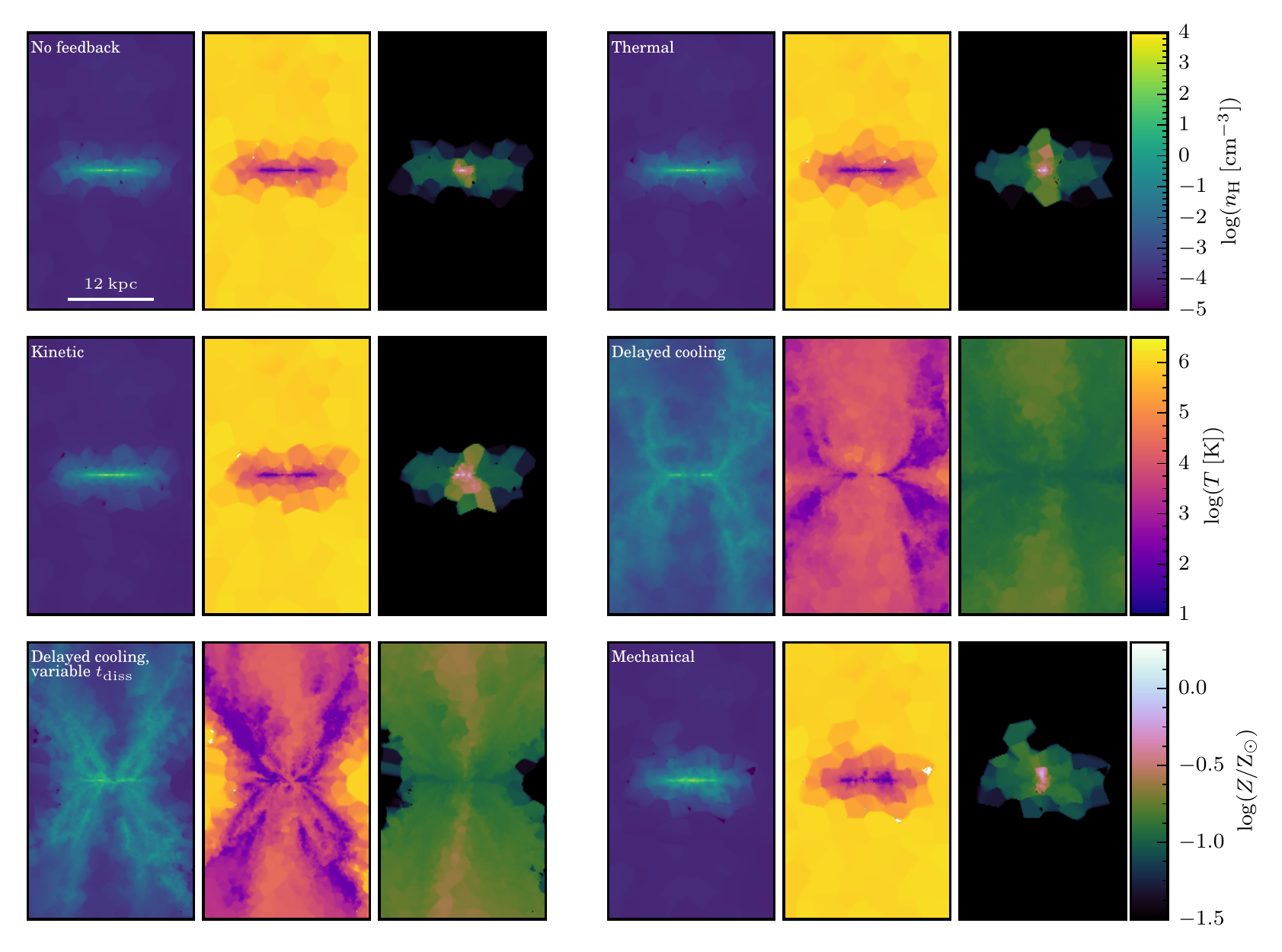}\\
$200\ \mathrm{M_\odot}$ resolution\\
\includegraphics[width=0.8\textwidth]{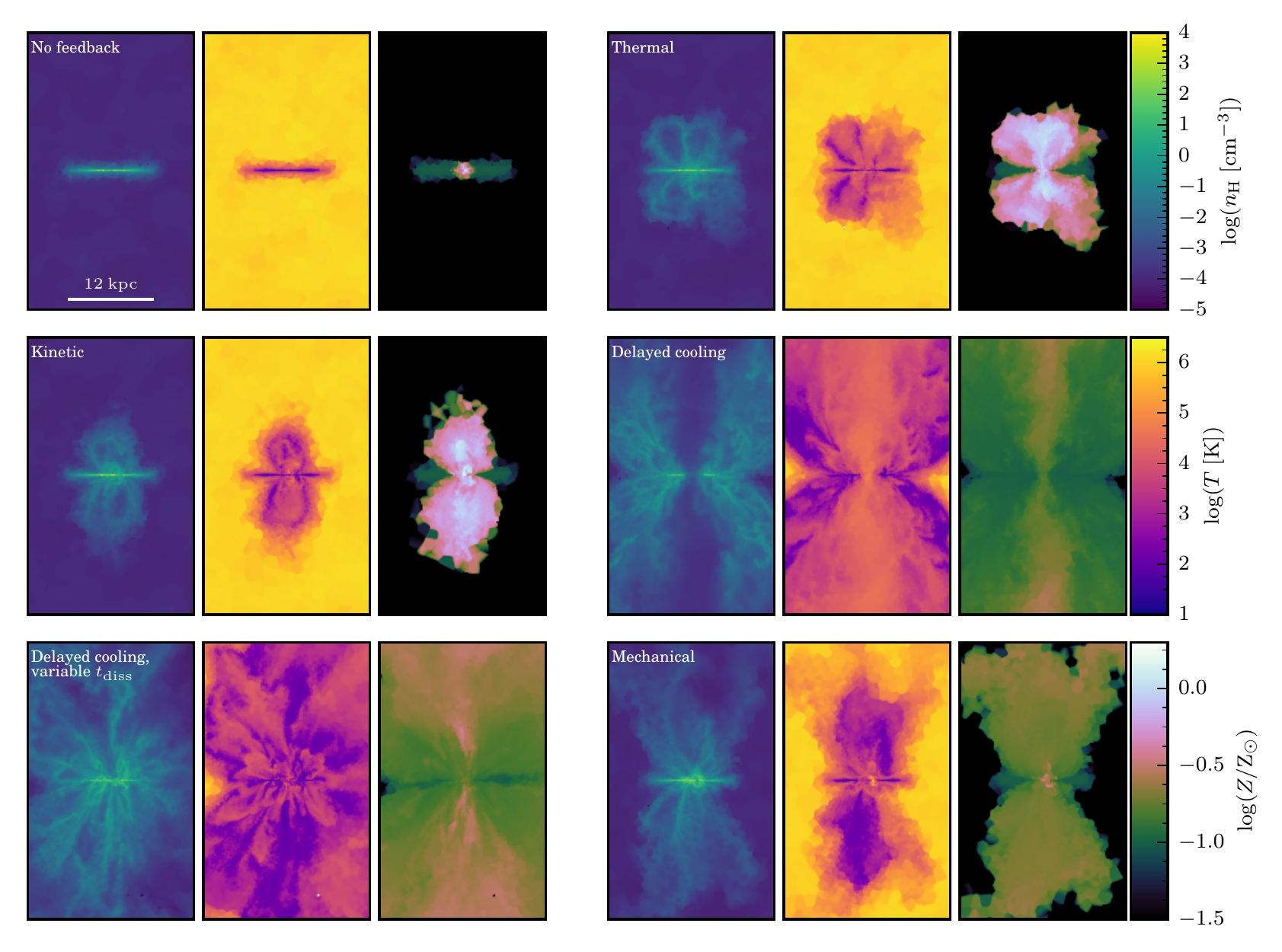}\\
\vspace{-3ex}
\caption{Density, temperature and metallicity slices at 250 Myr for $2000$ and $200\ \mathrm{M_\odot}$ 
resolution runs (top and bottom respectively). The mixed feedback simulation is not shown as the results
  are similar to the thermal and kinetic feedback simulations.
  Unsurprisingly, the no feedback simulations do not produce outflows.  
  Only the delayed cooling schemes drive outflows at both resolutions, ejecting the majority of material
  from the centre of the system.
  At the $200\ \mathrm{M_\odot}$ resolution, the classical feedback schemes produce outflows that reach a short distance above the disk. They are highly metal enriched because of the large number of SNe driving the outflows, due to
  the inefficiency of the feedback caused by overcooling. The mechanical feedback
  is able to drive a modest outflow at the $200\ \mathrm{M_\odot}$ resolution, though the outflow peaked $\sim100\ \mathrm{Myr}$ previously resulting in material returning to the disk in a galactic
  fountain (see Figs.~\ref{vel_slice} and \ref{massload}).}
\label{disk_slice_lr} 
\end{figure*}
\vspace{-4ex}
\section{Other resolutions} \label{appendix_other}
This appendix contains results for our lower resolution simulations for
comparison to the figures in the main text that show our highest resolution
simulations. Figs.~\ref{disk_projections_lr} and \ref{disk_projections_hr}
show face-on and edge-on density projections of gas and newly formed stars
after $250$~Myr (see Fig.~\ref{disk_projections_uhr} for the highest
resolution case). Fig.~\ref{disk_slice_lr} shows density, temperature and metallicity slices at
$250$~Myr to demonstrate how outflow properties change with resolution (see
Fig.~\ref{disk_slice_uhr} for the highest resolution case).
\end{document}